\documentclass[11pt,final]{article}

\usepackage[utf8]{inputenc}
\usepackage[T1]{fontenc}
\usepackage[english]{babel}
\usepackage{lmodern}
\usepackage{showlabels,soul}
\usepackage{xcolor}

\usepackage[bookmarks=false,colorlinks=False]{hyperref} 

\usepackage{amsmath} 
\usepackage{amsfonts,dsfont}
\usepackage{amssymb}
\usepackage{amsthm}
\usepackage{mathtools}
\usepackage{thmtools}
\usepackage{bm} 

\numberwithin{equation}{section} 

\usepackage{graphicx}
\usepackage{subcaption}
\usepackage{rotating}
\usepackage{multirow}

\newlength{\bredde}
\def\slash#1{\settowidth{\bredde}{$#1$}\ifmmode\,\raisebox{.15ex}{/}
\hspace*{-\bredde} #1\else$\,\raisebox{.15ex}{/}\hspace*{-\bredde} #1$\fi}
\DeclarePairedDelimiter{\floor}{\lfloor}{\rfloor} 

\renewcommand{\epsilon}{\varepsilon} 
\let\Re\relax\DeclareMathOperator{\Re}{Re} 

\newcommand*{\euler}{\mathrm{e}} 
\newcommand*{\iunit}{\mathrm{i}} 
\newcommand*{\dif}{\mathop{}\!\mathrm{d}} 

\newcommand*{\Cset}{\mathds{C}} 

\newcommand*{\detjk}{\underset{1\leq j,k\leq N}{\det}}
\newcommand*{\detjkTo}[1]{\underset{1\leq j,k\leq {#1}}{\det}}

\newcommand{\be}{\begin{equation}}
\newcommand{\ee}{\end{equation}}

\newcommand{\Tr}{{\rm Tr}}

\newcommand*{\esum}[2]{\scalebox{1.2}{e}_{#1}(#2)}

\usepackage{extpfeil}
\newextarrow{\xrightrightarrows}{{5}{8}{0}{0}}
{\bigRelbar\bigRelbar{\bigtwoarrowsleft\rightarrow\rightarrow}}

\DeclareMathOperator{\heaviside}{\Theta} 
\DeclareMathOperator{\erf}{erf} 
\DeclareMathOperator{\erfc}{erfc} 
\newcommand*{\cconj}[1]{\overline{#1}} 
\newcommand*{\sfrac}[2]{{#1}/{#2}} 

\DeclareMathOperator{\jpdf}{P} 

\makeatletter
\newcommand*{\doublerightarrow}[2]{\mathrel{
 \settowidth{\@tempdima}{$\scriptstyle#1$}
 \settowidth{\@tempdimb}{$\scriptstyle#2$}
 \ifdim\@tempdimb>\@tempdima \@tempdima=\@tempdimb\fi
 \mathop{\vcenter{
 \offinterlineskip\ialign{\hbox to\dimexpr\@tempdima+1em{##}\cr
 \rightarrowfill\cr\noalign{\kern.5ex}
 \rightarrowfill\cr}}}\limits^{\!#1}_{\!#2}}}
\makeatother

\begin{document}
\title{Three non-Hermitian random matrix universality classes of complex edge statistics: Spacing ratios and distributions
}
\author{{\sc Gernot Akemann\footnote{corresponding author: akemann@physik.uni-bielefeld.de}, Georg Angermann, Noah Ayg\"un, Adam Mielke,}\\ 
 {\sc Patricia P\"a{\ss}ler, Christoph Raitzig, and Tobias Winkler}\\~\\
Faculty of Physics, Bielefeld University, PO-Box 100131,\\ D-33501 Bielefeld, Germany
}

\date{}

\maketitle

\begin{abstract}
The conjectured three generic local bulk statistics amongst all non-Hermitian random matrix symmetry classes have recently been extended to three generic local edge statistics.  
We study analytically and numerically complex spacing ratios and nearest-neighbour (NN) spacing distributions that characterise such local statistics.  
We choose the three simplest representatives of these universality classes, given by the Gaussian ensembles of complex Ginibre, complex symmetric and complex self-dual matrices, denoted by class A, AI$^\dag$ and AII$^\dag$.
In the first part, we analytically study the complex spacing ratio in class A, at finite matrix size $N$. Introducing a conditional point process,
we simplify existing expressions and show why an uncontrolled approximation introduced earlier converges well in the large-$N$ limit in the bulk. 
When specifying to the elliptic Ginibre ensemble, we present a parameter-dependent $N=3$ surmise for the complex spacing ratio,  interpolating to that of the Gaussian unitary ensemble (GUE), where such a surmise is very accurate.
In the second numerical part, we compare complex spacing ratios, its moments, and NN spacing distributions for all three ensembles with that of uncorrelated points, the two-dimensional (2D) Poisson process, both in the bulk and at the edge.
The varying degree of repulsion within these different edge universality classes can be well understood in terms of an effective 2D Coulomb gas description, at different values of inverse temperature $\beta$. We find indications that the complex spacing ratio does not fully unfold the local statistics at the edge.
Finally we verify that for small argument, in all three symmetry classes the NN spacing distributions in the bulk and at the edge are consistent with the universal cubic repulsion.

\end{abstract}
\begin{center}
{\it Dedicated to the memory of Santosh Kumar}
\end{center}

\section{Introduction}\label{intro}

The application of non-Hermitian random matrix theory (RMT) to physics is a very active area of research. This was initiated by the generalisation of the quantum chaos conjecture to dissipative open quantum chaotic system{s} \cite{GHS}, and nowadays goes much beyond this area of research, see \cite{Ueda} for a review. Whilst in many applications the description of data by RMT is heuristic, there are examples for a precise map to an RMT description, where we will focus on non-Hermitian examples only.  Such a link exists for non-interacting Fermions in 2D in a rotating trap, or equivalently perpendicular magnetic field, where the ground state properties can be mapped to the complex Ginibre ensemble (GinUE) \cite{PdF,Forrester-Jankovici,BSG}. Extensions beyond the lowest Landau levels or zero temperature exist, going beyond standard RMT classes, cf. \cite{Kulkarni} and references therein. A second example relates non-Hermitian RMT  to effective field theory of Quantum Chromodynamics, with finite chemical potential \cite{James} or finite lattice spacing \cite{DSV}. The corresponding symmetry classes are the chiral partners of the complex elliptic Ginibre ensemble \cite{James}, and the Wilson--Dirac class \cite{MarioJac}. 

The application to dissipative open quantum chaotic systems \cite{GHS} has seen a strong revival recently, with many authors contributing, including very prominently Santosh Kumar and his group. These works include the dissipative spectral form factor \cite{LiProsenChan,SKSK}, the number variance \cite{CP}, the complex spacing ratio \cite{Eva,SSK,Lea} introduced in \cite{SaRibeiroProsen}, and the NN spacing distribution in radial distance in the complex plane
\cite{ABCPRS} as tools, with the latter two being local. They are applied to physical systems such as the kicked top or rotor \cite{GHS,CP,Eva,Lea} and open dissipative quantum spin chains \cite{AKMP,Hamazaki et al} to name few examples.
The main focus is on the transition between integrable and chaotic open quantum systems, characterised by 2D Poisson respectively random matrix statistics in the corresponding symmetry class---the content of the quantum chaos conjecture.

The study of symmetry classes in non-Hermitian RMT has been revisited recently. In 2019 Kawabata et al. \cite{Kawabata} reanalysed the classification of Bernard--LeClair \cite{Bernard LeClair} and Magnea \cite{Magnea} from 2002, finding a total number of 38 symmetry classes.\footnote{A revised version of \cite{Bernard LeClair} was published after the work \cite{Kawabata} appeared, it agrees with this total number of classes.} Shortly thereafter, it was proposed by Hamazaki et al. \cite{Hamazaki et al}, based on numerics of the NN spacing distribution and heuristic arguments, 
that only 3 generic local bulk universality classes exist amongst all symmetry classes. This conjecture was later extended to the spectral edge \cite{AAKP}.
For comparison, in the case of Hermitian random matrices, where 10 symmetry classes are identified \cite{Altland Zirnbauer}, only three different local bulk and three different local edge statistics are known which are generic. These can be represented by the three Dyson-ensembles \cite{Dyson}. 
In the non-Hermitian situation, the three simplest representatives are the classes A (complex Ginibre), AI$^\dagger$ (complex symmetric) and AII$^\dagger$ (complex self-dual) which are all Gaussian, as put forward by \cite{Hamazaki et al}. 
Much earlier, it has been conjectured on simple arguments from level crossings in quantum mechanical systems,  that in the small argument expansion of the different NN spacing distributions
all show a cubic vanishing that is universal \cite{Grobe Haake,Haake2010}.

What is known analytically about the NN spacing distribution and the complex spacing ratio in the three symmetry classes A, AI$^\dag$ and AII$^\dag$? The complex eigenvalues in class A represent a determinantal point process, which is well described by the universal exponential Ginibre kernel in the bulk, and the complementary error function kernel (also called Faddeeva plasma kernel) at the edge of the spectrum, both known at finite $N$ and in the limit of infinite matrix dimension $N\to\infty$.  Closed form expressions exist for the gap probabilities in terms of the Fredholm determinant of these two kernels, which are most explicitly known in the bulk \cite{GHS}. This leads to explicit expressions for the NN and next-to-nearest neighbour (NNN) spacing distribution \cite{GHS}. 
For a recent review on the Ginibre and related ensembles we refer to \cite{PeterSungsoo}.
In contrast, even in class A only approximate expressions exist for the complex spacing ratio \cite{SaRibeiroProsen,DusaWettig}.
They include an $N$-dependent determinantal expression and a surmise at $N=3$, which is not very accurate. This is in stark contrast to the spacing ratio in Hermitian RMT introduced by \cite{OH,Atas_et_al} in the bulk of the real spectrum, which is very well approximated by $N=3$. The idea behind the spacing ratio is that unfolding (see e.g. \cite{Haake2010}), that is necessary to compare local statistics from data to RMT, drops out in taking the ratio of the difference to the NN and NNN eigenvalue. Exact analytic limiting results are scarce for the spacing ratio, see however \cite{Nishigaki,Buijsman}. Higher order spacing ratios have been considered \cite{Bhosale,Ruth}, as well as transitions between different symmetry classes \cite{KKSG},  and more generally $\beta$-ensembles of a 1D Dyson gas \cite{Atas2}.

In classes AI$^\dag$ and AII$^\dag$ no such determinantal expressions are known, or even expected, that would at least in principle allow to formulate NN spacing statistics in terms of Fredholm determinants or Pfaffians. Approximate descriptions of the NN and NNN spacing distribution have been found in terms of a 2D Coulomb gas at specific values of an effective $\beta=1.4$ and 2.6, respectively \cite{ABCPRS, AMP}. Very recently, first results for the spectral density at finite-$N$ and its local limit at the spectral edge have been derived in class AI$^\dag$ \cite{AFS}. They are based on the Kac--Rice formalism \cite{SommersIida}, established rigorously in this context recently \cite{Yan25}. For class AII$^\dag$, based on universality arguments the limiting density at the edge is conjectured \cite{OK}. Complementary results exist for expectation values of characteristic polynomials in all three classes, cf. \cite{AAKP,AkemannVernizzi2003,Liu Zhang,Kulkarni Kawabata Ryu,Forrester24}.

It is the goal of this article to increase our analytical and numerical understanding of the NN, NNN spacing distribution and complex spacing ratio in all 3 symmetry classes. Our comparison includes 2D Poisson statistics of independent points in the complex plane, where limiting closed form expression exist for the NN, NNN spacing distribution and complex spacing ratio in the bulk of the spectrum \cite{SaRibeiroProsen,Haake2010}.
We will focus on the edge of the spectrum here, which has been less explored so far.

{This article contains two parts. In the first analytical part, for class A we present results for the complex spacing ratio at finite-$N$ including an improved surmise in Section \ref{Sec:AnalyticsA}, and the NN spacing distribution at the edge including its cubic behaviour at small argument in Appendix \ref{App:kernel}. Both are compared to our numerical findings in the second numerical part Section \ref{Sec:Numerics}, which covers all three symmetry classes for the complex spacing ratio, NN and NNN spacings distribution at the edge.}

{In more detail, in Section \ref{Sec:AnalyticsA}  
analytical expressions are derived} for the complex spacing ratio at finite-$N$ within class A, including the parameter-dependent elliptic or Ginibre--Girko ensemble (eGinUE) \cite{Girko,Sommers et al}. Following the idea of Grobe, Haake and Sommers \cite{GHS},  we consider the point process with one eigenvalue conditioned at the origin for general $N$ in Subsection \ref{sec:generalN}. 
Together with the known translational invariance of the GinUE in the large-$N$ bulk limit, this explains why the approximation proposed in \cite{SaRibeiroProsen,DusaWettig} converges well. 
In Subsection \ref{sec:surmise} we show that the surmise at $N=3$ in this conditional setup agrees with the leading order large-$N$ expansion of the unconditional ensemble, {thus providing a qualitative improvement, as we will see in comparison to the numerics in the next Section \ref{Sec:Numerics}.}  
For the eGinUE, in Subsection \ref{sec:comparisonGUE} the surmise is shown to interpolate to the GUE in the Hermitian limit  $\tau\nearrow 1$, where such a surmise is a very accurate. 
{Appendix \ref{App:kernel} concludes our analytical results for class A. Here, we expand the Fredholm determinant of the conditional kernel at the edge of the spectrum at small argument, to show analytically that it vanishes as a cubic. This behaviour is confirmed numerically at the end of the next Section \ref{Sec:Numerics}.}

In Section \ref{Sec:Numerics} we present numerical results at the spectral edge and compare to the bulk for all 3 symmetry classes A, AI$^\dag$ and AII$^\dag$. Starting with the definition of these ensembles, we first focus on the complex spacing ratio in Subsection \ref{sec:spacing-ratio2D} and moments thereof in Subsection \ref{sec:spacing-ratio-mom}. We discuss the effect and extension of the spectral edge, including a comparison to the 2D Poisson point process of independent points with Gaussian distribution, as a function of the effective repulsion $\beta$ {found in classes AI$^\dag$ and AII$^\dag$ \cite{ABCPRS, AMP}.} 
The NN and NNN spacing distributions are compared to {the} numerics {for all 3 classes} in Subsection \ref{sec:NN}. After establishing an unfolding procedure at the edge of the spectrum in Subsection \ref{App:unfold}, where the mean density varies on the scale of the mean level spacing, 
{we compare the NN and NNN spacing at the edge and in the bulk, finding distinct statistics for all 3 classes, respectively. In contrast, }for 2D Poisson bulk and edge statistics agree after unfolding. 
In Subsection \ref{sec:NNcubic} we address the question of the small argument expansion of the NN spacing, which we confirm to be cubic for all 3 ensembles, in the bulk and at the edge.
We close with conclusions and open problems in Section \ref{Sec:conclusion}.

\section{Analytic results on complex spacing ratios in class A} \label{Sec:AnalyticsA}
We consider the distribution of the complex spacing ratio between NN and NNN, as introduced in \cite{SaRibeiroProsen}, with respect to a general joint eigenvalue distribution function (jpdf) $\jpdf^{(N)}(z_1,\dots,z_N)$. Since the latter is invariant under reordering of the eigenvalues $z_j$, without loss of generality we can choose $z_1$ as a reference point, with NN $z_2$ and NNN $z_3$. Following \cite{SaRibeiroProsen}, the complex spacing ratio is defined as
\begin{eqnarray}
    \varrho^{(N)}(z) &=& \frac{1}{\alpha_{N}} \int_{\mathbb{C}^N} \dif^2z_1 \dots \dif^2 z_N \jpdf^{(N)}(z_1,\dots,z_N) \delta^{(2)}\left(z-\frac{z_2-z_1}{z_3-z_1}\right) \heaviside\left(|z_3-z_1|^2 -|z_2-z_1|^2 \right) \nonumber \\ &&\quad\quad\quad\times \prod_{j=4}^{N} \heaviside\left(|z_j-z_1|^2 -|z_3-z_1|^2 \right) ,
 \label{def.dist_ofSpacingRatio}
\end{eqnarray}
for a general ensemble with point process $\jpdf^{(N)}(z_1,\dots,z_N)$. This includes in principle the eigenvalues of the symmetry classes of complex symmetric (AI$^\dag$) or complex self-dual (AII$^\dag$) random matrices, viewed as point processes. These ensembles will be defined in Section \ref{Sec:Numerics} on a matrix level. Their jpdfs are unknown to date.
The Heaviside function $\heaviside(z)$ in the first line of \eqref{def.dist_ofSpacingRatio} enforces that 
$z_2$ respectively $z_3$ are NN respectively NNN of $z_1$. 
The remaining Heaviside functions ensure that all other points are further away. Notice that NN and NNN need not be unique in the complex plane, e.g. when two (or more) points share the same radial distance at a different angle. In the following we neglect this event, which has measure zero.

The normalisation constant\footnote{Notice that in \cite{SaRibeiroProsen} no normalisation was introduced in \eqref{def.dist_ofSpacingRatio}.} $\alpha_{N}$  ensures that 
\begin{align}
\int_{\mathbb{C}} \dif^2z \varrho^{(N)}(z) =1\ .
\label{eq.norm_alpha}
\end{align}
Based on the invariance of $\jpdf^{(N)}(z_1,\dots,z_N)$ with respect to reordering of the eigenvalues $z_j$, the normalisation is given by  $\alpha_{N}=(N-1)(N-2)=2\binom{N-1}{2}$, which is easy to check for $N=3$. By definition, the complex spacing ratio $\varrho^{(N)}(z)$ is only supported on the unit disc, as $|z_2-z_1|\leq|z_3-z_1|$ in \eqref{def.dist_ofSpacingRatio}.

After shifting the variables $z_j \rightarrow z_j+z_1$, for $j=2,\dots,N$, and evaluating the $z_2$-integral using the elementary property of the two-dimensional Dirac delta $ \delta^{(2)}(z=x+iy)=\delta(x)\delta(y)$,
\begin{equation}
    \delta^{(2)}\left(z-\frac{z_2}{z_3}\right) = |z_3|^2 \delta^{(2)}\left(zz_3 -z_2  \right) , 
    \label{eq.delta-z_3}
\end{equation}
we obtain  
\begin{align}
    \varrho^{(N)}(z) = \frac{\heaviside\left(1 -|z|^2 \right) }{\alpha_{N}} \int_{\mathbb{C}^{N-1}} \dif^2z_1\dif^2z_3 \dots \dif^2 z_N \; |z_3|^2 \prod_{j=4}^{N} \heaviside\left(|z_j|^2 -|z_3|^2 \right) \nonumber \\ \times \jpdf^{(N)}(z_1,z_1+z z_3,z_1+z_3,\dots,z_1+z_N)  .
 \label{eq.dist_ofSpacingRatio-z_2}
\end{align}
So far we have merely rederived the expression for the complex spacing ratio in \cite[eq. (C.4)]{SaRibeiroProsen}.

In this section we will study jpdfs of eigenvalues of the form
\begin{equation}
    \jpdf^{(N)}_{\rm A}(z_1,\dots,z_N) = \frac{1}{Z_N} |\Delta_N(z_1,\dots, z_N)|^\beta \prod_{j=1}^{N} \omega(z_j),\quad \beta=2,
    \label{eq.jpdf_gen.weight}
\end{equation}
which defines class A, a 2D Coulomb gas at inverse temperature $\beta=2$. Here, we allow for  a general weight function $\omega(z)$ on $\mathbb{C}$, that typically confines the complex eigenvalues to a compact support at large-$N$, after appropriate rescaling. 
We denote by 
\begin{equation}
    \Delta_N(z_1,\dots, z_N) = 
    \detjk \left[ z_{j}^{k-1} \right] = \prod_{1\leq j<k \leq N} (z_k - z_j)
    \label{def.vandermonde}
\end{equation}
the Vandermonde determinant, and $Z_N$ is the normalisation constant.

The jpdf eq.~\eqref{eq.jpdf_gen.weight} includes, in particular, the case of the GinUE with weight function $\omega_{\rm A}(z)=\exp\left[ - |z|^2 \right]$. It was introduced by Ginibre~\cite{Ginibre} and consists of $N \times N$ random matrices $J$ with independent and identically distributed complex normal matrix elements $J_{jk} \sim \mathcal{N}_{\mathbb{C}}(0,1)$, for $j,k=1,\ldots,N$, cf. Section \ref{Sec:Numerics}.
Its $N$ complex eigenvalues represent a one-component 2D Coulomb gas at inverse temperature $\beta=2$, with $\omega_{\rm A}(z)$ in \eqref{eq.jpdf_gen.weight}.

A one parameter extension, the eGinUE or Ginibre--Girko ensemble, depending on $\tau\in[0,1)$, interpolates between the GinUE at $\tau=0$ and the Hermitian GUE at $\tau \nearrow 1$. This interpolating ensemble was introduced independently in \cite{Girko,Sommers et al} and represents a 2D Coulomb gas in a quadrupolar magnetic field \cite{PdF,Forrester-Jankovici}.
Its jpdf of eigenvalues is given by eq.~\eqref{eq.jpdf_gen.weight}, with weight function 
\begin{equation}
    \omega_{\rm A}(z; \tau) = \exp\left[
    -\frac{1}{1-\tau^2}\left( |z|^2 -\frac{\tau}{2}(z^2+\cconj{z}^2) \right)\right] ,\quad \tau \in [0,1).
    \label{eq.weight-eGinUE}
\end{equation}

Motivated by the well-known translation invariance in the large-$N$ limit in the bulk of the spectrum in class $A$, we follow the idea of \cite{GHS} in computing the NN and NNN spacing distribution in the bulk from a point at the origin, $z_1=0$. Thus,  
we introduce the {\it conditional complex spacing ratio}
\begin{align}
    \varrho^{(N)}_{\rm C}(z) = \int_{\mathbb{C}^{N-1}}\frac{\dif^2z_2 \dots \dif^2 z_N}{\alpha_{N, {\rm C}}}\jpdf^{(N)}(0,z_2,\dots,z_N) \delta^{(2)}\left(z-\frac{z_2}{z_3}\right) \heaviside\left(|z_3|^2 -|z_2|^2 \right)
    \prod_{j=4}^{N} \heaviside\left(|z_j|^2 -|z_3|^2 \right) .
 \label{def.dist_ofCondSpacingRatio}
\end{align}
It is defined by the conditioning (labelled by C) that the reference eigenvalue $z_1=0$ is fixed. Given that in the ensemble we consider the origin represents a generic bulk point, the expressions \eqref{def.dist_ofSpacingRatio} and \eqref{def.dist_ofCondSpacingRatio} should agree in the large-$N$ limit. As we will show now the latter is much simpler at finite-$N$.

\subsection{Results for general matrix size $N$}\label{sec:generalN}
The purpose of this subsection is twofold. First, we show how for the eGinUE the conditioning at the origin leads to a simplification, compared to the unconditional case at finite-$N$. Second, we will rederive the five-diagonal determinantal expression of \cite{DusaWettig}, using the Andr\'eief formula instead of the supersymmetric method. Our proof holds for a general rotationally invariant weight function $\omega(z)=\omega(|z|)$. This formula is then suitable for a numerical evaluation, as it was shown in \cite{DusaWettig}.

For the eGinUE, with weight eq.~\eqref{eq.weight-eGinUE}, the conditional complex spacing ratio \eqref{def.dist_ofCondSpacingRatio} reads
\begin{eqnarray}
&&\varrho^{(N)}_{\rm A,C} (z;\tau) = \frac{\heaviside\left(1-|z|^2\right) |z|^2|1-z|^2 }{\alpha_{N,{\rm C}} Z_{N,{\rm C}}(\tau)} \int_{\mathbb{C}} \dif^2 z_3 \;  |z_3|^8  \euler^{ - \frac{1}{1-\tau^2} \left[ |z_3|^2(1+|z|^2) - \tau \Re\left(z_3^2 (1+z^2) \right) \right]  } 
\nonumber\\
&&\times \int_{\mathbb{C}^{N-3}} \prod_{j=4}^{N} \dif^2 z_j \; \heaviside \left(|z_j|^2 - |z_3|^2\right) |z_j|^2 |z_j-z_3|^2 |z_j-zz_3|^2\euler^{ -\frac{1}{1-\tau^2}  \left[ |z_j|^2 - \tau \Re ( z_j^2 ) \right]}\!\!
\prod_{4\leq \ell<k}^{N} |z_k-z_\ell|^2\!, \quad\quad
 \label{eq.dist-CondSpacingRatioeGinUE-Finite-N}
\end{eqnarray}
where the $z$-dependent prefactors and powers of $z_3$ result from the Vandermonde determinant, via the conditioning and delta constraint \eqref{eq.delta-z_3}, cf. \eqref{eq.Delta023} below. 

Let us compare to the unconditional complex spacing ratio~\eqref{eq.dist_ofSpacingRatio-z_2}, specified to the eGinUE weight \eqref{eq.weight-eGinUE}. The Gaussian integral over $z_1$ can be performed. The shift of all variables by $z_1$ in ~\eqref{eq.dist_ofSpacingRatio-z_2} leads to an $N$-dependent weight \eqref{eq.weight-eGinUE} in $z_1$, which after completing the square induces terms $\sim1/N$:
\begin{eqnarray}
&&    \varrho^{(N)}_{\rm A} (z;\tau) = \frac{\pi \sqrt{1-\tau^2}}{N \alpha_{N}Z_N(\tau)}\heaviside\left(1-|z|^2\right) |z|^2|1-z|^2  \int_{\mathbb{C}} \dif^2 z_3 \; |z_3|^8 
    \euler^{ - \frac{1}{1-\tau^2}|z_3|^2\left(1+|z|^2 - \frac{|1+z|^2}{N}\right)} \nonumber \\
  &&\times \euler^{ \frac{\tau}{1-\tau^2} \Re\left( z_3^2(1+z^2-\frac{(1+z)^2}{N})\right)}     
\int_{\mathbb{C}^{N-3}} \prod_{j=4}^{N} \dif^2 z_j \; \heaviside \left(|z_j|^2 - |z_3|^2\right) |z_j|^2  |z_j-z_3|^2 |z_j-zz_3|^2  \prod_{4\leq j<k}^{N} |z_k-z_j|^2 \nonumber \\ 
    &&\times
    \euler^{-\frac{1}{1-\tau^2} \sum_{j=4}^{N} \left[ |z_j|^2 - \tau \Re (z_j^2 ) - \frac{1}{N} \left(\Re \left( 2 \cconj{z}_j z_3(1+z)\right)  +\sum_{k=4}^{N} \cconj{z}_j z_k\right)  + \frac{\tau}{N}  \Re \left( 2 z_j z_3(1+z) + \sum_{k=4}^{N} z_j z_k \right)\right]} .
    \label{eq.distSpacingRatioeGinUE-Finite-N}
\end{eqnarray}
This expression generalises \cite[eq.~(C6)]{SaRibeiroProsen} to the eGinUE, to which it reduces  for $\tau=0$. Compared to the conditional complex spacing ratio eq.~\eqref{eq.dist-CondSpacingRatioeGinUE-Finite-N}, there are additional terms with a prefactor $\frac{1}{N}$ in the exponents in \eqref{eq.distSpacingRatioeGinUE-Finite-N}. 
In \cite{SaRibeiroProsen} and later in \cite{DusaWettig} it was proposed to neglect these terms proportional to  $\frac{1}{N}$, in order to obtain a leading order approximation.  While this seems plausible for the individual terms in the first and second line, as well as for the terms under a single sum over $j$ in the third line, this is certainly not the case for the terms including a second sum over $k$, which are a priori of the same order in the large-$N$ limit as the terms without prefactor $\frac{1}{N}$. The conditional expression \eqref{eq.dist-CondSpacingRatioeGinUE-Finite-N} thus explains why in the large-$N$ limit these terms can be neglected, despite representing an uncontrolled approximation in \eqref{eq.distSpacingRatioeGinUE-Finite-N}.

We turn to the expression obtained by Dusa and Wettig~\cite{DusaWettig} in this approximation. 
The remaining goal of this subsection is to generalise their result. After making the above-mentioned approximation, 
they  expressed $\varrho_{\rm A}^{(N)}(z;\tau=0)$ for the GinUE class A at $\tau=0$ as an average of a pentadiagonal $(N-3) \times (N-3)$ matrix, by using a supersymmetric
representation for the Vandermonde determinant. In their derivation they made explicit use of the fact that the weight function in the GinUE  is Gaussian, leading
to integrals of incomplete Gamma-function type\footnote{This result in \cite{DusaWettig} can most likely be extended to rotational invariant weight functions.}.
In this case they set up an efficient numerical algorithm for the resulting five-diagonal matrix, that allowed them to evaluate the complex spacing ratio very well for large matrix size, in this approximation.
Below, we will show that the five-diagonal structure continues to hold for
general rotationally invariant weight functions, using Andr\'eief's integral formula instead.

Using the  definition  eq.~\eqref{def.dist_ofCondSpacingRatio} in the conditional case and the Dirac-delta \eqref{eq.delta-z_3},  
the $z_2$-integral becomes trivial and 
the conditional complex spacing ratio is obtained as 
\begin{align}
    \varrho_{\rm A,C}^{(N)}(z) = \frac{\omega(0)}{\alpha_{N,{\rm C}}Z_{N,{\rm C}}} \int_{\mathbb{C}^{N-2}} \dif^2z_3 \dots \dif^2 z_N |\Delta_N(0,zz_3,z_3,z_4,\dots, z_N)|^2 |z_3|^2\omega(zz_3) \omega(z_3) 
    \nonumber \\ \times \heaviside\left(|z_3|^2 -|zz_3|^2 \right) \prod_{j=4}^{N} \omega(z_j) \heaviside\left(|z_j|^2 -|z_3|^2 \right). 
    \label{eq.rho-cond}
\end{align}
The idea is to apply Andr\'eief's integral identity~\cite{Andreief} to the integrations over $z_4, \dots, z_N$. For general $n$ it reads 
\begin{equation}
    \int \dif\mu(z_1) \dots \dif\mu(z_n) \det_{1\leq j,k\leq n} [\phi_j(z_k)] \det_{1\leq j,k\leq n} [\psi_j(z_k)] = n! \det_{1\leq j,k\leq n} \left[\int \dif\mu(\tilde{z}) \phi_j(\tilde{z}) \psi_k(\tilde{z}) \right],
    \label{eq.AndreiefIntId}
\end{equation}
which holds for general functions $\phi_j(z)$ and $\psi_j(z)$, as long as the integrals exist, with respect to some measure $\dif\mu$ on $\mathbb{R}$ or $\mathbb{C}$. As we will see, in our case the weight function $\omega(z_j)$ will receive some additional contributions. 

In order to apply Andr\'eief's integral identity we split the Vandermonde determinant in \eqref{eq.rho-cond} as
\begin{equation}
    \Delta_N(z_1,z_2,\dots, z_N) = \Delta_{N-3}(z_4,\dots, z_N) \Delta_{3}(z_1, z_2 , z_3)\prod_{j=4}^N(z_j-z_1)(z_j-z_2)(z_j-z_3) \ .
    \label{eq.Delta-split}
\end{equation}
In particular, for the conditional point process \eqref{def.dist_ofCondSpacingRatio}, together with the Dirac-delta \eqref{eq.delta-z_3} we obtain 
\begin{align}
\Delta_{3}(z_1, z_2 , z_3)
\left.\prod_{j=4}^N(z_j-z_1)(z_j-z_2)(z_j-z_3)
\right|_{z_1=0,\,z_2=zz_3}=
z(1-z)(z_3)^3 \prod_{j=4}^Nz_j(z_j-zz_3)(z_j-z_3).
\label{eq.Delta023}
\end{align}
Let us apply Andr\'eief's integral  formula \eqref{eq.AndreiefIntId} to the remaining Vandermonde determinant squared $|\Delta_{N-3}(z_4,\dots, z_N)|^2$ from \eqref{eq.Delta-split}, with $n=N-3$.  Thus, 
the product on the right-hand side of \eqref{eq.Delta023} leads to a modified weight function $|z_j|^2|z_j-zz_3|^2|z_j-z_3|^2\omega(z_j) \heaviside(|z_j|^2 -|z_3|^2) $ inside the $(N-3)$ integrals over $z_4,\ldots,z_N$ in \eqref{eq.rho-cond}, after taking the absolute value square of \eqref{eq.Delta023}. As a result we obtain the following expression
\begin{eqnarray}
    \varrho_{\rm A, C}^{(N)}(z) &=& \frac{\omega(0)\,(N-3)!}{\alpha_{N,{\rm C}} Z_{N,{\rm C}}} |z(1-z)|^2\heaviside\left(1 -|z|^2 \right) \int_{\mathbb{C}}\dif^2 z_3 |z_3|^8 \omega(zz_3) \omega(z_3) \nonumber \\ &&\times \detjkTo{N-3}\left[ \int_{\mathbb{C}} \dif^2\tilde{z}\ \tilde{z}^{j-1}\cconj{\tilde{z}}^{k-1} |\tilde{z}|^2|\tilde{z}-zz_3|^2|\tilde{z}-z_3|^2 \omega(\tilde{z}) \heaviside\left(|\tilde{z}|^2 -|z_3|^2 \right) \right] .
    \label{eq.rho-detA}
\end{eqnarray}
For $N=3$ the determinant is obviously absent. Let us now further investigate the following matrix inside the determinant for $N>3$:
\begin{equation}
    A_{jk} = \int_{\mathbb{C}} \dif^2 \tilde{z} \ \tilde{z}^{j-1}\cconj{\tilde{z}}^{k-1} |\tilde{z}|^2|\tilde{z}-z_3z|^2|\tilde{z}-z_3|^2\omega(\tilde{z})  \heaviside\left(|\tilde{z}|^2 -|z_3|^2 \right).
\end{equation}
In the special case of a radially symmetric weight function $\omega(\tilde{z})=\omega(|\tilde{z}|)$, by switching to polar coordinates $\tilde{z}=r\exp[\iunit \theta]$, we obtain
\begin{equation}
    A_{jk} = \int_{|z_3|}^{\infty} \dif r \; r^{j+k-1} \omega(r) \int_{0}^{2\pi} \dif \theta \;  |\tilde{z}|^2|\tilde{z}-z_3z|^2|\tilde{z}-z_3|^2 \euler^{\iunit \theta (j-k)}.
    \label{def.Matrix_A-radialSymmetric}
\end{equation}
It is clear that only those terms contribute where the angular dependence cancels inside the angular integral. From expanding each factor, e.g. $|\tilde{z}-z_3|^2=r^2+|z_3|^2+r(z_3\euler^{-\iunit \theta}+\cconj{z}_3\euler^{+\iunit \theta})$, it is clear that this leads to a five-diagonal matrix. To be explicit we now expand
\begin{equation}
    |\tilde{z}|^2|\tilde{z}-z_3z|^2|\tilde{z}-z_3|^2 = \sum_{l,m=1}^{3} M_{lm} \tilde{z}^l \cconj{\tilde{z}}^{m} ,
\end{equation}
using the same notation as in \cite{DusaWettig}, with the following Hermitian matrix  coefficients $M_{lm} = \cconj{M}_{ml}$:
\begin{align}
    \begin{array}{lll}
    M_{11} = |z_3|^4 |z|^2,  &  M_{12} =\cconj{M}_{21} = - z_3|z_3|^2 (|z|^2+z),  & M_{13} = \cconj{M}_{31} = z_3^2 z, \\
    M_{22} = |z_3|^2 |1+z|^2,  & M_{23} =\cconj{M}_{32} = -z_3(1+z),  & M_{33} = 1 .
    \end{array}
    \label{def.M}
\end{align}
This expansion allows us to evaluate the angular integral in eq.~\eqref{def.Matrix_A-radialSymmetric}, hence
\begin{equation}
    A_{jk} =2\pi \sum_{l,m=1}^{3} M_{lm} \delta_{l+j,m+k} \int_{|z_3|}^{\infty} \dif r \; r^{2(j+l)-1} 
    \omega(r) .
    \label{eq.Afinal}
\end{equation}
Since the Kronecker delta $\delta_{l+j,m+k} $ implies
\begin{equation}
    l+j=m+k \iff j-k = m - l \in \{-2,-1,0,1,2\},
    \label{eq.SelectionRule_Matrix_A}
\end{equation}
we obtain a pentadiagonal matrix $A_{jk}$ for a general rotationally invariant weight function $\omega(|z|)$. This pentadiagonal structure coincides with the approximation to the unconditional ensemble obtained by Dusa and Wettig~\cite{DusaWettig}, for the GinUE with $\omega(z)=\exp[-|z|^2]$.

It was claimed in \cite{DusaWettig} that the determinant of the matrix $A$ in \eqref{eq.Afinal} is independent of the phase of $z_3$, allowing the integral therein to be simplified. This property holds more generally for arbitrary rotationally invariant weight functions: the phase dependence of $z_3$ in \eqref{def.M} can be made explicit, $M_{lm}= C_{lm}(z,|z_3|)\exp\left[\iunit \arg(z_3)(m-l) \right]$.  Combining this with the Kronecker delta constraints in \eqref{eq.SelectionRule_Matrix_A}, each matrix element can be written as
\begin{equation}
    A_{jk}(z,z_3) = \euler^{\iunit \arg(z_3)(j-k)} \tilde{A}_{jk}(z,|z_3|),
\end{equation}
where $\tilde{A}$ depends only on the complex spacing ratio $z$ and the modulus $|z_3|$, being entirely independent of the phase of $z_3$.
By the multilinearity of the determinant, it follows that
\begin{equation}
    \det\limits_{1 \le j, k \le N-3} [A(z,z_3)] = \det\limits_{1 \le j, k \le N-3} [\tilde{A}(z,|z_3|)],
    \label{eq.detA}
\end{equation}
so that the integral over the phase of $z_3$ becomes trivial.

For further details on the efficient numerical evaluation of the remaining radial integral over \eqref{eq.detA} for large matrix size $N$ we refer to \cite{DusaWettig}.

\subsection{Surmise for $N=3$: Conditional and unconditional eGinUE}\label{sec:surmise}
We now investigate the complex spacing ratio at $N=3$ for the eGinUE, comparing the conditional and unconditional spacing ratio. This will allow us to interpolate from the GinUE ($\tau=0$) to the GUE ($\tau\nearrow1$) in the next Subsection \ref{sec:comparisonGUE}, where this surmise for the (real) spacing ratio is very accurate. Furthermore, the conditional expression at $\tau=0$ immediately gives the leading order limiting expression for the unconditional case (which however is not a good approximation \cite{SaRibeiroProsen}).
 
We start with eq. \eqref{eq.dist-CondSpacingRatioeGinUE-Finite-N}  respectively \eqref{eq.distSpacingRatioeGinUE-Finite-N} at $N=3$, where we parametrise $z_3=s+\iunit t$ and $z=x+\iunit y$. Both expressions (the conditional and unconditional (C)) take the following form
\begin{equation}
    \varrho_{\rm A,(C)}^{(N=3)}(z;\tau)=\frac{\heaviside\left(1-|z|^2\right) |z|^2|1-z|^2}{K_{\rm (C)}(\tau)}  \int_{\mathbb{R}^2} \dif s \dif t \; (s^2 + t^2)^4  
    \euler^{ -t^2 A_{\rm (C),+}(z;\tau)+2tsB_{\rm (C)}(z;\tau)- s^2 A_{\rm (C),-}(z;\tau) },
    \label{eq:N3ratioAB}
\end{equation}
with normalisation
\be
    K_{\rm C}(\tau) = \frac{1}{\alpha_{N=3,{\rm C}}\, Z_{N=3,{\rm C}}(\tau)}, \text{ and } K(\tau) = \frac{\pi \sqrt{1-\tau^2}}{3  \alpha_{N=3}\,Z_{N=3}(\tau)}.
\ee
For the conditional case \eqref{eq.dist-CondSpacingRatioeGinUE-Finite-N}  we obtain 
\begin{equation}
    A_{{\rm C},\pm}(z;\tau) = \frac{1}{1-\tau^2} \left[ x^2+y^2+1 \pm \tau ( x^2-y^2+1 ) \right] , \quad B_{\rm C}(z;\tau)= - \frac{2\tau}{1-\tau^2} xy .
    \label{eq.dist-CondSpacingRatioeGinUE-N=3}
\end{equation}
In the unconditional case \eqref{eq.distSpacingRatioeGinUE-Finite-N}, the corrections in $\frac{1}{N}=\frac{1}{3}$ lead to the following expressions
\begin{equation}
    A_{\pm}(z;\tau) = \frac{1}{1-\tau^2}\frac{2}{3} \left[ x^2+y^2-x+1 \pm \tau ( x^2-y^2 -x+1 ) \right] , \quad B(z;\tau) = \frac{\tau}{1-\tau^2} \frac{2}{3}y(1-2x). 
    \label{eq.dist-SpacingRatioeGinUE-N=3}
\end{equation}
The Gaussian two-fold integral \eqref{eq:N3ratioAB} can be evaluated by the following differentiation, where we suppress the arguments and subscript ${\rm C}$ here:
\begin{eqnarray}
 &&   \left(\frac{\partial}{\partial A_{+}} + \frac{\partial}{\partial A_{-}}  \right)^4\int_{\mathbb{R}^2}\dif s\dif t\; 
 \euler^{ -t^2 A_{+}+2tsB- s^2 A_{-}} 
    = \pi \left(\frac{\partial}{\partial A_{+}} + \frac{\partial}{\partial A_{-}}  \right)^4  (A_{+}A_{-}-B^2)^{-\frac{1}{2}} \nonumber \\ 
&&  =\frac{3\pi}{16 (A_{+}A_{-}-B^2)^{\frac{9}{2}}} \left[48  (A_{+}A_{-}-B^2)^2-120  (A_{+}A_{-}-B^2)(A_++A_-)^2+ 35 (A_++A_-)^4  \right]
\label{eq.intv1}\\    
&&    =  3 \pi \frac{  35 (A_{+}^4 + A_{-}^4) + 20(A_{+}^2+A_{-}^2)(6B^2+A_{+}A_{-}) + 6(3A_{+}^2A_{-}^2+24A_{+}A_{-}B^2+8B^4)}{16 (A_{+}A_{-}-B^2)^{\frac{9}{2}}} .\quad\quad\quad\quad
\label{eq.intv2}
\end{eqnarray}
In both the conditional and unconditional case, we have to check the positivity of the expression $ (A_{+}A_{-}-B^2)$ from the Gaussian integral before differentiation, and we obtain after some algebra
for the conditional case 
\begin{eqnarray}
A_{\rm C,+}(z;\tau)A_{\rm C,-}(z;\tau) - B_{\rm C}(z;\tau)^2 = \frac{1}{(1-\tau^2)^2}\left[(1-\tau^2)(x^2+y^2+1)^2+4\tau^2y^2
\right]>0.
\label{eq.A2C-BC2}
\end{eqnarray}
This is obviously positive, including the origin $x=y=0$. Similarly, the unconditional case yields
\begin{eqnarray}
A_{+}(z;\tau)A_{-}(z;\tau)-B(z;\tau)^2=\frac{1}{(1-\tau^2)^2}\frac{4}{9}\left[(1-\tau^2)(x^2+y^2+1-x)^2+3\tau^2y^2
\right]>0,
\label{eq.A2-B2}
\end{eqnarray}
which is also positive. 
In order to arrive at the final expression for the surmise at $N=3$, it is most convenient to use the intermediate expression for the Gaussian integral \eqref{eq.intv1}, inserting one of the respective two expressions \eqref{eq.dist-CondSpacingRatioeGinUE-N=3} and \eqref{eq.dist-SpacingRatioeGinUE-N=3}. 
We obtain the following final results, given in terms of real and imaginary part $x$ and $y$.  For the conditional case it holds
\begin{eqnarray}
   \varrho_{\rm A,C}^{(3)}(x+\iunit y;\tau) &=& 
    \frac{3\pi(1-\tau^2)^5}{K_{\rm C}(\tau)}\frac{\heaviside\left(1-x^2-y^2\right) (x^2+y^2)
    (g_{\rm C}(x,y)-2x) }{\left[(1-\tau^2)g_{\rm C}(x,y)^2+4\tau^2 y^2\right]^{\frac{9}{2}}} 
    \label{eq.rho3Cfinal}
    \\
&&     \times \left[8g_{\rm C}(x,y)^4 + 24 \tau^2 g_{\rm C}(x,y)^2 \left(g_{\rm C}(x,y)^2 -4y^2 \right) +3 \tau^4 \left(g_{\rm C}(x,y)^2 -4y^2 \right)^2\right],
\nonumber\\
\mbox{with}&&g_{\rm C}(x,y)=x^2+y^2+1\ ,
\end{eqnarray}
and for the unconditional case we have 
\begin{eqnarray}
\varrho_{\rm A}^{(3)}(x+\iunit y;\tau) &=& 
    \frac{3^6\pi(1-\tau^2)^5}{2^5 K(\tau)}\frac{\heaviside\left(1-x^2-y^2\right) (x^2+y^2)
    (g(x,y)-x) }{\left[(1-\tau^2)g(x,y)^2+3\tau^2 y^2\right]^{\frac{9}{2}}} 
    \label{eq.rho3final}
    \\
&&     \times \left[8g(x,y)^4 + 24 \tau^2 g(x,y)^2 \left[g(x,y)^2 -3y^2 \right] +3 \tau^4 \left[g(x,y)^2 -3y^2 \right]^2
\right],\quad\quad\quad
\nonumber\\
\mbox{with}&&g(x,y)=x^2+y^2+1-x\ .
\end{eqnarray}
The respective normalisation constants $\alpha_{N=3,{\rm (C)}}(\tau)$ have to satisfy  the condition \eqref{eq.norm_alpha}. They read
\begin{eqnarray}
K_{\rm C}(\tau)&=& \pi^2 (1-\tau^2)(2+\tau^2)\ ,
\label{eq.norm3C}\\
K(\tau)&=& 18\pi^2 (1-\tau^2)\ ,
\label{eq.norm3}
\end{eqnarray}
where the latter agrees with the expression found in \cite{SaRibeiroProsen} at $\tau=0$. 
While $K_{\rm C}(\tau)$ can be computed explicitly, see \cite{BScTW}, for $K(\tau)$ we have made a scaling Ansatz and have checked the correct normalisation numerically for various values of $\tau$.

\begin{figure}[h!]
\includegraphics[width=\linewidth]{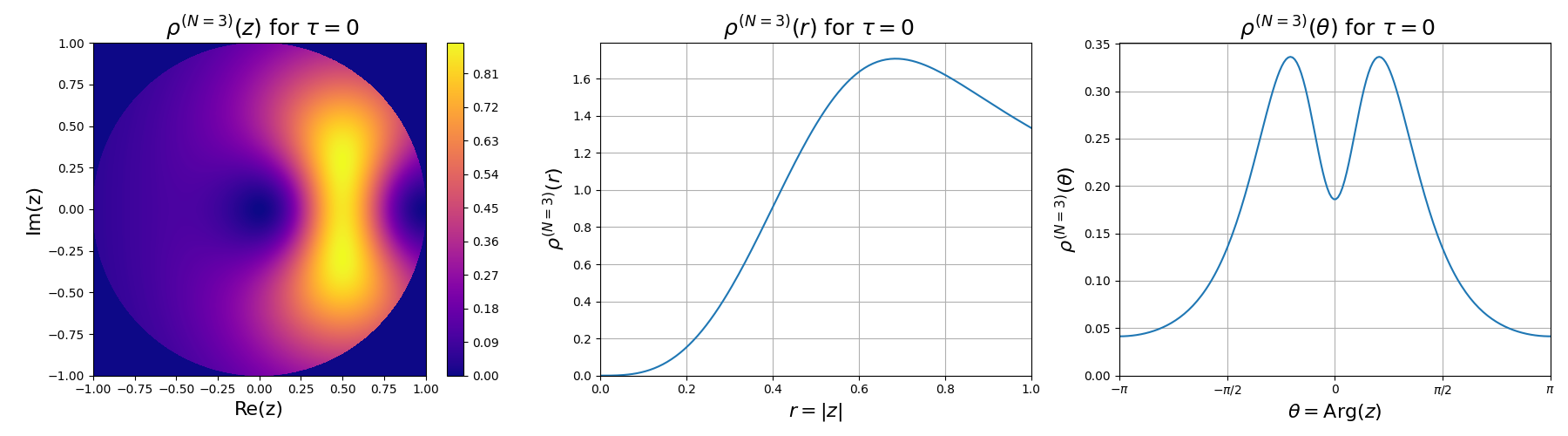} 
\includegraphics[width=\linewidth]{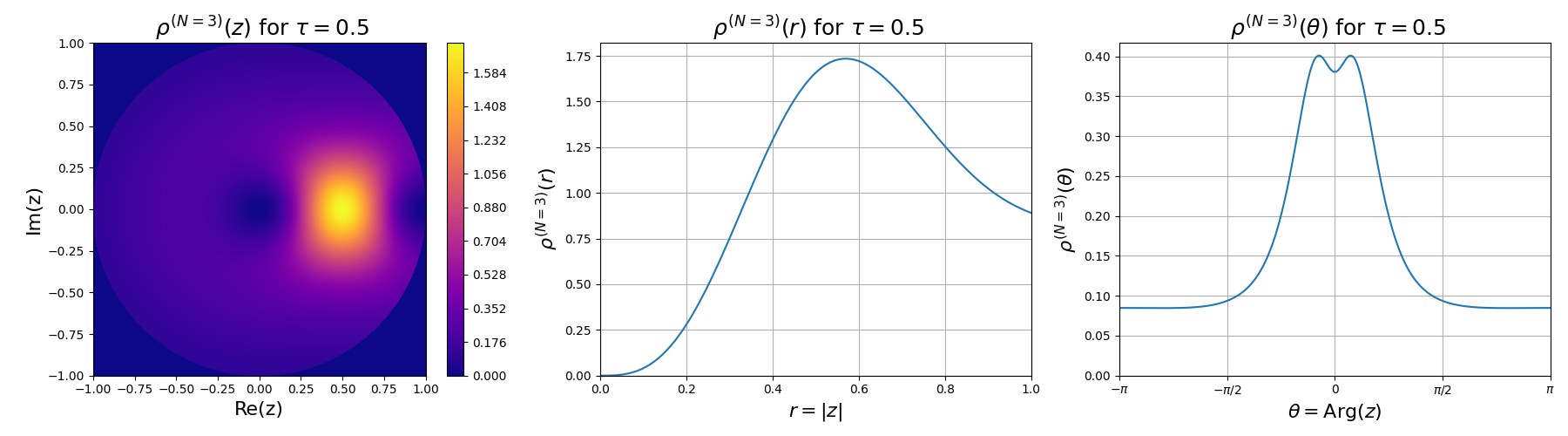}
\includegraphics[width=\linewidth]{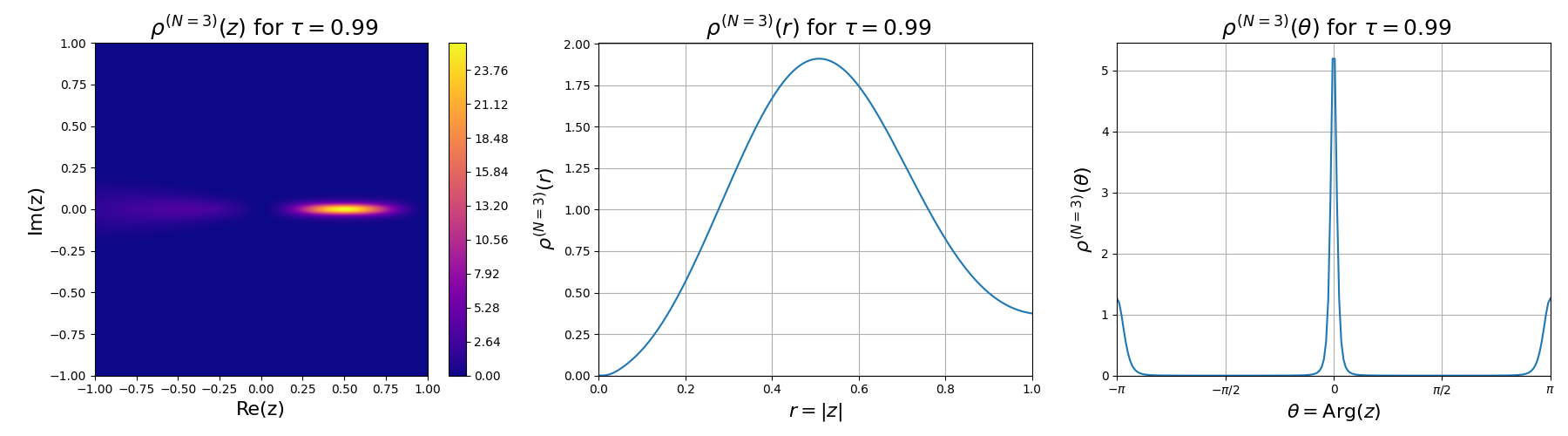}  
\caption{The unconditional $N=3$ surmise for the eGinUE $ \varrho_{\rm A}^{(N=3)}(x+\iunit y;\tau)$ from \eqref{eq.rho3final} for the complex spacing ratio (left), its radial (middle) and angular marginal distribution centred at $\theta=0$, with $\theta\in(-\pi,\pi]$ (right),  at 
 $\tau=0,0.5,0.99$ (top, middle bottom). Note that the colour coding of the 2D density in the left column strongly depends on $\tau$, as the eigenvalues concentrate more around the $x$-axis with increasing $\tau\nearrow 1$.}
\label{fig:rho3_uncond}
\end{figure}

\begin{figure}[h!]
\includegraphics[width=\linewidth]{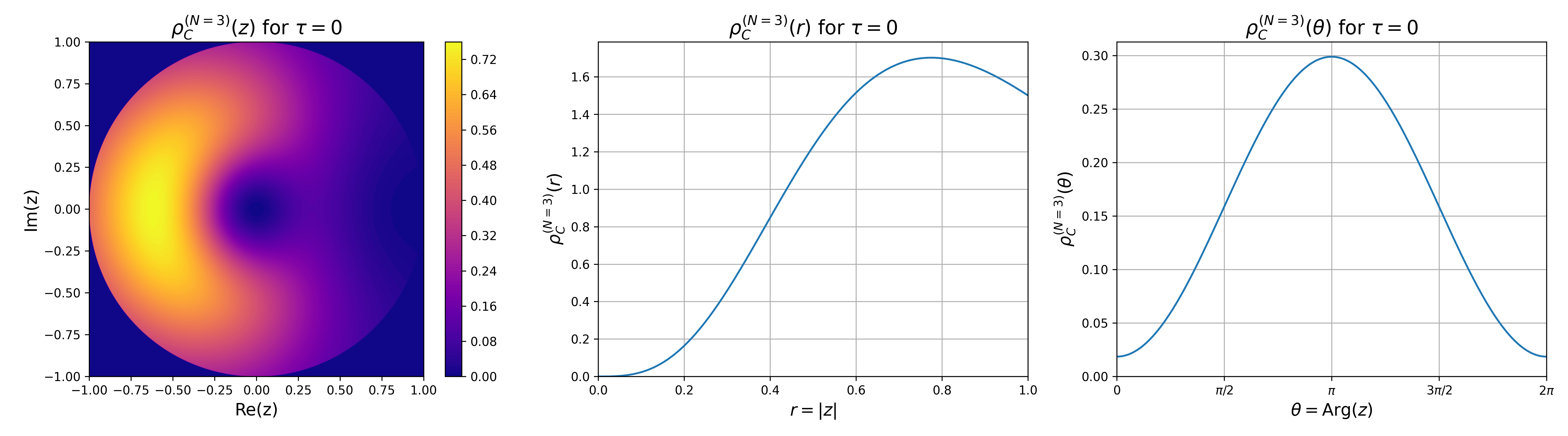}
\includegraphics[width=\linewidth]{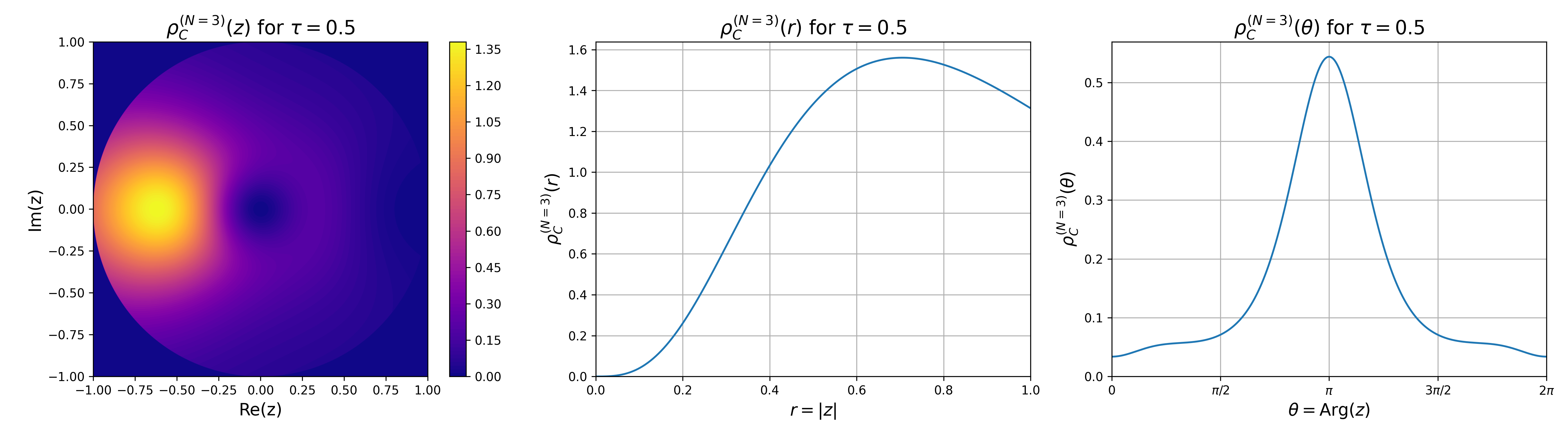}
\includegraphics[width=\linewidth]{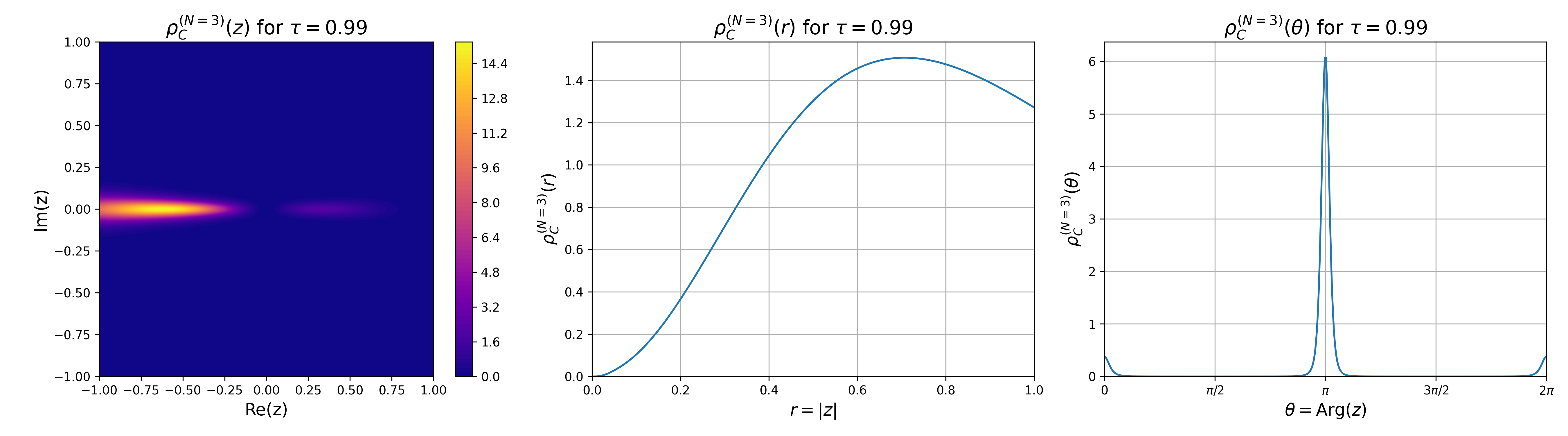}
\caption{The same plots as in Fig. \ref{fig:rho3_uncond} in the conditional eGinUE, $\varrho_{\rm A,C}^{(3)}(x+\iunit y;\tau)$ from \eqref{eq.rho3Cfinal}, with $\theta\in[0,2\pi)$. Notice that  here the angular marginal distribution is centred at $\theta=\pi$ (right column).}
\label{fig:rho3_cond}
\end{figure}

Let us define the radial marginal distribution
\be
    \rho^{(N)}_{\rm A,(C)}(r; \tau):= \int_{-\pi}^{\pi} \dif \theta \; r \varrho^{(N)}_{\rm A,(C)}(r\exp{[\iunit\theta]};\tau), 
    \label{eq.rad-marg}
\ee
and the angular marginal
\be
    \rho^{(N)}_{\rm A,(C)}(\theta; \tau)  := \int_{0}^{1} \dif r \; r \varrho^{(N)}_{\rm A,(C)}(r\exp{[\iunit\theta]};\tau). 
\label{eq.ang-marg}    
\ee
The radial distribution $ \rho^{(3)}_{\rm A,C}(r; \tau)$ can be written explicitly in terms of complete elliptic integrals of the first and second kind \cite{BScTW}. 
Together with the resulting spacing ratios, $\varrho_{\rm A,(C)}^{(3)}(z;\tau)$ in eq.~\eqref{eq.rho3final} and \eqref{eq.rho3Cfinal}, respectively,  
the two marginals are plotted for $\tau=0,0.5,0.99$ in Figs.~\ref{fig:rho3_uncond} and \ref{fig:rho3_cond}, respectively. For all values of $\tau$ we see a clustering around small angles $\theta\approx 0$ in the unconditional case in Fig. \ref{fig:rho3_uncond}. This is also seen despite the dip in the angular distribution at $\tau=0$ and 0.5. {This angular behaviour seen in Fig. \ref{fig:rho3_uncond} top left does not agree with the numerically observed large-$N$ behaviour in Fig.  \ref{fig:spacing ratio symmetry classes} top middle plot.} 
In contrast, there is a clustering at angle $\theta\approx\pi$ in the conditional case in Fig. \ref{fig:rho3_cond}. At least qualitatively the conditional case is much closer to the numerical behaviour of the complex spacing ratio for large values of $N$ 
in the Ginibre ensemble at $\tau=0$ to be discussed in the next Section \ref{Sec:Numerics}, 
{compare Fig. \ref{fig:rho3_cond} top left with}
Fig.  \ref{fig:spacing ratio symmetry classes} top middle plot. This shortcoming of the unconditional surmise was also discussed in \cite{SaRibeiroProsen}.

We come back to our analytical result and discuss the conditional case at $\tau=0$, compared to \cite{SaRibeiroProsen}. Introducing polar coordinates $x=r\cos\theta$, $y=r\sin\theta$, we have 
\begin{equation}
 \varrho_{\rm A,C}^{(3)}(r\exp[\iunit \theta];\tau=0) =
    \frac{12}{\pi}\frac{r^2(r^2+1-2r\cos\theta)}{(r^2+1)^5} \heaviside(1-r^2)\ .
    \label{eq.rhoN3Ctau0}     
\end{equation}
It agrees with the limiting leading order result in an expansion in powers of $r$ for the {\it unconditional} case as derived in \cite[eq. (C8)]{SaRibeiroProsen} for the GinUE  at $\tau=0$. 
As already noted there, {despite this improvement it does not provide a good quantitative approximation to large-$N$ yet.}  
For completeness we compare with the result for the eGinUE in the unconditional case, setting $\tau=0$ in eq.~\eqref{eq.rho3final}:
\begin{equation}
 \varrho_{\rm A}^{(3)}(r\exp[\iunit \theta];\tau=0) =
    \frac{81}{8\pi}\frac{r^2(r^2+1-2r\cos\theta)}{(r^2+1-r\cos\theta)^5} \heaviside(1-r^2)\ .
    \label{eq.rhoN3tau0}     
\end{equation}
It agrees with \cite[eq. (C7)]{SaRibeiroProsen} and serves as a consistency check.

\subsection{Hermitian limit vs.  consecutive and NN GUE spacing ratio}
\label{sec:comparisonGUE}

Before taking the Hermitian limit $\tau \nearrow 1$ of the conditional and unconditional surmises in the eGinUE, let us recall what is known analytically in the GUE for spacing ratios of real eigenvalues $x_1,\ldots,x_N$ of a complex $N\times N$ Hermitian random matrix $H$ with independent Gaussian matrix elements distributed as $P(H)\sim\exp[-\Tr H^2]$. 
Its joint density is proportional to (see e.g. \cite{Haake2010})
\be
    \jpdf_{\rm GUE}^{(N)}(x_1,\dots,x_N) \propto \Delta_N(x_1,\dots,x_N)^2 \prod_{j=1}^{N} \exp{\left[-x_j^2\right]}.
\label{eq:GUEjpdf}
\ee
In contrast to complex eigenvalues $z_j$, the real eigenvalues $x_j$ can be put into consecutive order, 
\be
x_1<x_2<\ldots<x_N ,
\ee
where the probability of two eigenvalues to be equal is vanishing. Consequently, several different spacing ratios can be defined. First, for every fixed $k$, with $1<k<N$, we can define a {\it consecutive spacing ratio}, following \cite{OH,Atas_et_al}:
\be 
r_k =\frac{x_{k+1}-x_k}{x_k-x_{k-1}}>0, \quad k=2,\ldots,N-1.
\label{eq:consec-rk}
\ee 
Notice that by definition this ratio is always positive. In \cite{OH}, the minimum value $\tilde{r}_k=\min\{r_k,1/r_k\}\in[0,1]$ was considered. In case of additional symmetries, the distributions of $r_k$ and $\tilde{r}_k$ can be related, cf. \cite{Atas_et_al}.
The $N=3$ surmise (with $k=2$) for the consecutive spacing ratio in the GUE is thus obtained as \cite{Atas_et_al}:
\begin{eqnarray}
 \varrho_{{\rm GUE}}^{(3)}(x) &\propto& 
 \int_{-\infty}^\infty\dif x_2 \int_{-\infty}^{x_2}\dif x_1 \int_{x_2}^\infty \dif x_3
\jpdf_{\rm GUE}^{(N=3)}(x_1,x_2,x_3)\ \delta^{(1)}\left(x-\frac{x_3-x_2}{x_2-x_1}\right)  \nonumber \\
    &=&\frac{1}{\alpha_{3,\rm GUE}}\frac{ x^2(1+x)^2 }{(1+x+x^2)^4},\quad x\geq 0, 
    \label{eq:GUEcons3}
\end{eqnarray}
with the correct normalisation reading  
$\alpha_{3,\rm GUE}= \frac{4 \pi}{81\sqrt{3}}$ \cite{Atas_et_al} . 
This surmise is an accurate approximation to the consecutive spacing ratio in the bulk of the spectrum for large $N$, see Fig.~1~in~\cite{Atas_et_al}. Corresponding surmises exist for the orthogonal, symplectic and in fact $\beta$-ensembles, too \cite{Atas2}\footnote{For the large-$N$ limit in the bulk of the spectrum one has to exclude edge eigenvalues from \eqref{eq:consec-rk}.}.

\begin{figure}
\setlength{\unitlength}{1mm}
\begin{picture}(150,10)
\put(0,0){\vector(1,0){60}}
\put(10,0){\circle*{2}}
\put(8,5){$x_{k-2}$}
\put(24,0){\circle*{2}}
\put(21,5){$x_{k-1}$}
\put(31,0){\circle*{2}}
\put(29,5){$x_{k}$}
\put(36,0){\circle*{2}}
\put(34,5){$x_{k+1}$}
\put(50,0){\circle*{2}}
\put(48,5){$x_{k+2}$}
\put(78,0){\vector(1,0){60}}
\put(82,0){\circle*{2}}
\put(80,5){$x_{k-2}$}
\put(90,0){\circle*{2}}
\put(88,5){$x_{k-1}$}
\put(104,0){\circle*{2}}
\put(102,5){$x_{k}$}
\put(108,0){\circle*{2}}
\put(106,5){$x_{k+1}$}
\put(112,0){\circle*{2}}
\put(114,5){$x_{k+2}$}
\end{picture}
\caption{Left: normal NN and NNN ordering of consecutive eigenvalues on the real line, with labels $k_0=k+1$, and  $k_1=k-1$ in \eqref{def.k0k1}, right: NN and NNN both to the right with labels $k_0=k+1$, $k_1=k+2$.}
\label{fig:examplespacings}
\end{figure}
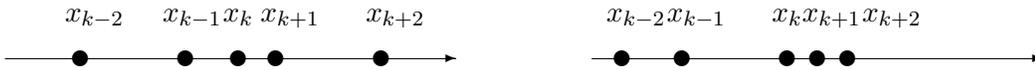

As a second possibility, also for real eigenvalues a {\it NN spacing ratio} can be defined,  as it was proposed in \cite{SaRibeiroProsen}:
\be 
s_k =\frac{x^{\rm NN}_{k}-x_k}{x^{\rm NNN}_k-x_{k}}, \quad k=2,\ldots,N-1.
\label{eq:NN-sk}
\ee 
Here, the NN and NNN eigenvalues are defined as 
\be 
x^{\rm NN}_{k}=\min_{j\neq k}|x_j-x_k|=x_{k_0}, \quad x^{\rm NNN}_{k}=\min_{j\neq k,k_0}|x_j-x_k|=x_{k_1}.
\label{def.k0k1}
\ee
Notice that the variable of this NN spacing ratio can be negative or positive. For a given $k$, in some situations the consecutive and NN spacing ratio can be related, i.e. in  Fig. \ref{fig:examplespacings} (left) we have that $r_k=-s_k$. On the other hand, in the case when both NN and NNN are on the same side of $x_k$, see  Fig. \ref{fig:examplespacings} (right), the two spacing ratios are unrelated.

For general $N$, the NN spacing ratio on the real line can be defined in exact analogy as for complex eigenvalues in \eqref{def.dist_ofSpacingRatio}, when choosing again as reference point $x_1$, with NN $x_2$ and NNN $x_3$, here without ordering according to the eigenvalue's label.  
In \eqref{def.dist_ofSpacingRatio} we need to 
replace the area measure $\dif^2z\to\dif x$ and the Dirac delta $\delta^{(2)}(z)\to\delta^{(1)}(x)$, leading to a different Jacobian
\begin{equation}
    \delta^{(1)}\left(x-\frac{x_2}{x_3}\right) = |x_3| \delta^{(1)}\left(xx_3 -x_2  \right) .
    \label{eq.delta-x-2}
\end{equation}
The modulus square trivially reduces to the ordinary square $|\cdot|^2\to(\cdot)^2$. Following the same logic as in the complex case,  the resulting $N=3$ surmise of the NN spacing ratio was derived in \cite[(B.8)]{SaRibeiroProsen},
\begin{eqnarray}
    \varrho_{{\rm GUE,\, NN}}^{(3)}(x)
    &=&\frac{1}{\alpha_{3,\rm GUE\,NN}}\frac{ x^2(1-x)^2 }{(1-x+x^2)^4}\Theta (1-x^2) ,\quad x\in\mathbb{R},
    \label{eq.GUENN3}
\end{eqnarray}
including its normalisation $\alpha_{3,\rm GUE\,NN}=\frac{2\pi}{243\sqrt{3}}$. In contrast to the surmise for the consecutive spacing ratio \eqref{eq:GUEcons3} above, \eqref{eq.GUENN3} is not a good approximation to the large-$N$ bulk limit, see \cite{SaRibeiroProsen} Fig.\ 2 (b). 
Notice that from the surmise of the NN spacing ratio \eqref{eq.GUENN3} the consecutive one \eqref{eq:GUEcons3} can be recovered, by taking only the {\it negative part} of {\eqref{eq.GUENN3}}
\be
\varrho_{{\rm GUE,\, NN}}^{(3)}(-x)\propto  \varrho_{{\rm GUE}}^{(3)}(x), \quad x\geq 0,
\label{eq.NNconsec-rel}
\ee
and then normalising it properly. 

Last but not least, in the same approach as for complex eigenvalues we can also define a {\it conditional NN spacing ratio}, here for the GUE. We condition again  {on} our reference point $x_1=0$, using the same definition as in \eqref{def.dist_ofCondSpacingRatio}, with all eigenvalues on the real line instead. For the corresponding $N=3$ surmise of the conditional (C) NN spacing ratio in the GUE we thus obtain, in analogy to \eqref{eq.rho-detA} at $N=3$:
\begin{eqnarray}
    \varrho_{{\rm GUE,\, CNN}}^{(3)}(x)
&\propto&x^2(1-x)^2 \Theta (1-x^2)  \int_{\mathbb{R}} \dif x_3|x_3|^7  \euler^{-x_3^2 (1+x^2)} \nonumber\\
    &=&\frac{1}{\alpha_{3,\rm GUE,CNN}}\frac{ x^2(1-x)^2 }{(1+x^2)^4}\Theta (1-x^2) ,\quad x\in\mathbb{R}, 
    \label{eq;GUECNN3}
\end{eqnarray}
with normalisation $\alpha_{3,\rm GUE,CNN}=\frac{\pi}{16}$. It agrees with the large-$N$ approximation of the unconditional NN spacing ratio proposed in  
\cite[eq. (B14)]{SaRibeiroProsen}. Even if it is a better approximation than the unconditional one \eqref{eq.GUENN3},  see \cite{SaRibeiroProsen} Fig. 2 (b), it is still not accurate quantitatively.
\\

We are now prepared to discuss the  limit $\tau \nearrow 1$ of the eGinUE surmises \eqref{eq.rho3Cfinal} respectively \eqref{eq.rho3final}, and to compare with the various quantities in the GUE as introduced above.
In the limit $\tau \nearrow 1$ the complex eigenvalues of the eGinUE collapse onto the real line. 
Thus, in the limit, the probability mass concentrates at $y=0$.
This can be seen explicitly in the surmises \eqref{eq.rho3Cfinal} respectively \eqref{eq.rho3final}, as the numerator vanishes as $\sim(1-\tau^2)^4$, whereas the denominator remains finite and proportional to $\sim y^9$, for $y\neq 0$. This leads to a weak limit of the probability distribution, here for the surmises, 
\be
    \lim_{\tau \nearrow 1} \varrho_{\rm A,(C)}^{(3)}(x+\iunit y ; \tau) = f_{\rm (C)}(x) \Theta (1-x^2)  \delta^{(1)}(y) .
    \label{eq.rho3t=1}
\ee
The limiting marginal distribution of the real part $f_{\rm (C)}(x)$ is determined by 
\be
    f_{\rm (C)}(x) \Theta (1-x^2) = \lim_{\tau \nearrow 1} \int_{\mathbb{R}} \dif y \; \varrho_{\rm A,(C)}^{(N=3)}(x+\iunit y ; \tau).
    \label{def.realMarginal,t=1,N=3}
\ee
In order to resolve the dependence of the denominators in $\varrho_{\rm A,(C)}^{(N=3)}(x+\iunit y ; \tau)$ on $\tau$, we make an expansion in powers of $(1-\tau^2)$, and at the same time make a rescaling of the imaginary part as follows:
\be 
y=\tilde{y}\sqrt{1-\tau^2}.
\label{eq.y-rescale}
\ee
In the conditional case we have 
\begin{align}
    \left[(1-\tau^2)g_{\rm C}(x,y)^2 +4\tau^2 y^2 \right]^{-\sfrac{9}{2}} = (1-\tau^2)^{-\sfrac{9}{2}} \left[(1+x^2+\tilde{y}^2(1-\tau^2))^2 +4\tau^2 \tilde{y}^2 \right]^{-\sfrac{9}{2}} \nonumber \\
    = (1-\tau^2)^{-\sfrac{9}{2}} \left[(1+x^2)^2 +4 \tilde{y}^2 \right]^{-\sfrac{9}{2}} \left[1 + \frac{(x^2-1)\tilde{y}^2}{(1+x^2)^2+4\tilde{y}^2} + \mathcal{O}\left((1-\tau^2)^2\right) \right],
\end{align}
and analogously for the unconditional case
\begin{align}
    \left[(1-\tau^2)g(x,y)^2 +3\tau^2 y^2 \right]^{-\sfrac{9}{2}}
    = (1-\tau^2)^{-\sfrac{9}{2}} \left[(1-x+x^2)^2 +3 \tilde{y}^2 \right]^{-\sfrac{9}{2}} \left[1 + \mathcal{O}\left(1-\tau^2\right) \right].
\end{align}
Therefore, the surmise for the complex spacing ratio can be expanded in both cases as
\begin{align}
    \varrho_{\rm A,(C)}^{(3)}(x+\iunit y ; \tau) = \Theta (1-x^2-y^2) x^2(1-x)^2 \sum_{k=0}^{\infty} (1-\tau^2)^{\frac{2k-1}{2}} A_{k,{\rm (C)}}(x,\tilde{y}) , 
    \label{eq.tauexp}
\end{align}
where the leading order expansion coefficients read 
\be
    A_{0,{\rm C}}(x,\tilde{y}) = \frac{1}{\pi} \frac{35 (x^2+1)^4}{\left[(1+x^2)^2 +4 \tilde{y}^2\right]^{\sfrac{9}{2}}}, \quad     A_{0}(x,\tilde{y}) = \frac{3^4 }{2^6\pi} \frac{35 (x^2-x+1)^4}{\left[(1-x+x^2)^2 +3 \tilde{y}^2 \right]^{\sfrac{9}{2}}} .
\ee
In the limit $\tau \nearrow 1$ only the term $k=0$ contributes to $f_{\rm (C)}(x)$ in eq.\eqref{def.realMarginal,t=1,N=3}, with the additional power $1/\sqrt{1-\tau^2}$ being absorbed by the rescaling \eqref{eq.y-rescale} of the $\tilde{y}$-integration:
\begin{align}
    f_{\rm (C)}(x) \Theta(1-x^2)= \Theta(1-x^2) x^2(x-1)^2  \int_{\mathbb{R}} \dif \tilde{y}\;  A_{0,{\rm (C)}}(x,\tilde{y}) .
\end{align}
Consequently we obtain
\be
    f_{\rm C}(x)  =\frac{1}{\pi}  x^2(x-1)^2 \int_{\mathbb{R}} \dif \tilde{y}\; \frac{35 (x^2+1)^4}{\left[(1+x^2)^2 +4 \tilde{y}^2\right]^{\sfrac{9}{2}}} = 
    \frac{16}{\pi}  \frac{x^2(x-1)^2  }{(1+x^2)^4} \ ,
\ee
and
\be
    f(x)  =  \frac{3^4 }{2^6\pi}  x^2(x-1)^2 \int_{\mathbb{R}} \dif \tilde{y}\; \frac{35 (x^2-x+1)^4}{\left[(1-x+x^2)^2 +3 \tilde{y}^2 \right]^{\sfrac{9}{2}}} = \frac{3^3\sqrt{3}}{2\pi}  \frac{ x^2(x-1)^2 }{ (x^2-x+1)^4} .
\ee
We thus have derived the following relations between the limiting surmises of the eGinUE and the corresponding quantities in the GUE:
\begin{align}
    \lim_{\tau \nearrow 1}\varrho_{\rm A,C}^{(3)}(x+\iunit y ; \tau) =& \frac{16}{\pi}  \frac{x^2(x-1)^2  }{(1+x^2)^4} \Theta (1-x^2) \delta^{(1)}(y) 
    =     \varrho_{{\rm GUE,\, CNN}}^{(N=3)}(x)\delta^{(1)}(y)
    ,\label{eq.rhoC3,t=1final} \\
    \lim_{\tau \nearrow 1}\varrho_{\rm A}^{(3)}(x+\iunit y ; \tau) =& \frac{3^3\sqrt{3}}{2\pi}  \frac{ x^2(x-1)^2 }{ (x^2-x+1)^4} \Theta (1-x^2) \delta^{(1)}(y)
     =     \varrho_{{\rm GUE,\, NN}}^{(N=3)}(x)
    \delta^{(1)}(y)\ .
    \label{eq.rho3,t=1final}
\end{align}
As it can be seen at the bottom of Fig. \ref{fig:rho3_cond} for the conditional, and of Fig. \ref{fig:rho3_uncond} for the unconditional case, at $\tau=0.99$ both distributions show a concentration at real positive values with angle $\theta\approx0$ and negative values $\theta\approx \pi$, where the (un)conditional spacing ratio concentrates on the (former) latter. For the conditional case the mass is distributed as follows for the angular marginal \eqref{eq.ang-marg} \cite{BScTW}
\be
    \rho_{\rm A, C}^{(3)}(\theta = 0;1)=
    \int_{0}^1 \dif x \,
    \varrho_{{\rm GUE,\, CNN}}^{(N=3)}(x)
    =
    \frac{3\pi-8}{6\pi}, \quad \rho_{\rm A, C}^{(3)}(\theta = \pi;1) =
    \int_{-1}^0 \dif x \,
    \varrho_{{\rm GUE,\, CNN}}^{(N=3)}(x) = \frac{3\pi+8}{6\pi},
\ee
compared to the unconditional case
\be
    \rho_{\rm A}^{(3)}(\theta = 0;1)=
    \int_{0}^1 \dif x \,
    \varrho_{{\rm GUE,\, NN}}^{(N=3)}(x)
    =
    \frac{2}{3}, \quad \rho_{\rm A}^{(3)}(\theta = \pi;1)=
    \int_{-1}^0 \dif x \,
    \varrho_{{\rm GUE,\, NN}}^{(N=3)}(x)
    =
    \frac{1}{3}\ .
\ee
This comes from the fact that for $N=3$, for each configuration the two outer eigenvalues contribute only to $\theta=0$ and the middle eigenvalue to $\theta=\pi$. 
From the relation \eqref{eq.NNconsec-rel} we deduce, that only the {\it negative part} of the unconditional complex spacing ratio \eqref{eq.rho3,t=1final}, which only makes up a small fraction of $\frac{3\pi-8}{6\pi}\approx 0.08$ of the total mass in the limit, leads to the consecutive spacing ratio of the GUE in the limit $\tau\nearrow 1$.


\section{Numerical edge results: Complex spacing ratios and NN spacings}
\label{Sec:Numerics}

In this section we numerically study  complex spacing ratios and its moments in Subsections \ref{sec:spacing-ratio2D}
and \ref{sec:spacing-ratio-mom}, respectively, and the NN and NNN spacing distribution{s} in Subsection \ref{sec:NN}.
The three simplest representatives for the conjectured generic local universality classes, class A  (complex Ginibre), AI$^\dagger$ (complex symmetric) and AII$^\dagger$ (complex self-dual) to be defined below, are investigated and compared with the corresponding Poisson distribution in 2D. 
We compare the expressions in the bulk of the spectrum, known analytically in class A and 2D Poisson only, to new spacing ratios, NN and NNN spacing distributions $p_{\rm NN(N)}(s)$ at the edge of the complex spectrum for all symmetry classes.
In  Subsection \ref{sec:NNcubic} we investigate the behaviour of $p_{\rm NN}(s)$ at small argument, both in the bulk and at the edge, as this has not been done for classes AI$^\dagger$ and AII$^\dagger$. It has been conjectured early on, based on general arguments, that the  cubic behaviour at small argument, $p_{\rm NN}(s)\sim s^3$, is universal \cite{Grobe Haake,Haake2010}, 
{which we will confirm. In particular, we will also compare with the cubic vanishing of $p_{\rm NN}(s)$ for class A at the edge, derived analytically in 
Appendix \ref{App:kernel}.}
\\

As in Hermitian RMT, the edge of the spectrum hosts universality classes different from the bulk.
The complex spacing ratio can be defined numerically as in \eqref{eq.def spacing ratio} below for the complex eigenvalues in any non-Hermitian ensemble, and we will use the matrix representation of the 3 ensembles to generate these. This is also because for classes AI$^\dag$ and AII$^\dag$ no analytical results are known so far for their jpdfs of complex eigenvalues, see \eqref{eq.jpdf_gen.weight} for class A. For simplicity, we will choose a Gaussian distribution of matrix elements for all 3 ensembles as follows.
Let $J$ be an $N\times N$ (respectively $2N\times 2N$ in the case of AII$^\dagger$) random matrix, where we assign a Gaussian weight $\propto \exp\left[-N\Tr JJ^\dagger \right]$ with it.\footnote{Hamazaki et al. \cite{Hamazaki et al} numerically tested the local bulk universality of the NN spacing distribution, with respect to Bernoulli distributed matrix elements. We will deal with Gaussian matrix ensembles in this work only.} Here, the Gaussian weight is normalised such that the limiting eigenvalue density will be supported on the unit disc. This is in contrast to the previous Section \ref{Sec:AnalyticsA}. 
The 3 symmetry classes are defined by the following symmetry relations:
\begin{itemize}
    \item Class A: $J\neq J^\dagger$ ,
    \item Class AI$^\dagger$: $J=J^T\neq J^\dagger$ ,
    \item Class AII$^\dagger$: $J=\Sigma J^T \Sigma\neq \Sigma \cconj{J} \Sigma$, where $\Sigma=\begin{pmatrix}
			0 & -i\mathds{1}_{N\times N}\\
			i\mathds{1}_{N\times N} & 0
		\end{pmatrix}$.
\end{itemize}
Here, $J^T$ denotes the transposition of $J$, $\cconj{J}$ the complex conjugation and $J^\dagger$ the Hermitian conjugate. It is convenient to parametrise the $2N\times 2N$ matrix $J$ of class AII$^\dag$ via three non-Hermitian $N\times N$ matrices $A,B,C$, where the matrices $B$ and $C$ are anti-symmetric, i.e. $B=-B^T$ and $C=-C^T$ as
\begin{align}
	J=\begin{pmatrix}
		A & B\\
		C & A^T
	\end{pmatrix},
\end{align}
and $A$ has no symmetry. {Here, it is important to scale the variance of the elements of matrices $A,B,C$ to unity (up to symmetries for $B,C$), as the variance of $J$ cannot be fixed by an overall factor as for the other matrix ensembles under consideration, cf. \cite[eq.~(2.9)]{AAKP}.}
It is well-known that the eigenvalues of a complex self-dual matrix are doubly degenerate \cite{hastings}. The distribution for 2D Poisson can be formulated in the same manner in matrix form, with $J=$diag$(z_1,\ldots,z_N)$ being diagonal. 

{Let us give some details about our numerical simulations. We used Python and C++ to study the complex spacing ratios.
The random matrices are simulated in Python (version 3.13) using the library NumPy (version 2.2.5). The Python standard library and NumPy provide functions for normal distributions and eigenvalue computation.
Normal distributions are internally generated from uniform distributions by the Box--Muller transform.
Python and the library ultimately use the system pseudorandom number generator. 
NumPy uses the OpenBLAS library to compute the eigenvalues; this library reduces the matrices to the Hessenberg form and subsequently performs a QR decomposition.
For reading the data in our C++ programs, we make use of the library cnpy.
We provide our data for the complex eigenvalues of the random matrices from Poisson statistics, respectively the symmetry classes AI$^\dagger$, A, and AII$^\dagger$ for
720~000 ensembles each and $N=1024$, respectively $2N=1024$ for AII$^\dagger$ in \cite{data repository}.
For the comparison of the NN and NNN spacings in Subsection \ref{sec:numNN}, where we perform unfolding, we use Mathematica and C++. }

\begin{figure}[h]
\includegraphics[width=0.49\linewidth]{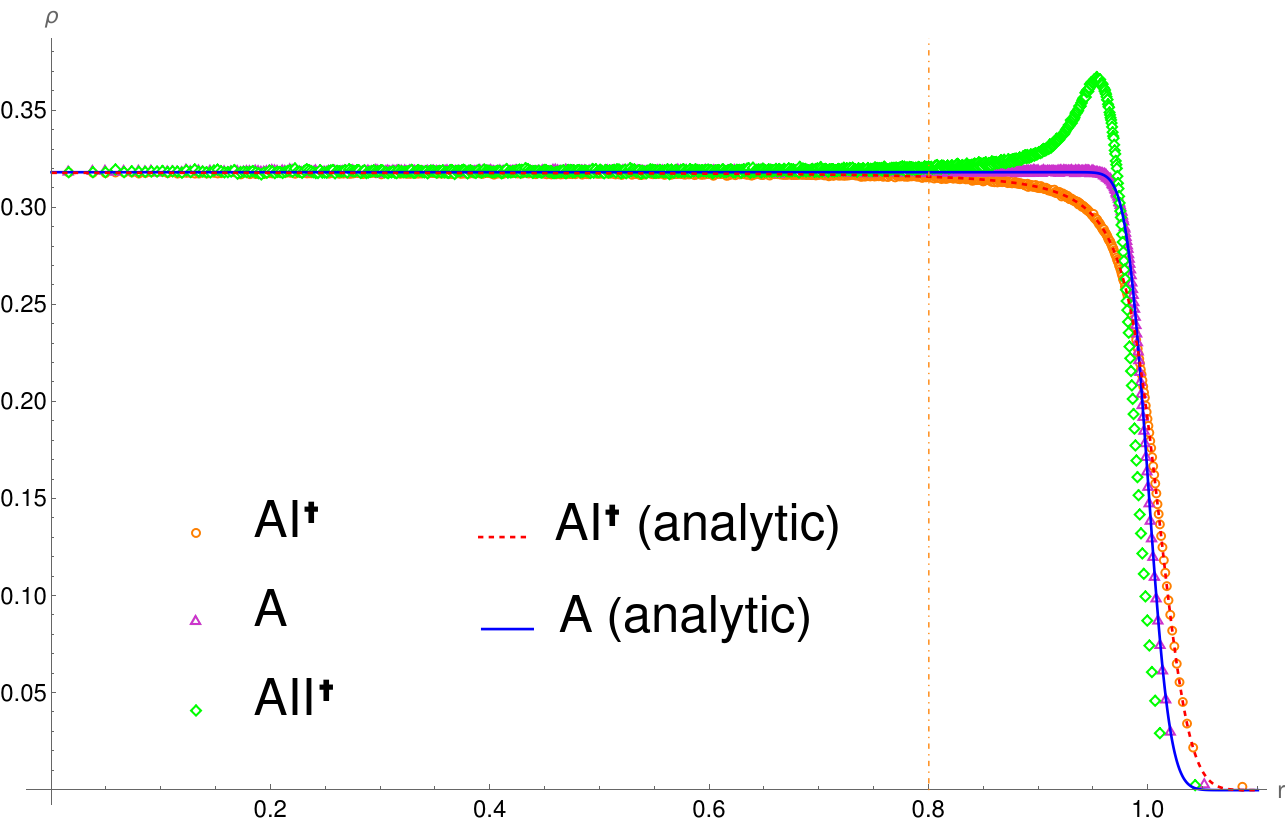}
\includegraphics[width=0.49\linewidth]{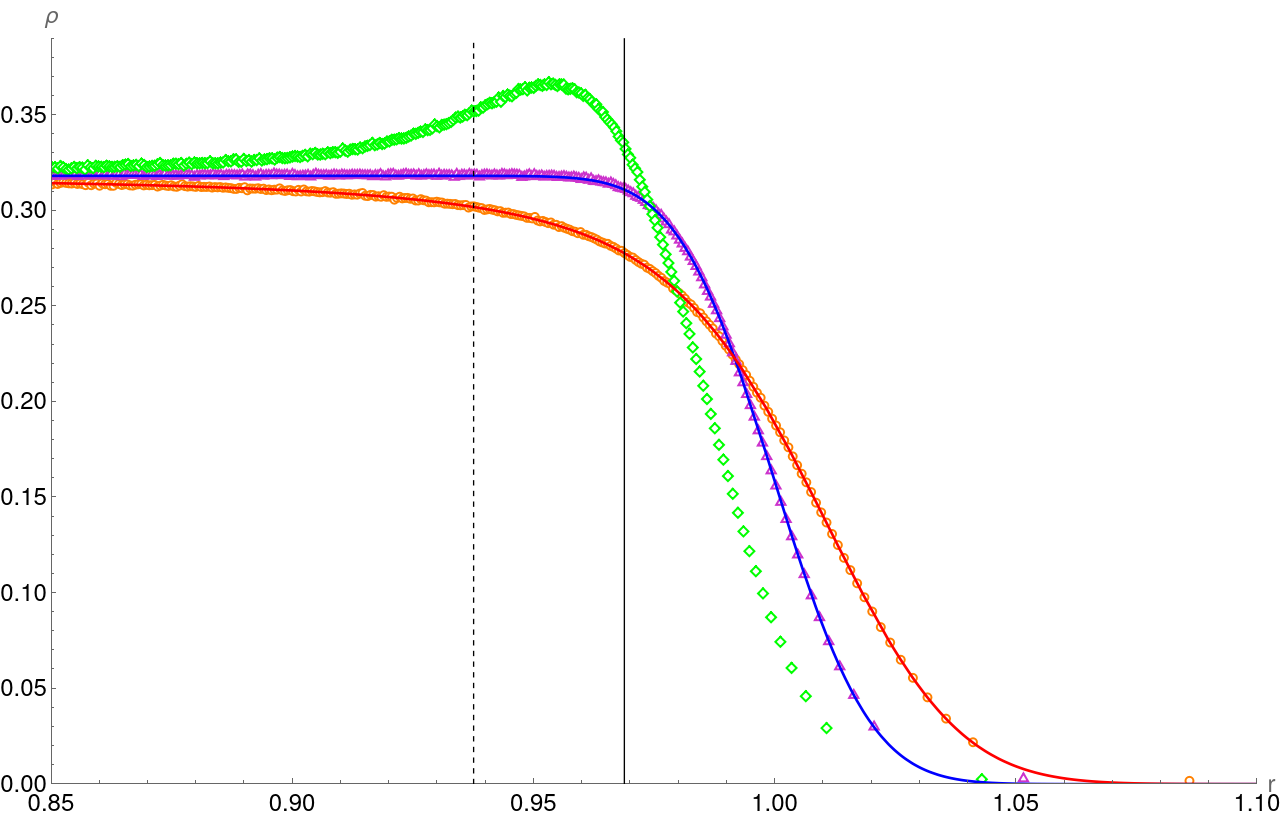}
\caption{The radial spectral density of class A \eqref{R1A} (blue full curve: analytical; purple triangle: numerical), class AI$^\dagger$ \eqref{R1AI+} (red dashed curve: analytical; orange circle: numerical), both 
at $N=1024$, and of class AII$^\dagger$ (green diamond: numerical only) with $2N=1024$, with $720\,000$ samples for all ensembles.
All 3 ensembles share a constant density in the bulk and show differences at the edge. For our value of $N$ we define the bulk at radius $r_b=0.8$ (orange dash-dotted vertical line, left).
The right picture shows details at the edge, with the region to the right of the solid vertical line at radius $r_-=1-1/\sqrt{1024} = 0.96875$ defined as edge, and the dashed vertical line at 
$r_-^\prime=1-2/\sqrt{1024} = 0.9375$ as extended edge.
}
\label{fig:radial density plot}
\end{figure}
\begin{figure}[h]
    \includegraphics[width=0.50\linewidth]{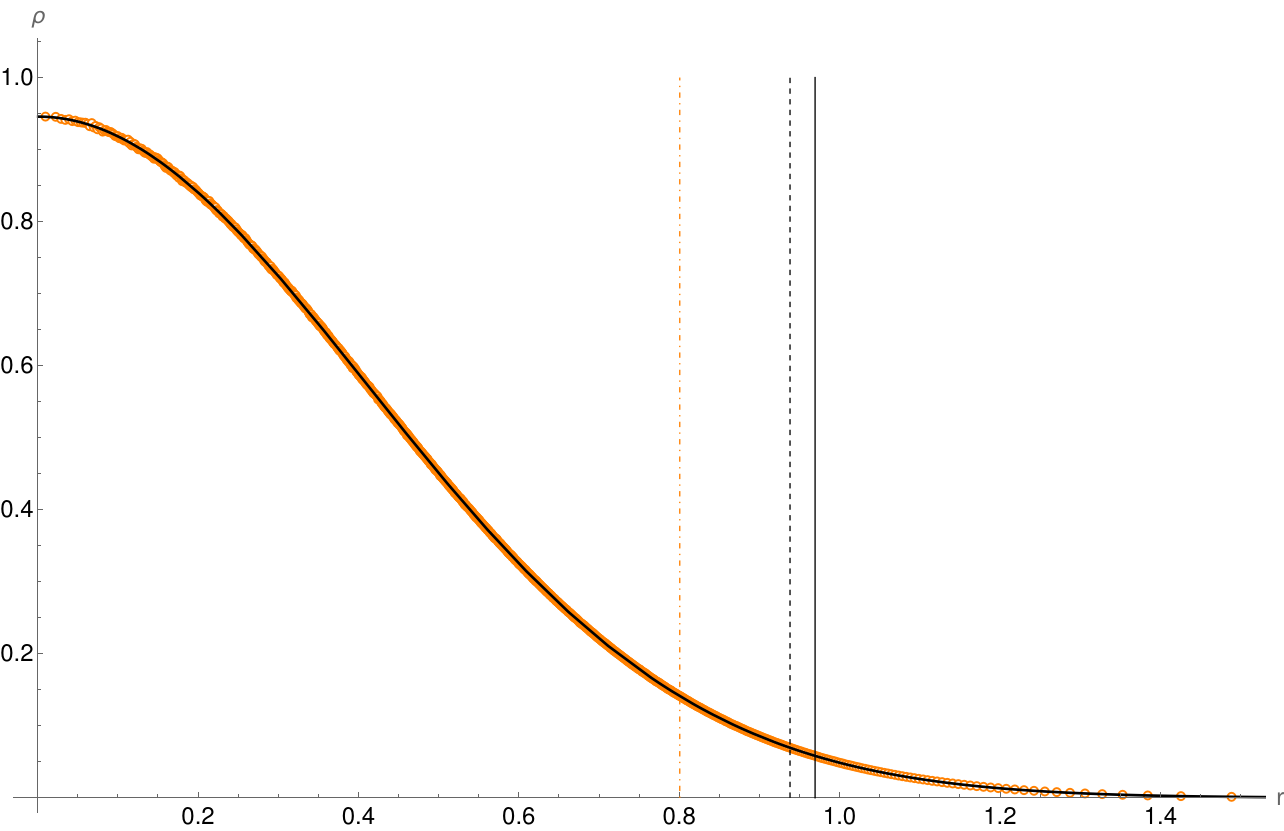}
    \caption{The radial spectral density of the 2D Poisson process for $N=1024$, analytically (black full line) and from $720\,000$ ensembles (orange circles). Notice the excellent agreement between the two overlapping curves. 
    The vertical lines show our choice for the bulk $r_b=0.8$ (orange dash-dotted), the edge $r_-=1-1/\sqrt{N}$ (black full line) and the extended edge $r_-^\prime=1-2/\sqrt{N}$ (black dashed).
}
\label{fig:radialDensityPoisson}
\end{figure}

In order to determine which region belongs to the bulk, and where does the edge start for the numerical analysis of a large but finite matrix size $N$ ($2N$), we first give a plot of the radial spectral density in  Fig. \ref{fig:radial density plot} (left).   All 3 classes have a rotationally symmetric spectral density. Thus we can fully characterise the bulk and edge region by the respective radius. 
In class A and AI$^\dag$ the spectral density $R^{(N)}(z)$ is known analytically for finite-$N$ and we briefly quote the results. In the GinUE the density follows from the Ginibre kernel of its determinantal point process, recalled in Appendix \ref{App:kernel}, cf. \eqref{eq.R1GinUE}, after rescaling with $N$. It is given by \cite{Ginibre}
\begin{equation}
R_{\rm A}^{(N)}(|z|=r)=\frac{1}{\pi} e^{-Nr^2}\sum_{j=0}^{N-1}\frac{(Nr)^{2j}}{j!}=\frac{\Gamma(N,Nr^2)}{\pi \Gamma(N)},
\label{R1A}
\end{equation}
where we also introduce the upper incomplete Gamma-function $\Gamma(N,x)$. The density is normalised to integrate to unity.  In class AI$^\dag$ the spectral density was obtained most recently \cite{AFS} and reads 
\begin{eqnarray}
R_{\rm AI^\dag}^{(N)}(|z|=r)&=&
\frac{\sqrt{2N}}{\pi r \Gamma(N+2)}\left[\sqrt{\frac{Nr^2}{2}}\left( (Nr^2)^Ne^{-Nr^2}+(N-\frac{Nr^2}{2})\Gamma[N,Nr^2]
\right) \right.
\nonumber\\
&&+\left.2^{\frac{N}{2}}\gamma\left(\frac{N+3}{2},\frac{Nr^2}{2}\right)\left(
(Nr^2)^{\frac{N}{2}}e^{-Nr^2/2} +\frac{(N-1-Nr^2)\Gamma(N,Nr^2)}{2(Nr^2)^{\frac{N}{2}}e^{-Nr^2/2}}
\right)
\right].\quad  
\label{R1AI+}
\end{eqnarray}
Note that compared to \cite{AFS} we have rescaled the edge to unity and give{n} the density without radial factor $r$ from the radial measure $\dif r\,r$,
corresponding to the circular (and not triangular) law here. Both densities are shown in  Fig. \ref{fig:radial density plot} and compared to the numerics.

All 3 ensembles share the same constant density in the bulk, with a sharp drop at the edge. Different features of the 3 classes are magnified in  Fig. \ref{fig:radial density plot} (right). The agreement between the analytical predictions at finite-$N$ \eqref{R1A} for class A and \eqref{R1AI+} for class AI$^\dag$ and our numerics is excellent.  
For class A it is long known \cite{Girko, Ginibre} that the global density approaches the circular law. For class AI$^\dag$ this has been shown very recently \cite{AFS}, while for class AII$^\dag$ the evidence so far is numerical. Our normalisation is such that in all classes we obtain $r=1$ as limiting radius for $N\to\infty$.

In \cite{ABCPRS, AMP} it was shown that in the bulk of the spectrum the local NN and NNN spacing distributions in class AI$^\dag$ and AII$^\dag$ 
are well approximated by a 2D Coulomb gas \eqref{eq.jpdf_gen.weight} at $\beta=1.4$ and $\beta=2.6$, respectively. In comparison, class A is known to be an exact realisation of $\beta=2$.
The observed behaviour of the 3 global densities at the edge in  Fig. \ref{fig:radial density plot} (right) is 
consistent with the numerically obtained $\beta$-dependence of the edge behaviour of the density of a true 2D Coulomb gas \cite[Fig. 1]{Edge behaviour beta>2}. There, for $\beta\leq2$ the radial density decreases monotonously towards the edge, where increasing $\beta$ makes the transition more sharp. For $\beta>2$, a local maximum develops at the edge, with its height increasing with $\beta$.
For this reason, in the following we will always plot and discuss the 3 classes in the order AI$^\dag$, A and AII$^\dag$ of increasing {\it effective} $\beta=1.4,\ 2$ and 2.6, respectively.

For classes A and AI$^\dag$ \cite{AFS} it is known analytically, that the width of the edge region scales as $O(1/\sqrt{N})$. The same scaling has been found in a symmetry class related to AII$^\dag$ \cite{OK}, too. Therefore, we will make a simple choice for the edge region as $r_-=1-1/\sqrt{N}$, with $r=1$ being the limiting radius of support (for AII$^\dag$ we choose $r_-=1-1/\sqrt{2N}$, to have the same numerical value). 
In order to test our choice, how far the edge extends for the given value of $N$ ($2N$), we will also compare to a larger choice called extended {edge}, with a twice as large deviation from unity, $r_-^\prime=1-2/\sqrt{N}$.

In  Fig. \ref{fig:radial density plot} (right) zooming into the edge behaviour, we also indicate by horizontal lines, from which radius $r_-=1-1/\sqrt{1024} = 0.96875$ (full line) onwards we count the eigenvalues with $|z_k|>r_-$ as edge eigenvalues, for the given value of $N=1024$ ($2N=1024$ for AII$^\dag$). For simplicity we take the same edge value for all 3 symmetry classes, as well as for 2D Poisson. 
As a check, in Subsection \ref{sec:spacing-ratio-mom} 
we compare with results obtained for a different, larger choice of the edge region $r_-' =1-2/\sqrt{1024} = 0.9375$ (dashed line). This takes into account the deviation of the radial density from the constant bulk value at smaller radii in classes  AI$^\dag$ and AII$^\dag$, compared to class A.

In all subsequent subsections, we will also compare to the spacing ratio, NN and NNN spacings of the 2D Poisson point process of uncorrelated, Gaussian random variables.
In line with the argument of interpreting different classes as 2D Coulomb gases at different effective values of $\beta$, our choice for 2D Poisson
aims at \eqref{eq.jpdf_gen.weight} at $\beta=0$. We choose a Gaussian weight $\omega(z)=\exp[-|z|^2/(2 \sigma^2)]$, such that the probability inside the
typical edge radius ($r_- < 1-1/\sqrt{N}$) is the same as the ratio of areas of edge disc and unit disc. This results in the variance \cite{BScCR}
\be
    2 \sigma^2 = -\frac{(1 - \frac{1}{\sqrt{N}})^2}{\log(1-(1 - \frac{1}{\sqrt{N}})^2)}.
\ee
The corresponding density is shown in Fig. \ref{fig:radialDensityPoisson}. Being given by a radially symmetric Gaussian, it is less obvious how to choose an
appropriate value of $r_b$ for the bulk and of $r_-$ for the edge. In order to avoid any bias we will choose the same values for $r_b$ and $r_-(r_-^\prime$) for the bulk and (extended) edge as for the other 3 ensembles.

An alternative choice for Poisson random variables in 2D  would be independent points with uniform probability on the unit disc. 
Although one might expect a strong effect through this hard cut-off at the edge, this realisation of 2D Poisson leads to rather similar results, cf.  \cite{BScCR}.


\subsection{Complex spacing ratios in 2D: Edge vs. bulk}\label{sec:spacing-ratio2D}

Let us begin by defining the complex spacing ratio as obtained numerically from the $N$ complex eigenvalues $z_1,\ldots,z_N$  an $N\times N$ non-Hermitian random matrix. The complex spacing ratio $\lambda_k$ of the $k$-th eigenvalue $z_k$, where no ordering is implied, is defined as 
\begin{align}\label{eq.def spacing ratio}
    \lambda_k=\frac{z_k^{\textsc{NN}}-z_k}{z_k^{\textsc{NNN}}-z_k}.
\end{align}
Here, $z_k^{\textsc{NN(N)}}$ denotes the NN(N) of $z_k$ in radial distance.
Despite the eigenvalue repulsion of the complex eigenvalues we consider, such a NN or NNN may not be unique, as two different complex eigenvalues may lie on the same circle of radius $|z_k^{\textsc{NN}}-z_k|$ sharing the same distance. In practice, the probability for this event is negligible, and choosing one of them as NN (and the other as NNN) will not affect our results.  

We observe that $\lambda_k$ is a complex variable in 2D.
We can assign a radius $r_k\geq0$ and an angle\newline $\phi_k\in(-\pi,+\pi]$ to it, $\lambda_k=r_k\exp[\iunit \phi_k]$. As the NNN is always further away (or at best at equal distance) than the NN, we have $r_k\leq 1$. One attractive feature of the complex spacing ratio is that it is claimed that no unfolding is needed. In 2D the unfolding procedure, i.e. extracting the local statistics from the averaged ones, is non-trivial, see \cite{CP,AKMP,MPW}. As the average spectral density at $z_k^{\textsc{NN}}$, and at $z_k^{\textsc{NNN}}$ are supposedly close, the unfolding will be (almost) the same in the numerator and denominator of eq. \eqref{eq.def spacing ratio}. We will see if this remains true at the edge of the spectrum, where the spectral density varies on the order of the mean level spacing.

\begin{figure}[h]

\begin{subfigure}{\textwidth}
\hspace{80pt}{AI$^\dagger$} \hspace{140pt}{A}\hspace{140pt}{AII$^\dagger$} \hfill\\
	\begin{turn}{90}\hspace{60pt}{bulk}\end{turn}
\includegraphics[width=0.3\linewidth]{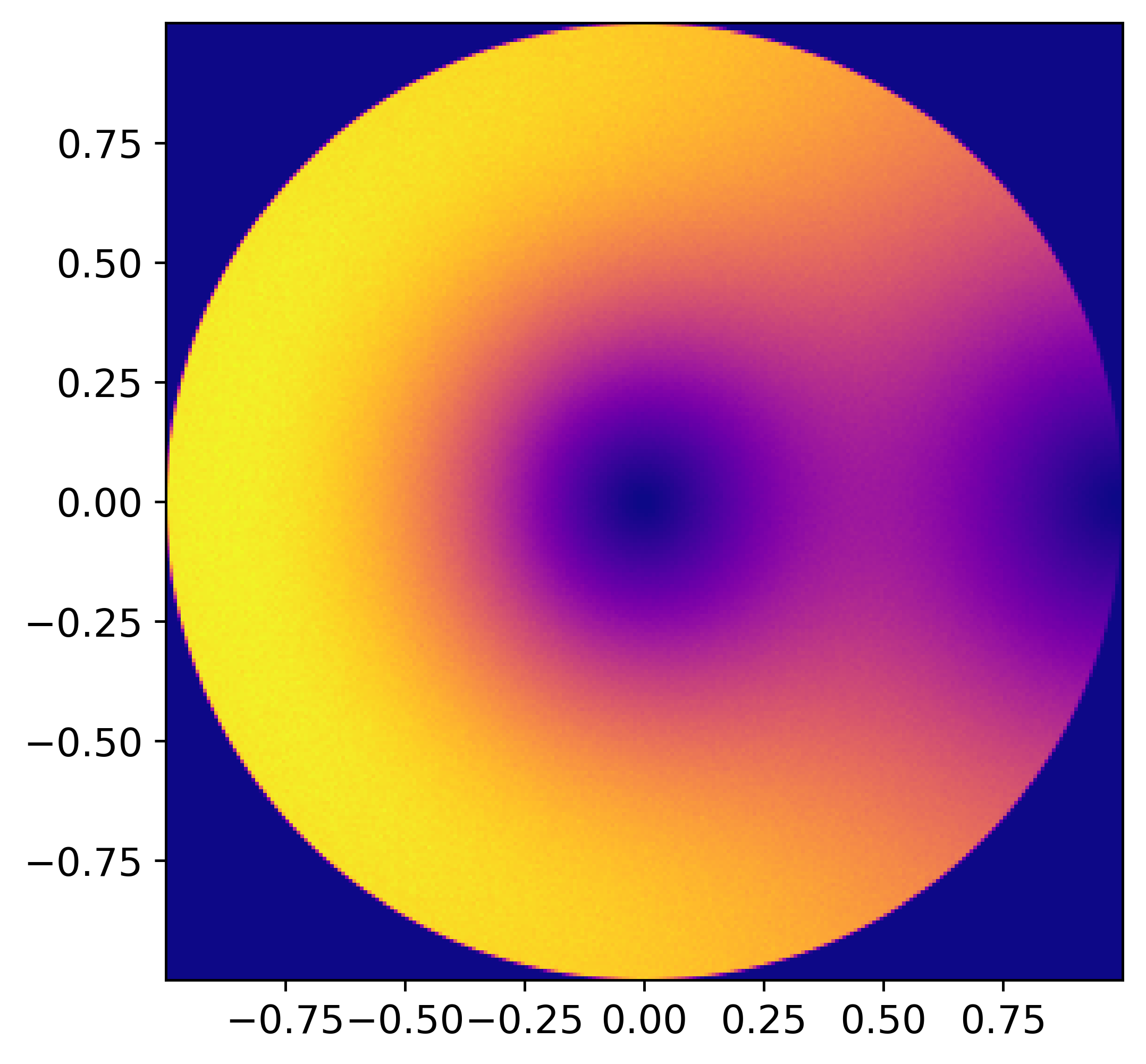} 
\includegraphics[width=0.3\linewidth]{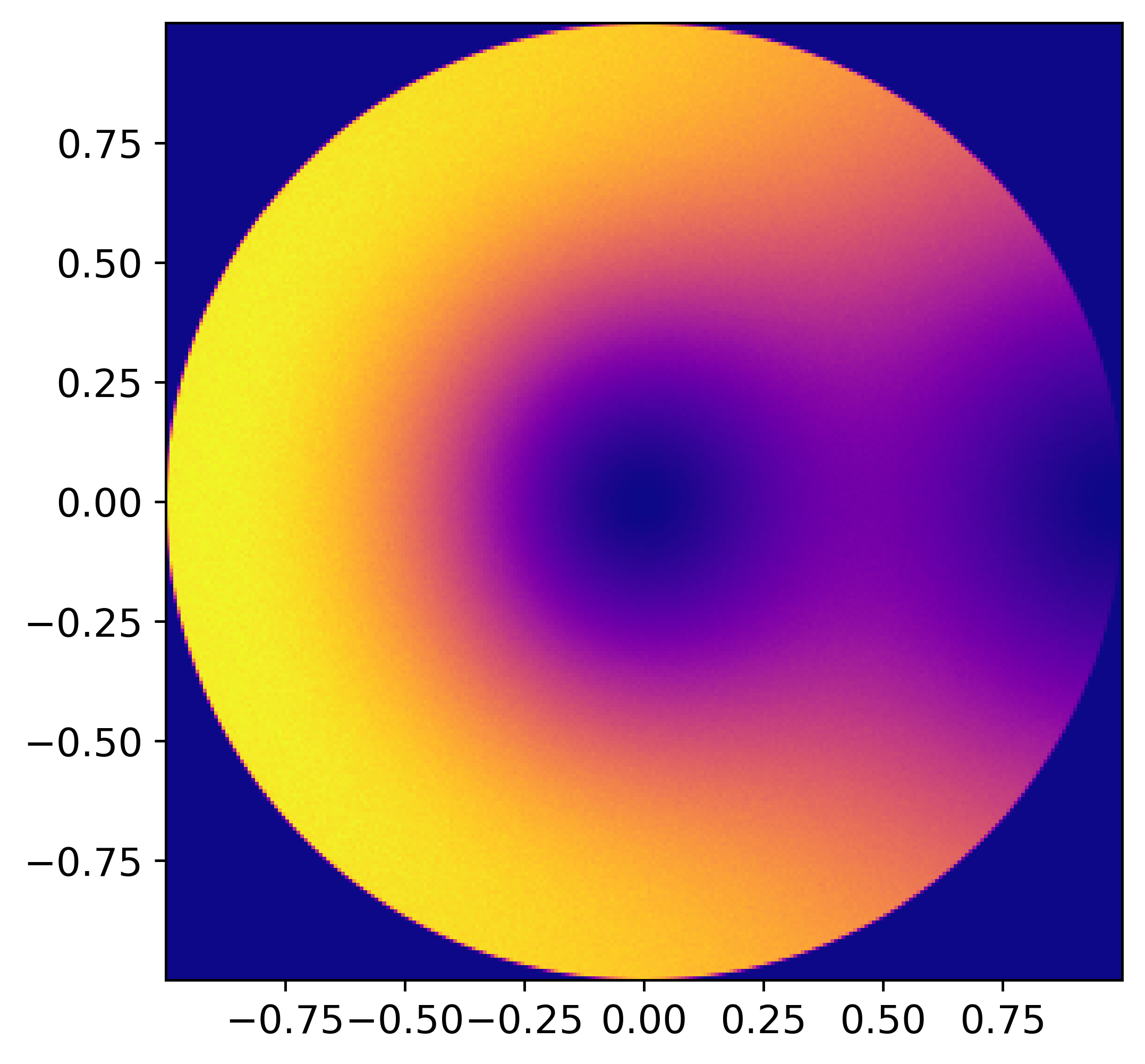}
\includegraphics[width=0.3\linewidth]{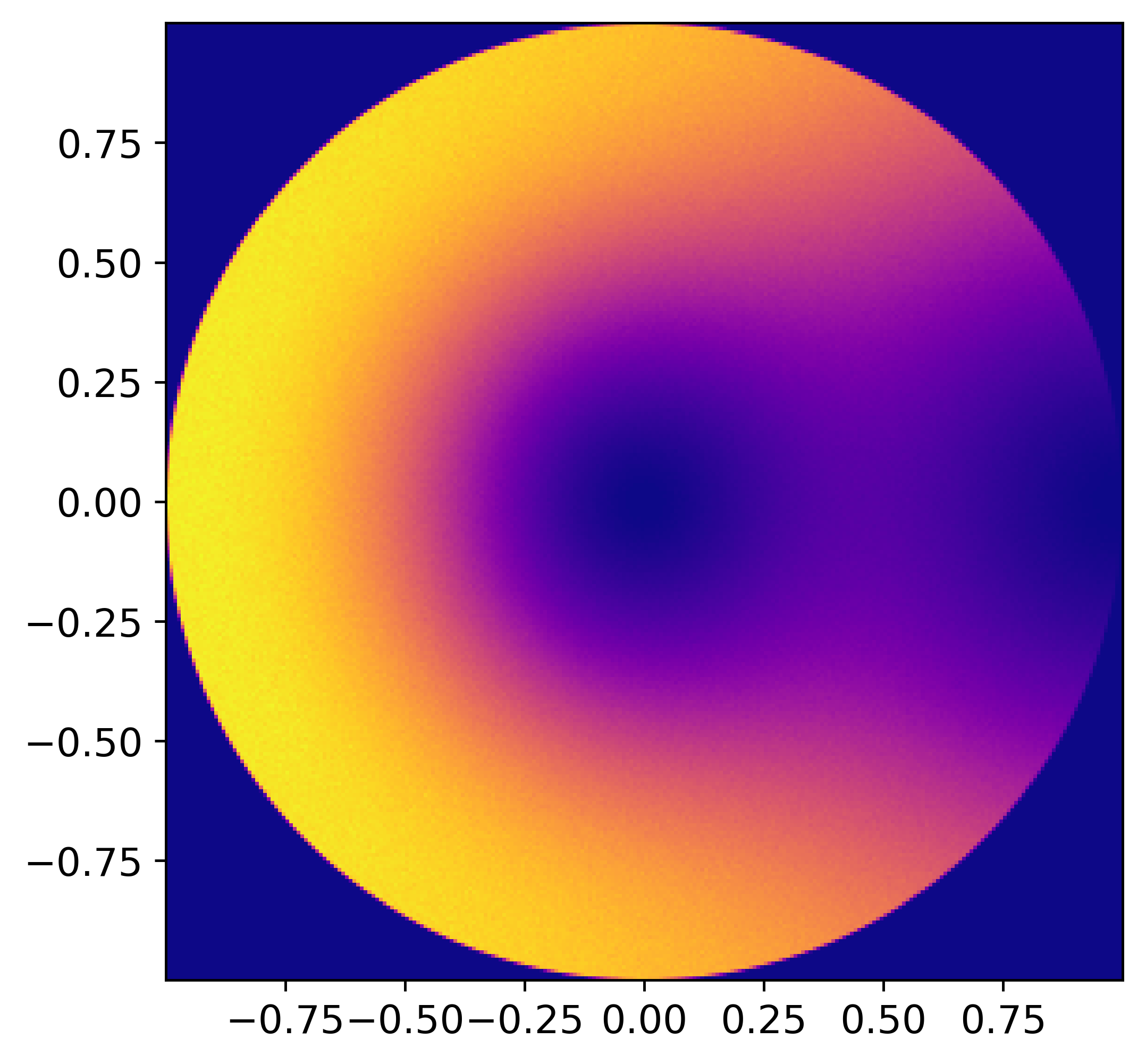}  
\raisebox{5pt}{\includegraphics[scale=0.46]{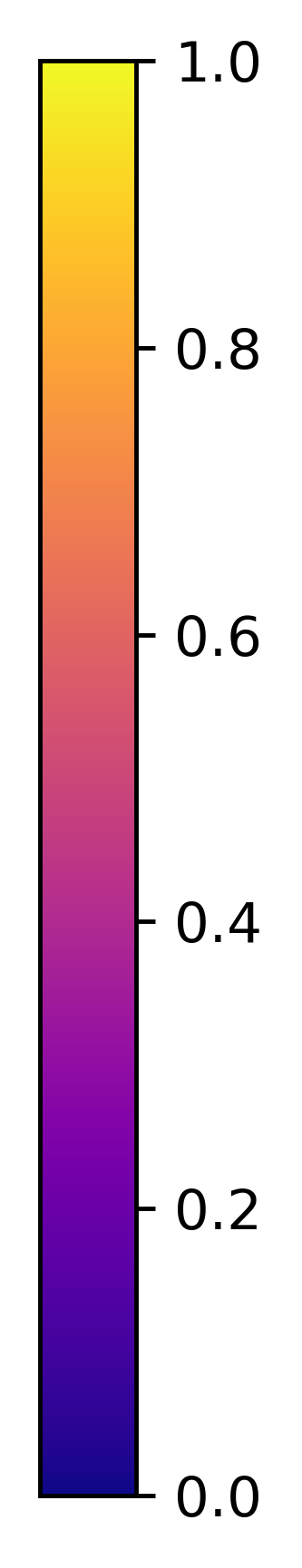}}
\label{fig:spacing ratio bulk}
\end{subfigure}
\begin{subfigure}{\textwidth}
\begin{turn}{90}\hspace{60pt}{edge}\end{turn}
\includegraphics[width=0.3\linewidth]{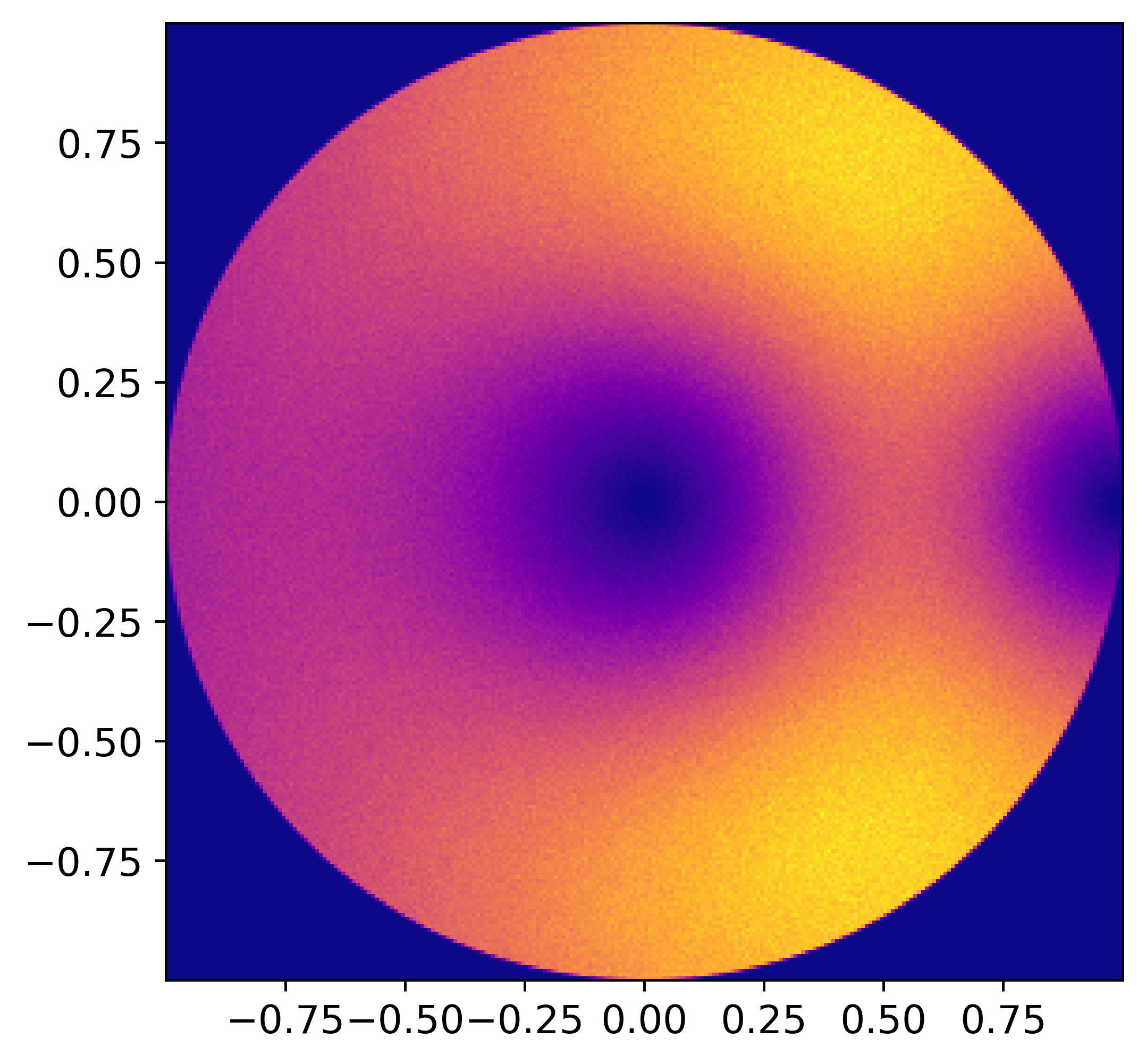}
\includegraphics[width=0.3\linewidth]{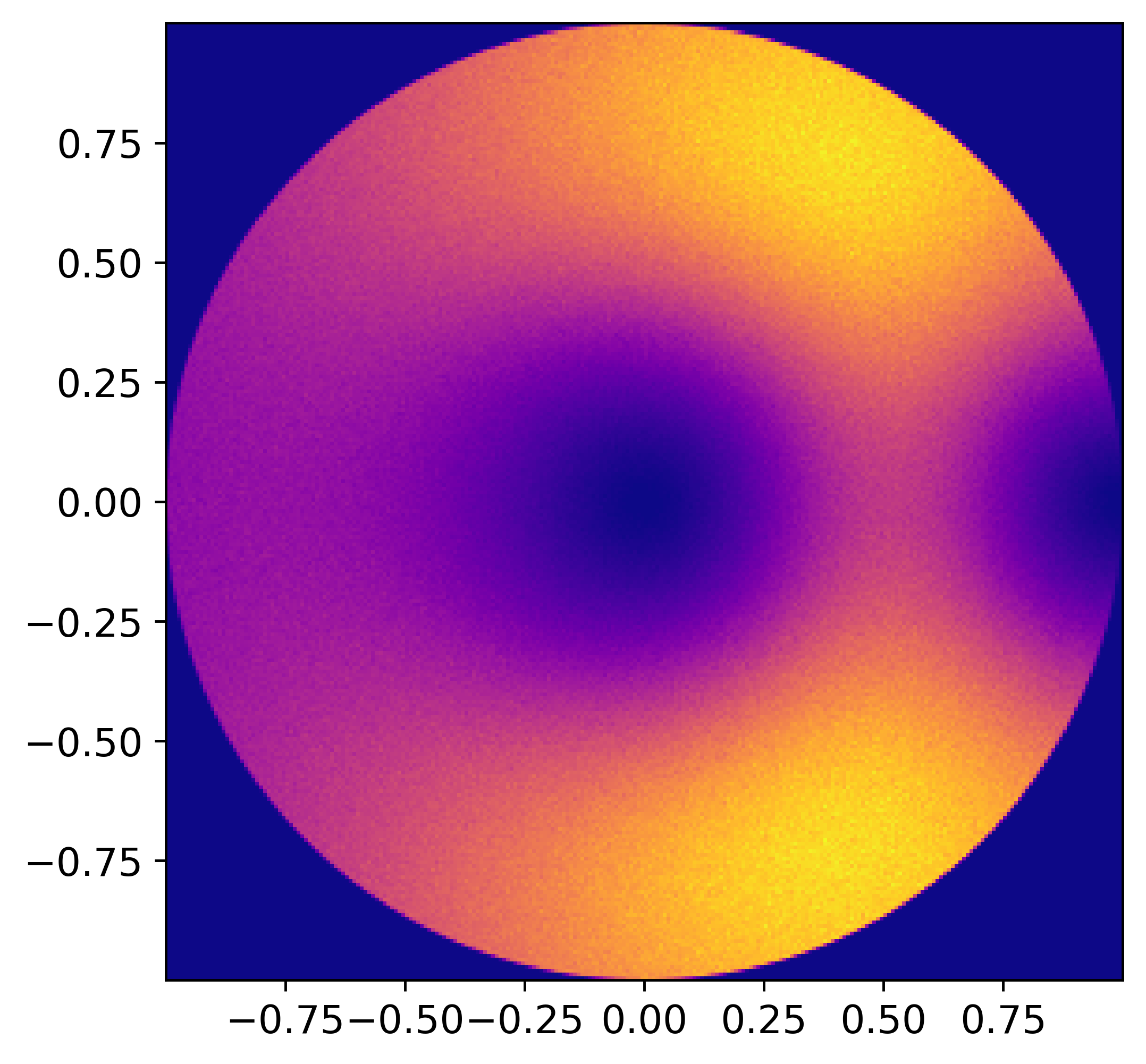}
\includegraphics[width=0.3\linewidth]{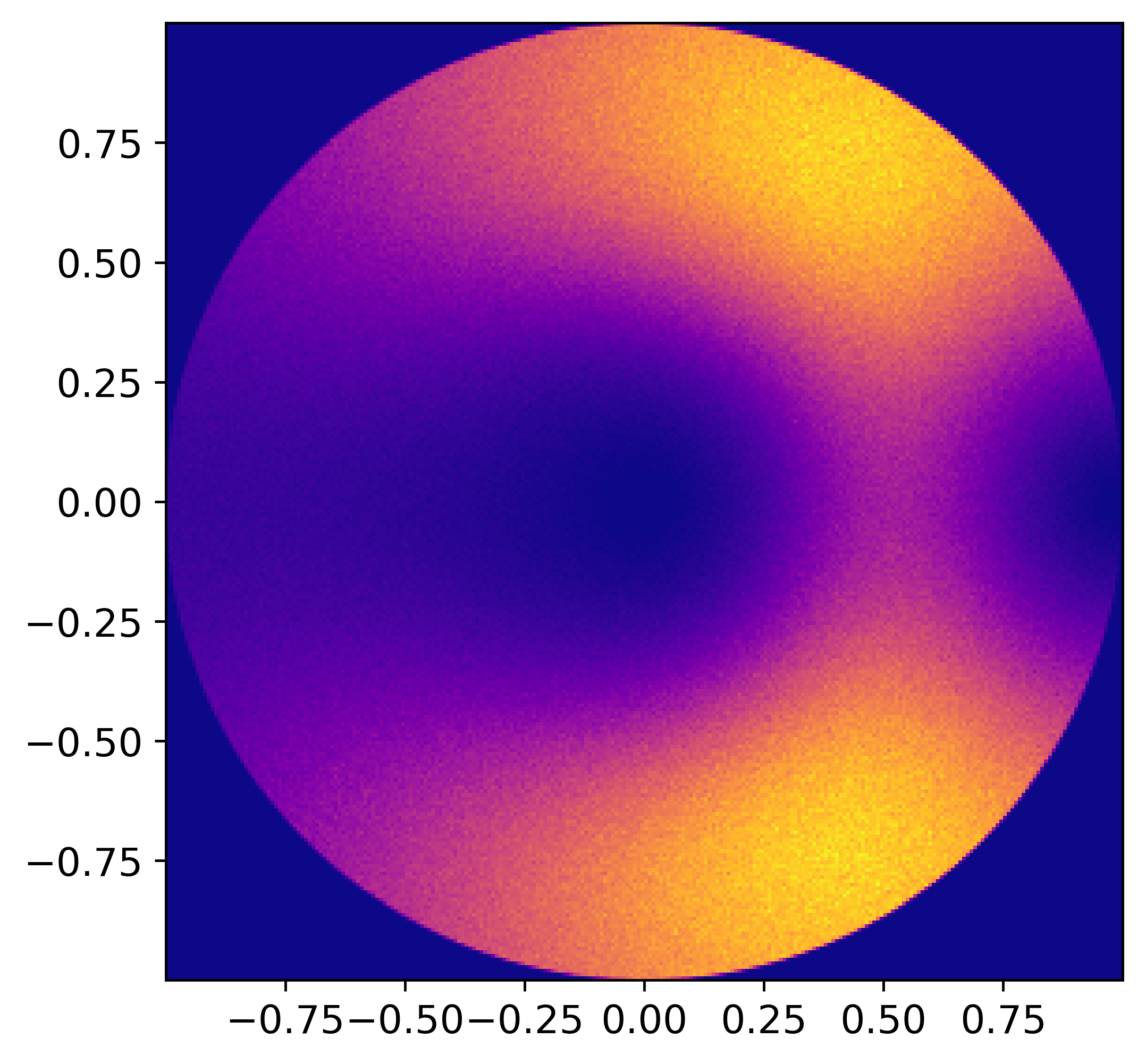}
\raisebox{5pt}{\includegraphics[scale=0.46]{figures/colorbar.png}}
\label{fig:spacing ratio edge}
\end{subfigure}
\caption{Density plots of the complex spacing ratio in the { bulk (top rows)} and at the { edge (bottom rows)} for classes AI$^\dag$, A, AII$^\dag$, from left to right. 
We show $720\,000$ ensemble realisations of size $N=1024$ (classes AI$^\dag$, A) and  $2N=1024$ (class AII$^\dag$). The colour code gives the density from low (dark, blue) to high (bright, yellow).}
\label{fig:spacing ratio symmetry classes}
\end{figure}

In the following we will call the complex spacing ratio $\lambda_k$ to belong to the bulk, if the corresponding point satisfies $|z_k|<r_b$, having a radius less than what we chose to belong to the bulk. On the other hand it belongs to the (extended) edge, if it holds $|z_k|>r_-(r_-^\prime)$. Obviously this removes some eigenvalues with $r_b\leq |z_k|\leq r_-(r_-^\prime)$ from our statistics, where it is not clear a priori, if they belong to the bulk or to the edge. Choosing two different values $r_-^\prime <r_-$ for the edge tests how far the edge actually extends.
It is remarkable that even in class A only approximate expressions exist for the complex spacing ratio in the bulk, as we have seen in the previous Section \ref{Sec:AnalyticsA}.

We now present the 2D density plots of the distribution of the complex spacing ratio \eqref{eq.def spacing ratio} for classes AI$^\dag$, A, AII$^\dag$, from left to right in Fig. \ref{fig:spacing ratio symmetry classes}, both in the bulk (top) and the edge (bottom). Here, a brighter yellow intensity indicates a higher density as a function of $z=r e^{\iunit \theta}=x+\iunit y$.
Our results
in the bulk agree with earlier studies by \cite{SaRibeiroProsen,DusaWettig}, where we add the following interpretation. 
The low density in the centre of the plot (dark central hole) represents the repulsion of the eigenvalues, i.e. the probability that two eigenvalues are getting very close to each other is going to zero. We observe that this region increases {from} class AI$^\dag$ (top left)  over class A (top middle) to class AII$^\dag$ (top right).
This is in agreement with an effective Coulomb gas description at $\beta=1.4$, $\beta=2$ and $\beta=2.6$ from left to right, as found in the local bulk statistics  \cite{ABCPRS,AMP}. 

As a second feature, notice that, on the horizontal $x$-axis with the origin in the middle, points on the positive $x$-axis correspond to an angle $\theta=0$, that is NN and NNN are aligned. In contrast, points on the negative $x$-axis correspond to $\theta=\pi$, that is, NN and NNN are anti-aligned, cf. Fig. \ref{fig:alignedspacings} below.
We observe, that due to the repulsion between points, the density for aligned NN and NNN with $\theta\approx 0$ is also suppressed, in particular close to radius $r=1$ (dark right hole). Again the effect increases with the repulsion in $\beta$.

\begin{figure}[h]
\includegraphics[width=0.3\linewidth]{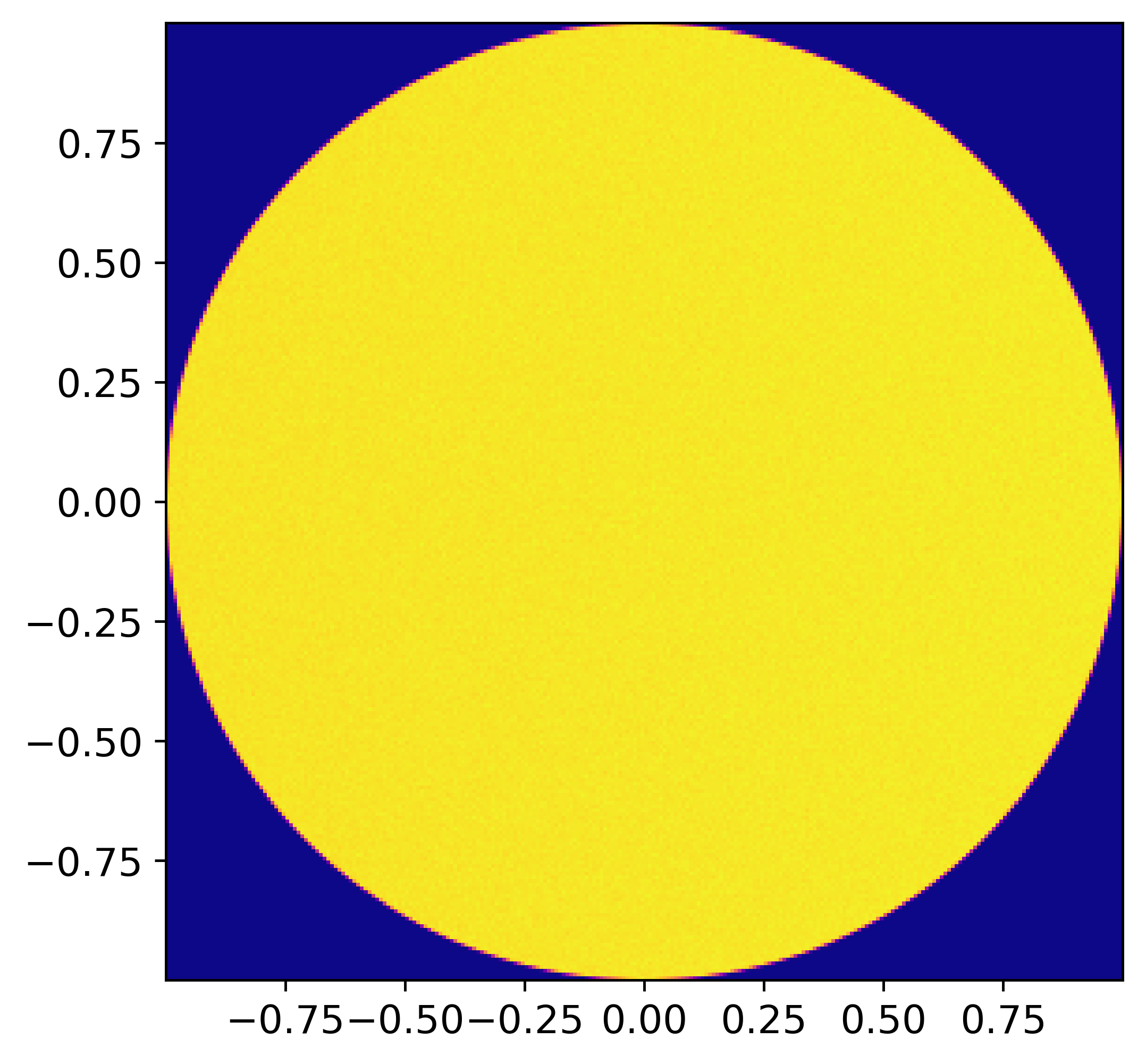} 
\includegraphics[width=0.3\linewidth]{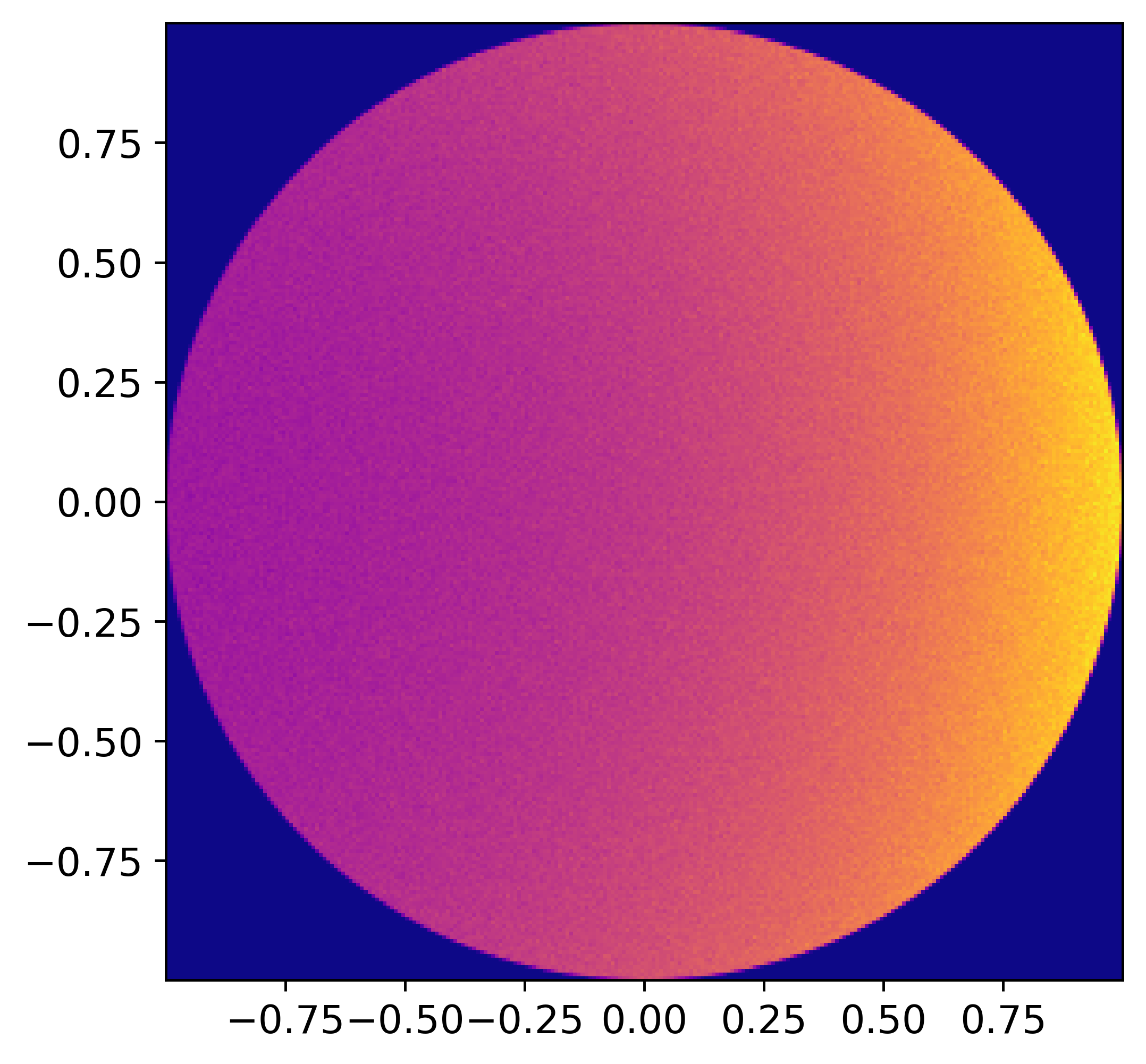}
\raisebox{5pt}{\includegraphics[scale=0.46]{figures/colorbar.png}}
\caption{Density plots of the complex spacing ratio in the bulk (left) and at the edge (right) for the 2D Poisson ensemble. We consider $720\,000$ ensembles of size $N=1024$.
}
\label{fig:spacing ratio Gaussian Poisson}
\end{figure}

At the edge, in the bottom row in Fig. \ref{fig:spacing ratio symmetry classes}, we first of all observe the same repulsion pattern and ordering in $\beta$ as in the bulk at the origin and small angle, with a weakening repulsion at $\theta\approx0$. In addition, also the case of anti-aligned NN and NNN at $\theta\approx \pm\pi$ is suppressed, again increasing with the repulsion in $\beta$. How can we understand this additional feature at the edge?

For comparison, let us first look at the edge effect in the complex spacing ratio of the 2D Poisson ensemble.  In Fig. \ref{fig:spacing ratio Gaussian Poisson} we present the bulk (left) and edge spacing ratio distribution (right). Due to vanishing repulsion, the complex spacing ratio is flat in the bulk. 
At the edge (right), we observe a suppression except for small to intermediate angles close to radius $r\lessapprox 1$. 
As drawn schematically in Fig. \ref{fig:alignedspacings},
the reason for that is when a point $z_1\ (\Delta)$ is in the edge region, it is favourable to have both NN $z_2\ (\bullet)$ and NNN $z_3\ (\circ)$ aligned in the direction of the increase of the density 
towards the bulk,  
as there is no repulsion between them. For anti-aligned points, most likely one of the points is at a larger radius, $r_k\gtrapprox1$, where the density is highly suppressed---unless they anti-align parallel to the edge. 

Coming back to the effect at the edge spacing ratio for classes AI$^\dag$, A, AII$^\dag$, in the bottom row in Fig. \ref{fig:spacing ratio symmetry classes} there is a competition between repulsion (in favour of $\theta\approx \pm\pi$) and an almost vanishing density outside the edge (in favour of $\theta\approx0$). As a consequence, an angle in around $\theta\lessapprox\pm\pi/2$ in the first and fourth quadrant is favoured in these plots, balancing repulsion and a vanishing spectral density outside the support (cf. sketch in Fig. \ref{fig:alignedspacings} right).
This is most pronounced for class AII$^\dag$, with the largest effective value of $\beta$. 
The same behaviour can be seen  much clearer when considering radial moments, see Fig. \ref{fig:spacing ratio 1D dist} (bottom middle), as discussed in the next subsection.

The above comparison of the complex spacing ratio in the bulk and at the edge assumes, that the unfolding drops out also at the edge in the ratio of the difference with the NN and NNN eigenvalue \eqref{eq.def spacing ratio}, respectively. We notice that at the edge the density varies strongly on the scale of the mean level spacing $\sim O(1/\sqrt{N})$, see Fig. \ref{fig:radial density plot}, and we will get back to this question in Subsection \ref{sec:NN}.

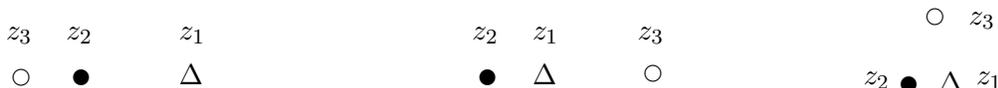
\begin{figure}
\setlength{\unitlength}{1mm}
\begin{picture}(150,15)
\put(10,1){\circle{2}}
\put(18,1){\circle*{2}}
\put(31,0){$\Delta$}
\put(8,6){$z_3$}
\put(16,6){$z_2$}
\put(31,6){$z_1$}
\put(72,1){\circle*{2}}
\put(78,0){$\Delta$}
\put(94,1.5){\circle{2}}
\put(70,6){$z_2$}
\put(78,6){$z_1$}
\put(92,6){$z_3$}
\put(128,0){\circle*{2}}
\put(132,-1){$\Delta$}
\put(131.5,9){\circle{2}}
\put(122,0){$z_2$}
\put(137,0){$z_1$}
\put(136,8){$z_3$}
\end{picture}
\caption{Left: {\it aligned} NN $z_2\ (\bullet)$ and NNN $z_3\ (\circ)$ of point $z_1\ (\Delta)$ with angle $\theta\approx 0$ in the complex plane; Middle:  {\it anti-aligned} NN and NNN with angle $\theta\approx \pm\pi$; Right: NN and NNN with angle $\theta\approx \pi/2$.}
\label{fig:alignedspacings}
\end{figure}


\subsection{Moments of complex spacing ratios}\label{sec:spacing-ratio-mom}

We turn to the study of 1D and 0D moments of the complex spacing ratio, by integrating over the radial, respectively angular part of the distribution, or both. Such 1D curves or real values are much more favourable for a comparison to data. Let us recall the definitions 
\eqref{eq.rad-marg} and \eqref{eq.ang-marg}  of the radial respectively angular marginal for all symmetry classes (for class A we consider $\tau=0$ only):
\begin{eqnarray}
 \rho^{(N)}(r)&=& \int_{-\pi}^{\pi} \dif \theta \; r \varrho^{(N)}(r\exp{[\iunit\theta]}). 
 \label{eq.rad-margN}\\
    \rho^{(N)}(\theta)  &=& \int_{0}^{1} \dif r \; r \varrho^{(N)}(r\exp{[\iunit\theta]}) ,
\label{eq.ang-margN}
\end{eqnarray}
We integrate with respect to polar coordinates, $\dif^2z=\dif r\;r \dif\theta$, {\it including} the radial part $r$ in the angular integral. For a constant density  $\varrho^{(N)}(r\exp{[\iunit\theta]})=const.$, as in 2D Poisson, see Fig. \ref{fig:spacing ratio Gaussian Poisson} (left), we thus obtain a linear radial density from \eqref{eq.rad-margN}.

For the scalar, 0D moments to be discussed later, see Tables \ref{tab:scalar moments bulk} and \ref{tab:scalar moments edge} and Fig. \ref{fig:scalar moments spacing ratio}, we define
\begin{eqnarray}
\langle r^{k} \cos(\phi)^l\rangle &:=&  \int_{-\pi}^{\pi} \dif \theta\int_{0}^{1} \dif r \, r^{k+1}
\cos(\theta)^l
\varrho^{(N)}(r\exp{[\iunit\theta]}) ,\quad k,l=0,1,2.
\label{eq.r2mom}
\end{eqnarray}
\begin{figure}[h]
\hspace{80pt}{bulk} \hspace{130pt}{edge}\hspace{130pt}{(bulk-edge)} \hfill\\
	\begin{turn}{90}\hspace{50pt}{radial}\end{turn}
\includegraphics[width=0.32\linewidth]{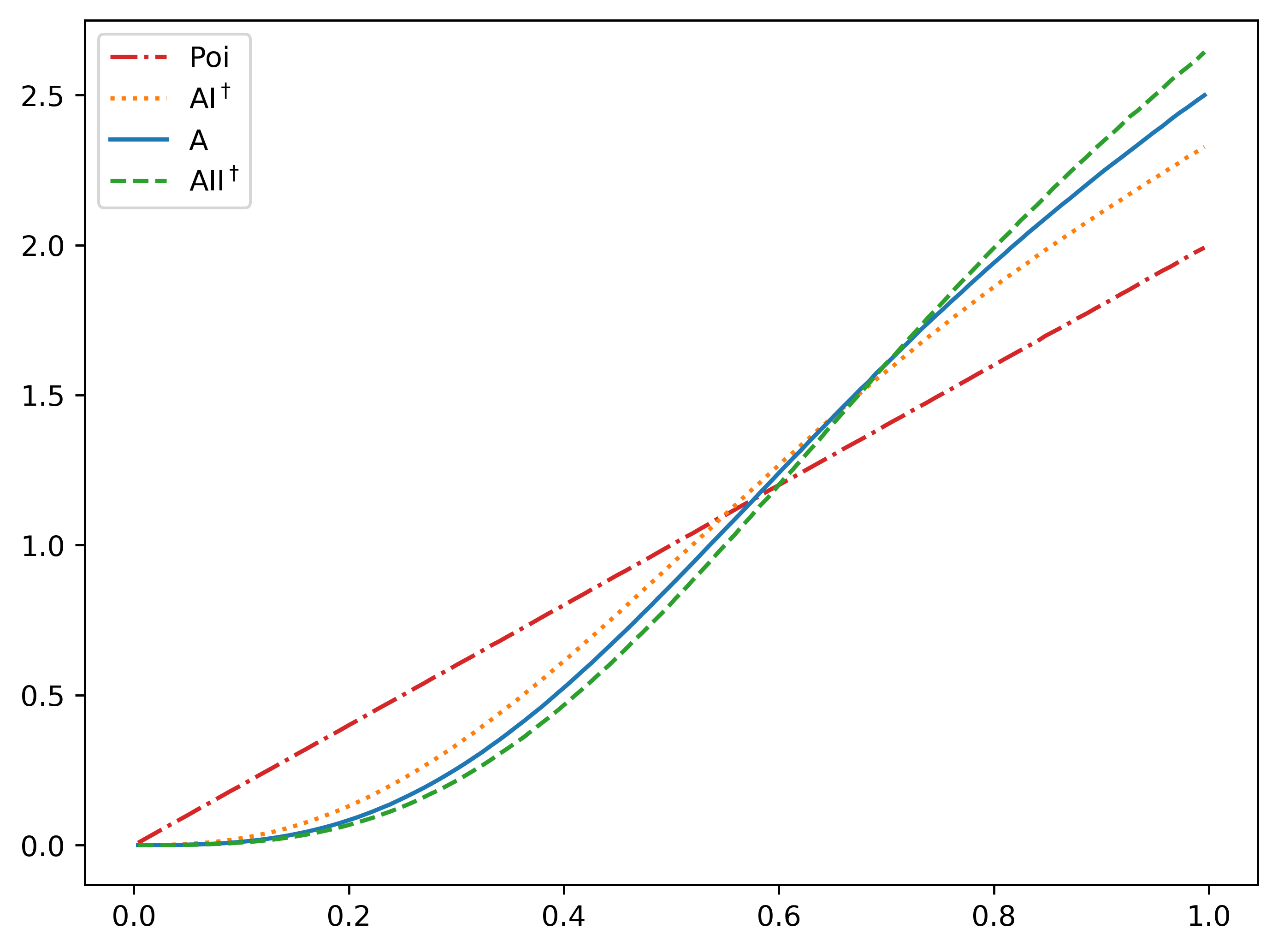} 
\includegraphics[width=0.32\linewidth]{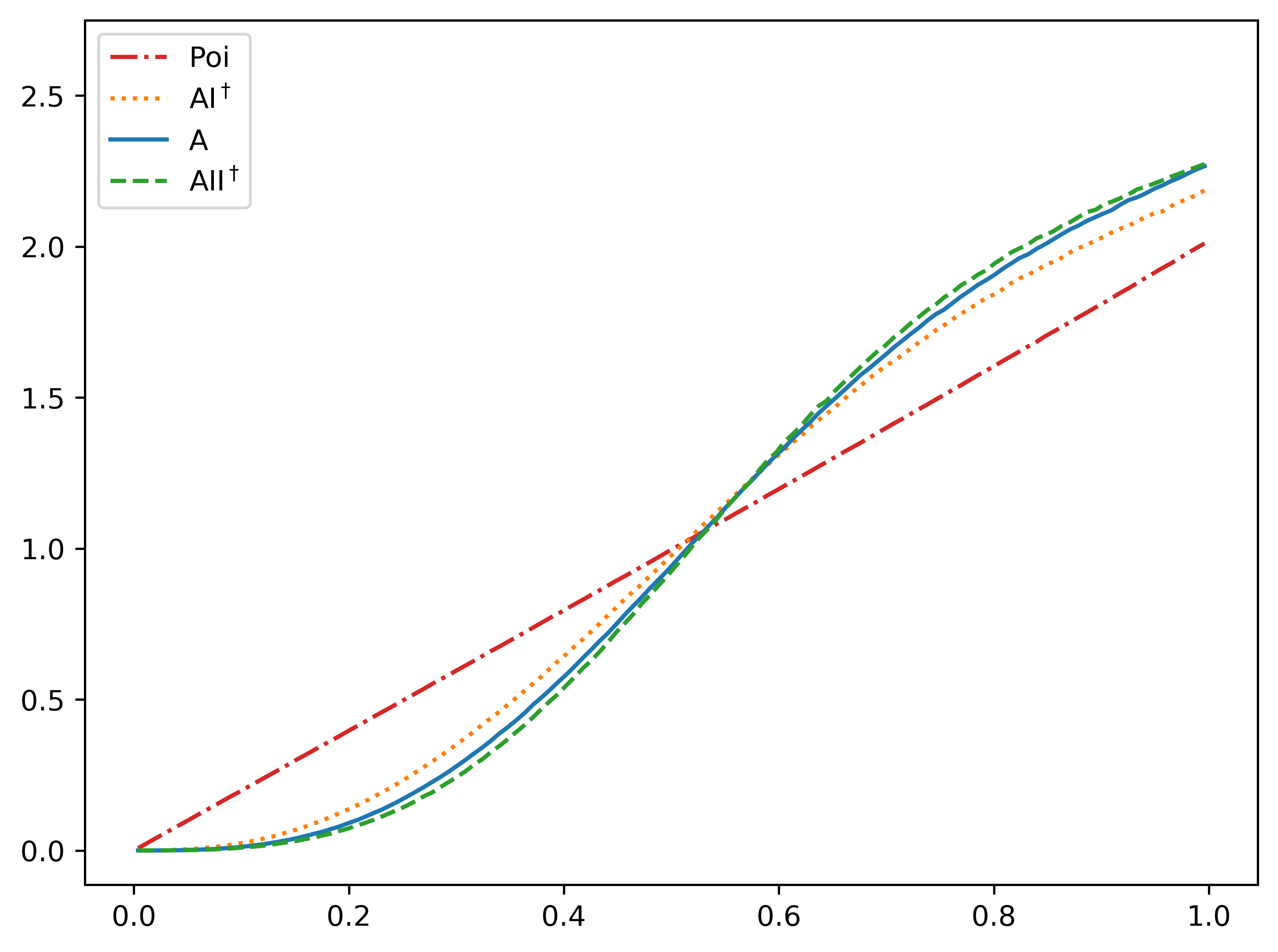}
\includegraphics[width=0.32\linewidth]{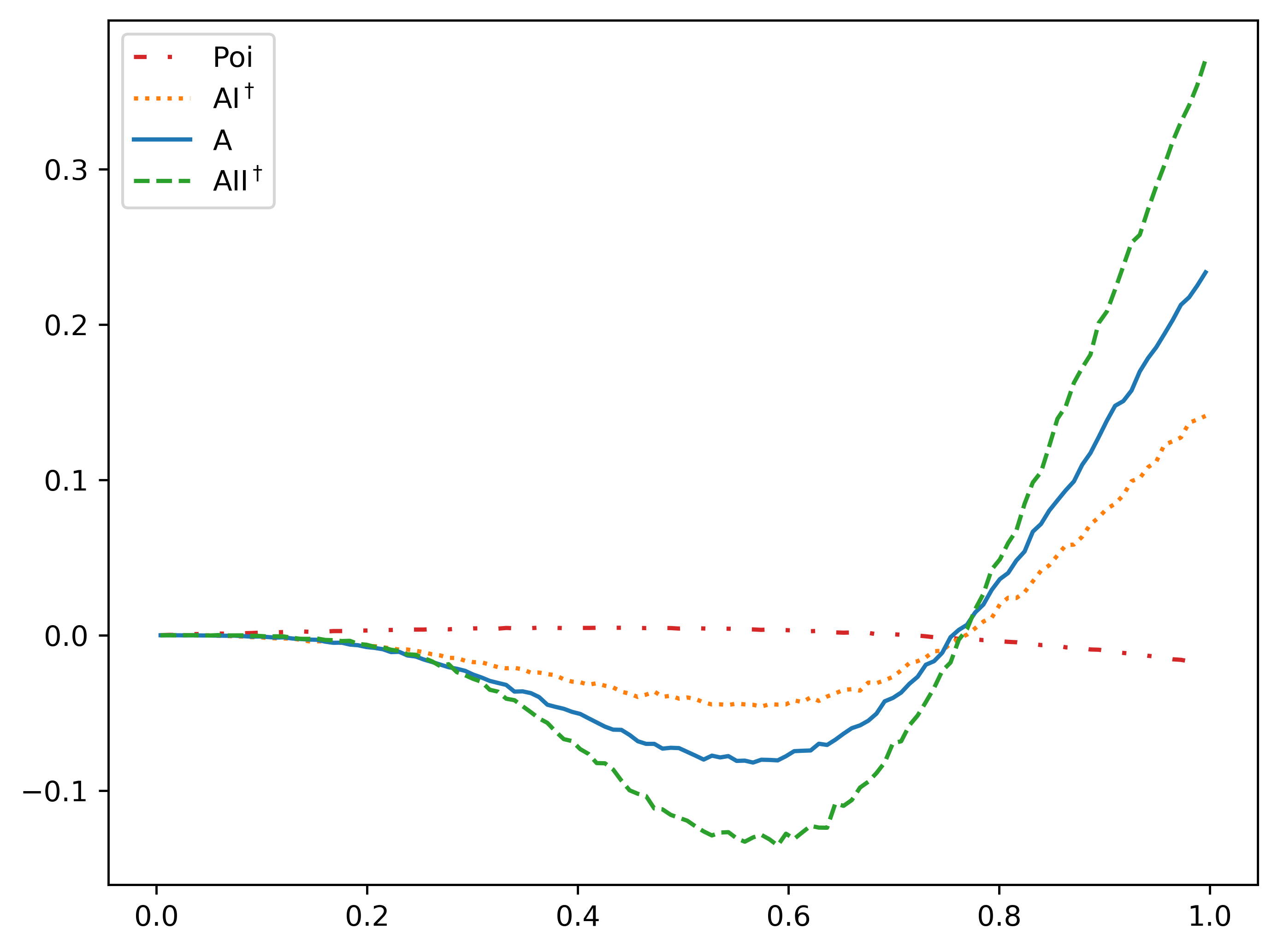}\\
\begin{turn}{90}\hspace{50pt}{angular}\end{turn}
\includegraphics[width=0.32\linewidth]{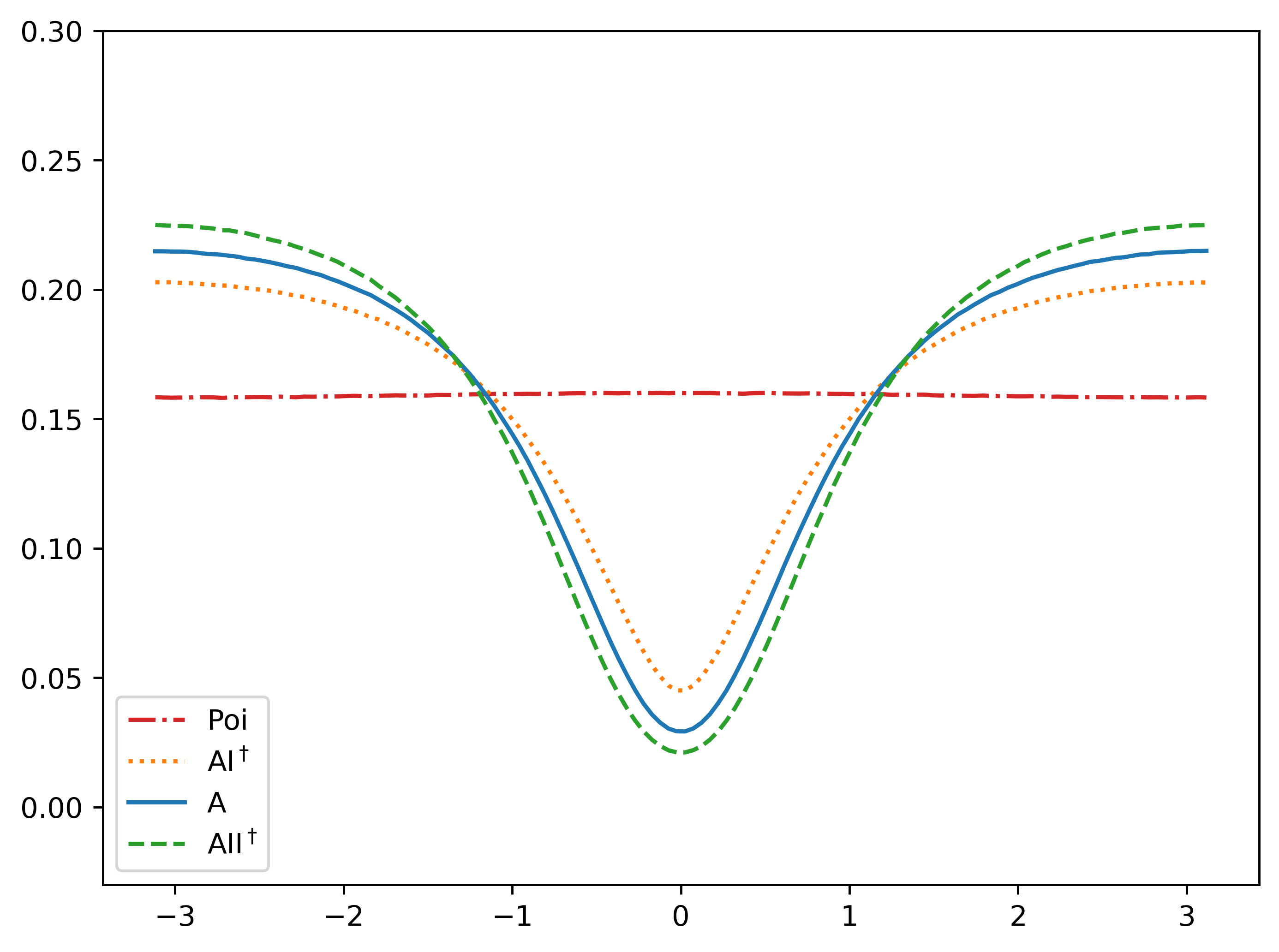}
\includegraphics[width=0.32\linewidth]{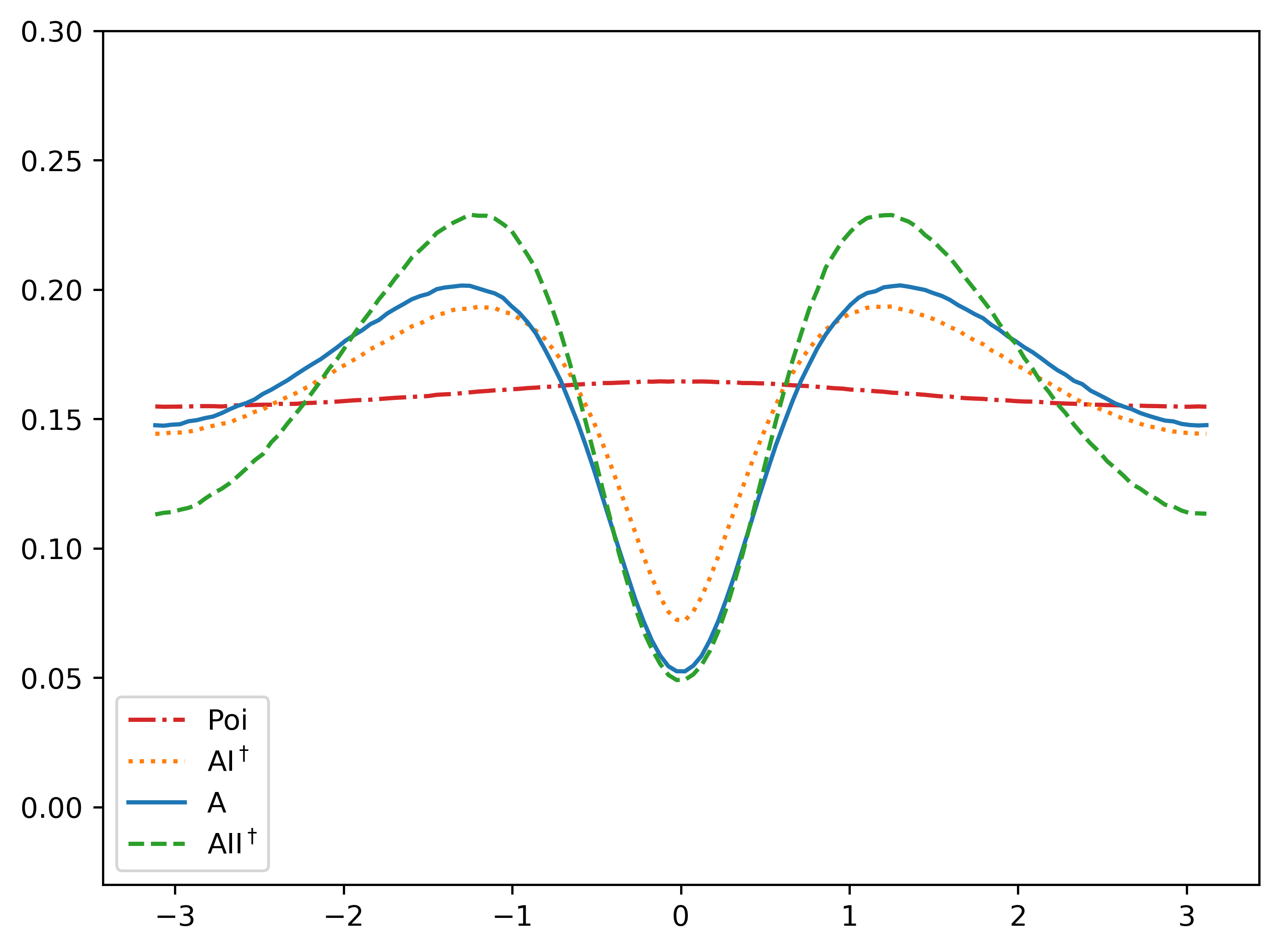}
\includegraphics[width=0.32\linewidth]{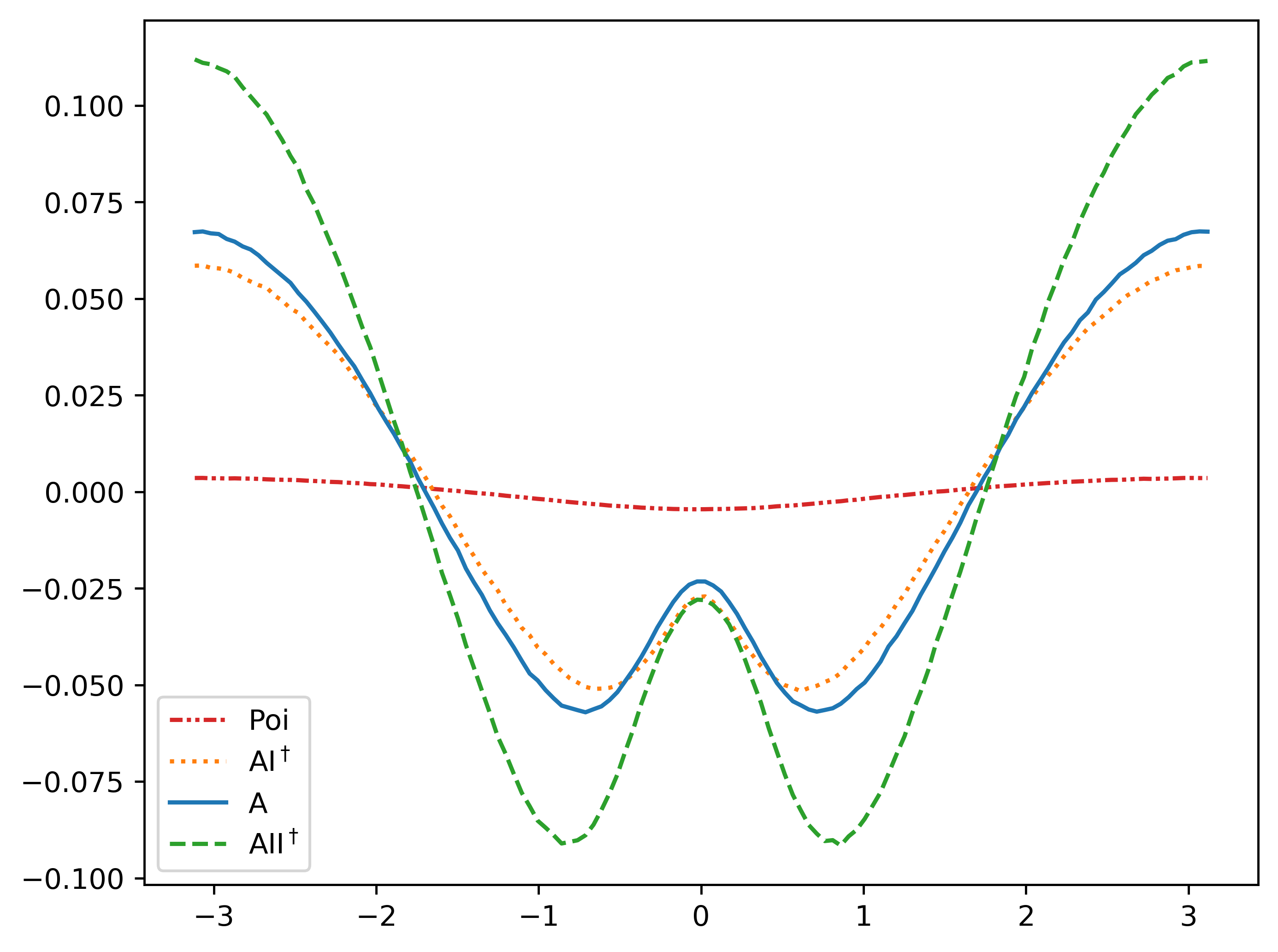}
\caption{Radial (top row) and angular (bottom row) marginal distributions \eqref{eq.rad-margN} 
and  \eqref{eq.ang-margN}
in the bulk (left) for $r<r_b$, at the edge (middle) for $r>r_-$, and their difference bulk minus edge (right).
The curves denote 2D Poisson (red dot-dashed), class AI$^\dag$ (orange dotted), A (blue full) and AII$^\dag$(green dashed),
each with $720\,000$ copies of $N=1024$ matrices, respectively $2N=1024$ for class AII$^\dag$.}
\label{fig:spacing ratio 1D dist}
\end{figure}

The results for the marginal densities defined above are shown in Fig. \ref{fig:spacing ratio 1D dist}: 
The left plots give the radial (top) and angular (bottom) bulk distribution, agreeing with previous studies (cf. \cite{SaRibeiroProsen, KanazawaWettig}).
The same marginals at the edge are given in the middle column, and their difference to the bulk in the right column. In all plots the same coloured line code is used:
2D Poisson (red dot-dashed), complex symmetric AI$^\dag$ (orange dotted), complex Ginibre A (blue full) and complex self-dual AII$^\dag$ (green dashed). 
We briefly recall the findings in the bulk.
For the 2D Poisson process there is no correlation, with a linear rise in the radial (top) and flat curve in angle (bottom)---in agreement with the flat density in Fig. \ref{fig:spacing ratio Gaussian Poisson} (left).  
The 3 symmetry classes AI$^\dag$, A, AII$^\dag$ show a repulsion increasing with the effective $\beta$ from the origin in radius (top) and from angle $\theta=0$ in the bulk plots.

Let us turn to the new results for the radial and angular distribution at the edge, 
in the second column of Fig. \ref{fig:spacing ratio 1D dist}.   
While the radial distribution (top middle) is qualitatively very similar to the bulk results, see top right for the quantitative differences (in different units),  the situation for the angular distribution differs significantly at the edge, see the  bottom middle plot of Fig. \ref{fig:spacing ratio 1D dist}. First, in 2D Poisson a small angle is now slightly favoured, as we already saw in Fig. \ref{fig:spacing ratio Gaussian Poisson} right. 
For the other three symmetry classes, in addition to the minimum at $\theta=0$ (i.e. suppressed aligned NN and NNN),
there is a second minimum suppressing anti-aligned NN and NNN at $\theta=\pm\pi$. In between, the distributions develop two maxima at approximately $\theta\approx \pm\pi/2$, symmetrical with respect to $\theta \rightarrow -\theta$. The height/depth of the maxima/minima increases with the repulsion in the effective $\beta$ values. 
This makes the picture we obtained from the previous subsection from the 2D plots more quantitative.


The next step is to study the 0D scalar moments of the complex spacing ratio, see \eqref{eq.r2mom}.
These quantities certainly depend on the dimension and ensemble size of the studied class, as discussed in \cite[Fig. 1]{DusaWettig}. This has to be kept in mind when comparing to the values in the literature. We choose a sufficiently large enough $N=1024$ (respectively $2N=1024$) 
to allow for a large number of ensembles within each class. This serves as a good compromise between large $N$ (which is computationally intensive) and the number of ensembles
 to minimise the statistical error.

 \begin{table}[h]\centering
	\caption{Various scalar moments in the bulk (up to $r<r_b=0.8$), for $720\,000$ matrices of size $N=1024$ (respectively $2N=1024$). 
We reproduce the known results \cite{SaRibeiroProsen, DusaWettig, KanazawaWettig} for 2D Poisson (analytical and numerical) and symmetry classes AI$^\dag$, A, AII$^\dag$.
	}
\begin{tabular}{|c|c || c | c | c | c | c|}
    \hline
&{\bf bulk moments}	 & $ \langle r \rangle $ & $ \langle r^2 \rangle $ & $ \langle \cos(\phi) \rangle $ & $ \langle \cos(\phi)^2 \rangle $ & $ \langle r\cos(\phi) \rangle $ \\ \hline
	 & 2D Poisson exact & $2/3$ & $1/2$ & $0$ & $1/2$ & $0$\\ \hline
	\multirow{4}{*}{$r<0.8$}&2D Poisson & $0.66666 $ & $ 0.50000 $ & $\ 0.00273 $ & $ 0.50000 $ & $\ 0.00204 $  \\ \cline{2-7}
	& class AI$^\dag$ & $ 0.72231 $ & $ 0.56065 $ & $ -0.19532 $ & $ 0.45780 $ & $ -0.14868 $ \\ \cline{2-7}
	& class A & $ 0.73867 $ & $ 0.58086 $ & $ -0.24685 $ & $ 0.44991 $ & $ -0.18884 $\\ \cline{2-7}
	& class AII$^\dag$ & $ 0.74893 $ & $ 0.59431 $ & $ -0.28580 $ & $ 0.44684 $ & $ -0.21928 $\\ \hline
	
\end{tabular}
\label{tab:scalar moments bulk}
\end{table}
\begin{table*}[h]\centering
	\caption{The same scalar moments as in Tab. \ref{tab:scalar moments bulk} at the edge ($r>r_-\approx 0.96875$) (top) and the enlarged edge ($r>r_-'\approx0.9375$) (bottom) of the spectrum.
	}
\begin{tabular}{|c|c || c | c | c | c | c|}
	 \hline
&{\bf edge moments}	 & $ \langle r \rangle $ & $ \langle r^2 \rangle $ & $ \langle \cos(\phi) \rangle $ & $ \langle \cos(\phi)^2 \rangle $ & $ \langle r\cos(\phi) \rangle $ \\ \hline
\multirow{4}{*}{$r>r_-$ }&
	2D Poisson & $ 0.68393 $ & $ 0.52171 $ & $ \ 0.17029 $ & $ 0.51219 $ & $\  0.13014 $\\ \cline{2-7}
	& class AI$^\dag$ & $ 0.71576 $ & $  0.55120 $ & $\  0.09769 $ & $ 0.43501 $ & $ \ 0.06779 $\\ \cline{2-7}
	& class A & $ 0.72735 $ & $ 0.56449 $ & $ \ 0.10872 $ & $ 0.40441 $ & $ \ 0.07473 $\\ \cline{2-7}
	& class AII$^\dag$ & $ 0.74438 $ & $ \ 0.58542 $ & $ 0.22132 $ & $ 0.35678 $ & $\  0.15712 $\\ \hline \hline
	\multirow{4}{*}{$r>r_-'$ }&
	2D Poisson & $ 0.68154 $ & $ 0.51870 $ & $ \ 0.14995 $ & $ 0.51025 $ & $\  0.11447 $\\ \cline{2-7}
	& class AI$^\dag$ & $ 0.71543 $ & $  0.55110 $ & $ -0.02321 $ & $ 0.44471 $ & $ -0.02115 $\\ \cline{2-7}
	& class A & $ 0.72735 $ & $ 0.56502 $ & $ -0.05943 $ & $ 0.42718 $ & $ -0.04899 $\\ \cline{2-7}
	& class AII$^\dag$ & $ 0.73233 $ & $ \ 0.57065 $ & $ 0.00686 $ & $ 0.39679 $ & $ -0.00022 $\\ \hline
\end{tabular}\label{tab:scalar moments edge}
\end{table*}

\begin{figure}[h!]
\hspace{120pt}{$\langle r\rangle$} \hspace{230pt}{$\langle r^2 \rangle$} \hfill\\
\includegraphics[width=0.49\linewidth]{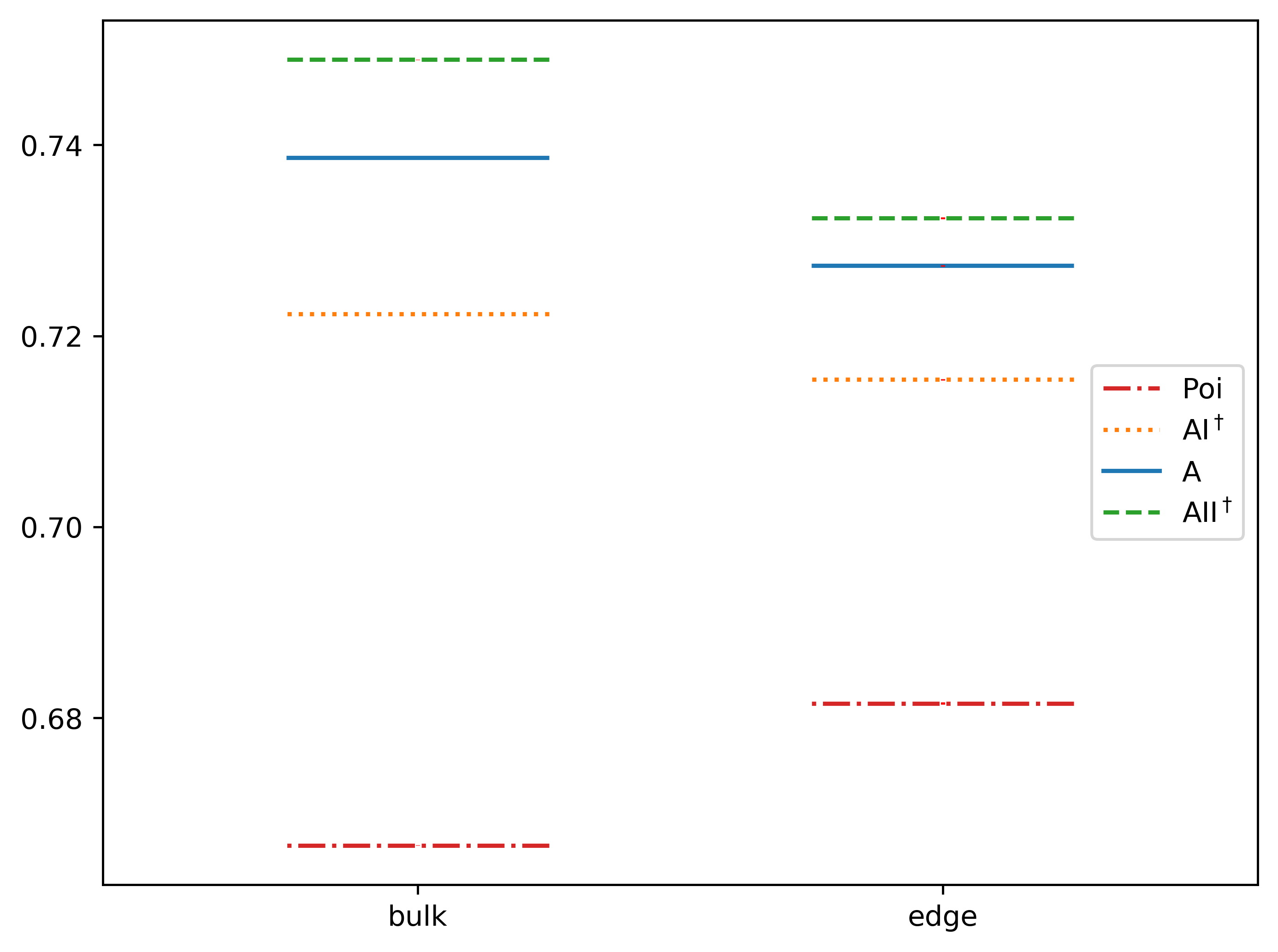} 
\includegraphics[width=0.49\linewidth]{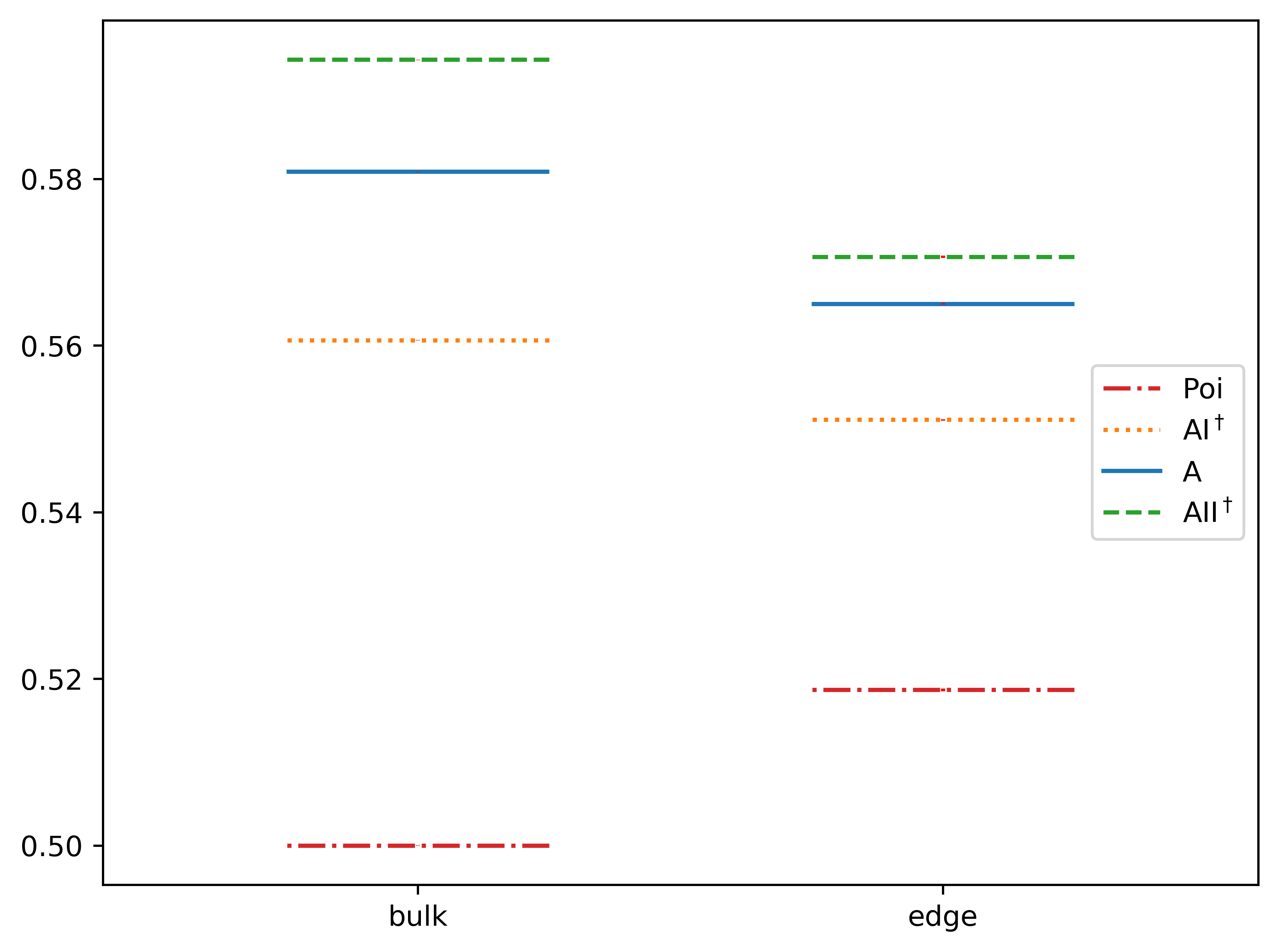}
\includegraphics[width=0.49\linewidth]{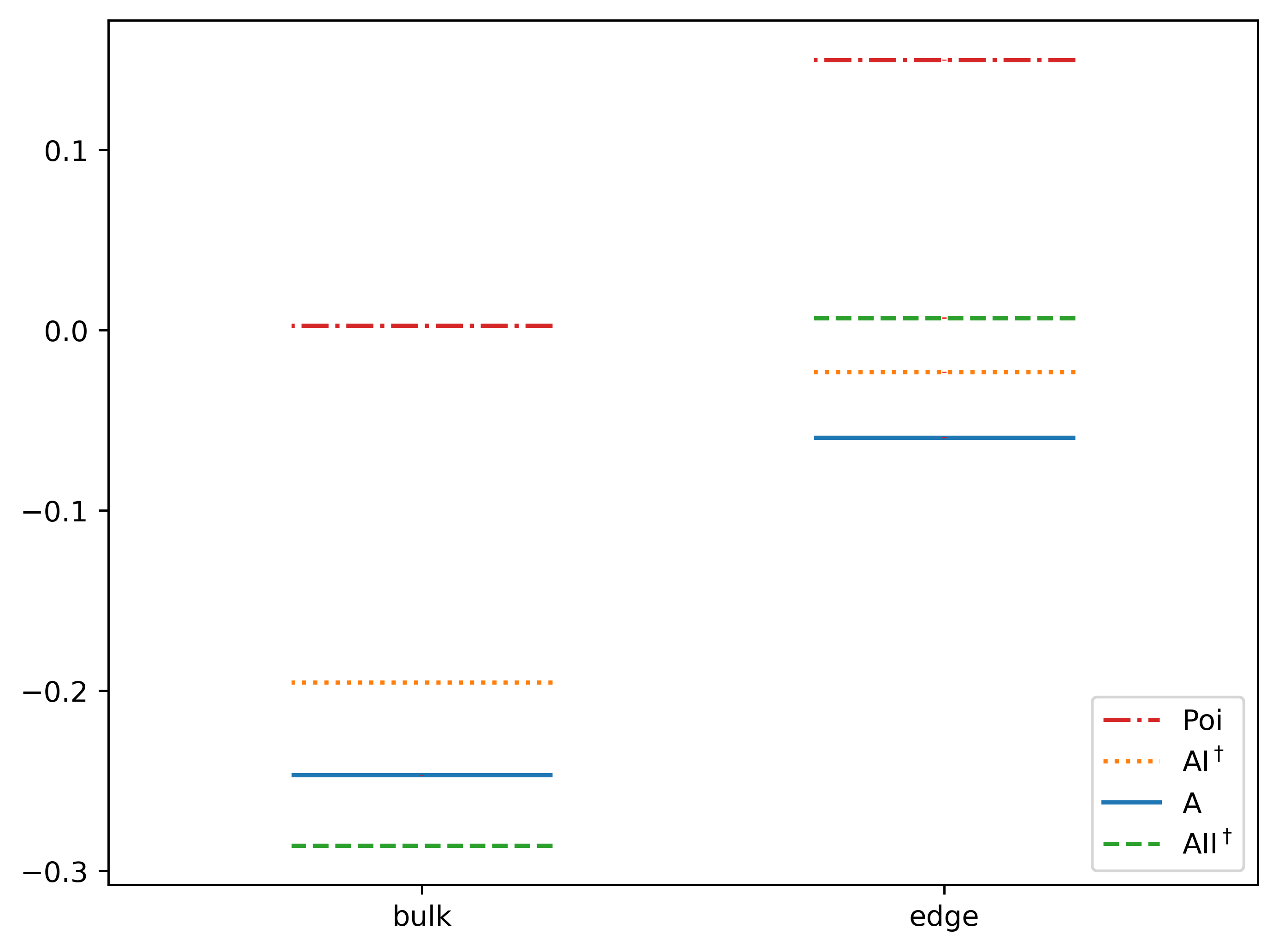}\hspace{8pt}
\includegraphics[width=0.49\linewidth]{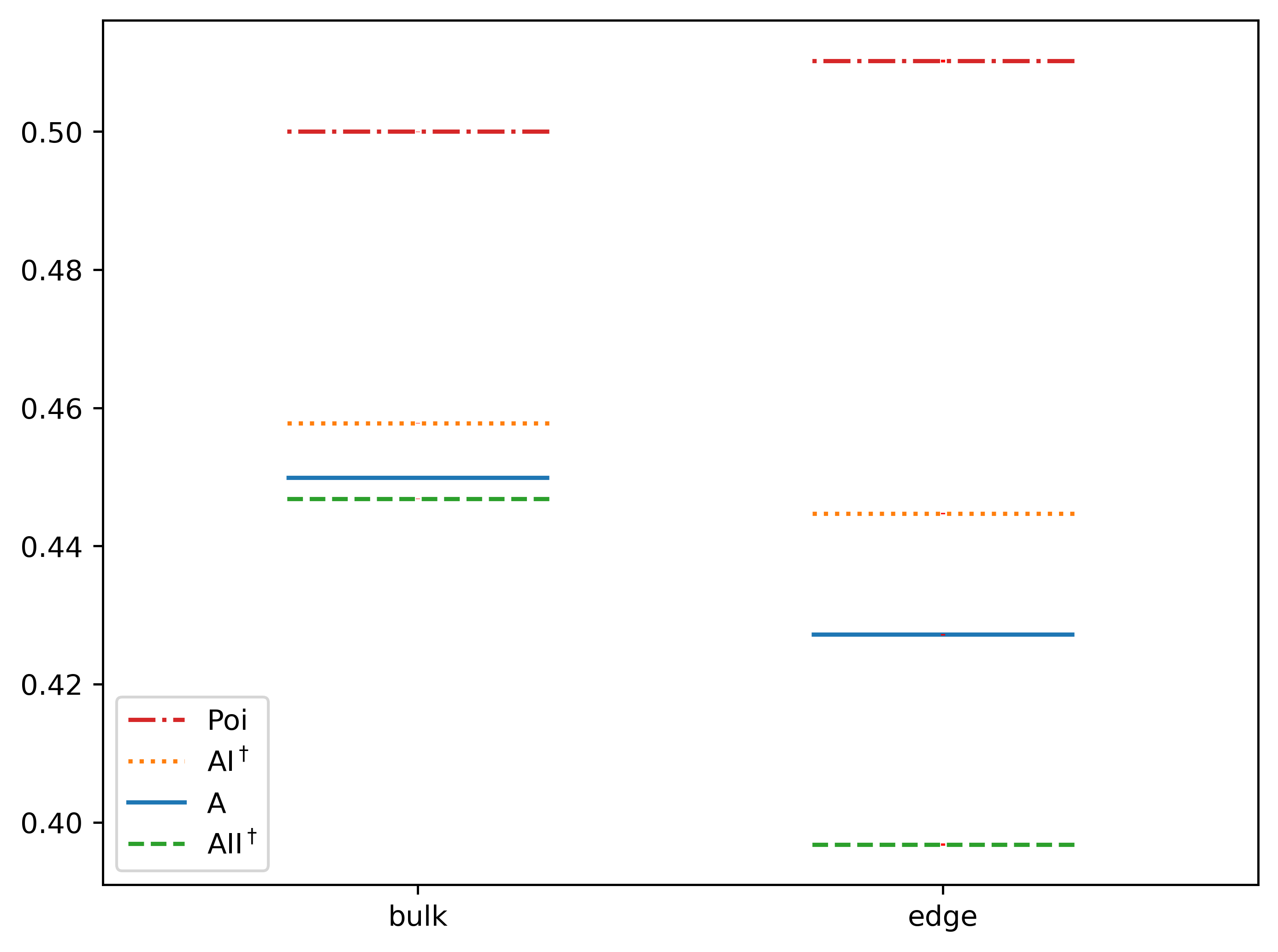}\\
\hspace*{110pt}{$\langle \cos(\phi)\rangle$} \hspace*{210pt}{$\langle \cos(\phi)^2 \rangle$} \hfill
\caption{{Top row}: Radial scalar moments $\langle r\rangle$ (left) and $\langle r^2 \rangle$ (right); {Bottom row}:  Angular scalar moments
$\langle \cos(\phi)\rangle$ (left) and $\langle \cos(\phi)^2 \rangle$ (right). We present the bulk for $r<r_b$ on the left sides, and the
edge for $r>r_-$ on the right sides of each plot.
We use $720\,000$ ensembles of size $N=1024$ (respectively $2N=1024$), denoting the 2D Poisson process (red dot-dashed), class AI$^\dag$ (orange dotted), A (blue full) and AII$^\dag$ (green dashed). While for radial moments $\langle r\rangle$, $\langle r^2\rangle$ and $\langle \cos(\phi)^2 \rangle$
the ordering with respect to $\beta$  is the same in the bulk (left sides) and the edge region (right sides),
it is inverted at the edge for $\langle \cos(\phi) \rangle$. Also, relative gap sizes are more stable for radial moments.
}
\label{fig:scalar moments spacing ratio}
\end{figure}

The results on the scalar moments $ \langle r \rangle,\  \langle r^2 \rangle,\  \langle \cos(\phi) \rangle, \  \langle \cos(\phi)^2 \rangle,$ and $ \langle r\cos(\phi) \rangle $ for $720\,000$ ensembles of size $N=1024$ (respectively $2N=1024$ for class AII$^\dag$) are given in Tab.\ref{tab:scalar moments bulk} for the bulk and in Tab.\ref{tab:scalar moments edge} in the edge region. Let us first discuss the bulk results, which were previously known (cf. \cite{SaRibeiroProsen, DusaWettig, KanazawaWettig}). For the 2D Poisson process, the moments of the complex spacing ratio can be calculated analytically in the bulk of the spectrum, cf. \cite{SaRibeiroProsen}. The results are given in the second row in Tab. \ref{tab:scalar moments bulk}. The numerically calculated results in the row below are in good agreement with the theory, up to the second or third digit. The vanishing moments $\langle \cos(\phi)\rangle$ and $\langle r\cos(\phi)\rangle$ reflect the rotational symmetry of the distribution, observed in Figs. \ref{fig:spacing ratio Gaussian Poisson} left, and Fig. \ref{fig:spacing ratio 1D dist} left. The radial moment $\langle r \rangle$ is a measure of distance between NN and NNN eigenvalue. Its value increases from top to bottom in the second column of Tab. \ref{tab:scalar moments bulk}, i.e. with the strength of the local repulsion in terms of $\beta$.
A similar increase (in absolute values) is observed for $\langle r^2\rangle,\ \langle \cos(\phi) \rangle$ and $\langle r\cos(\phi)\rangle$. Notice that the results for $\langle \cos(\phi) \rangle$ and $\langle r\cos(\phi)\rangle$ are negative in the bulk for all 3 symmetry classes AI$^\dag$, A and AII$^\dag$.

Next we discuss the edge results, presented in Tab. \ref{tab:scalar moments edge} for the same ensemble sizes as before. 
We begin with the results for a smaller choice for the edge, with $r>r_-=1-1/\sqrt{N}$ in Tab. \ref{tab:scalar moments edge} top,  where all spectral densities are in monotonous decay, see Fig. \ref{fig:radial density plot} (right).
First, we note that the relative ordering of the radial moments $\langle r \rangle$ and $\langle r^2\rangle$ {is} consistent with the one in the bulk, i.e. we find an increase from top to bottom, with changes only in the second digit. Comparison with Tab.\ref{tab:scalar moments bulk} gives us that the mean radius of the complex spacing ratio is strictly smaller in the edge region than in the bulk, for all 3 ensembles under consideration. The confining potential in each matrix ensemble results into a finite support. When considering eigenvalues at the edge, these are more sensitive to boundary effects than in the bulk. This is reflected by the probability to find NN and NNN of an edge eigenvalue outside of the support to be significantly suppressed. Hence, the mean radius of the complex spacing ratio is smaller in the edge region than in the bulk. For the moments $\langle \cos(\phi) \rangle$ and $\langle r \cos(\phi)\rangle$, we find an inversion of the ordering of the symmetry classes compared to the bulk. Moreover, the values change sign for classes AI$^\dag$, A and AII$^\dag$.
For a better visual  comparison, we present the moments given in Tab. \ref{tab:scalar moments bulk} and Tab. \ref{tab:scalar moments edge} top in Fig. \ref{fig:scalar moments spacing ratio}. It summarises the effects we just described.

Let us turn to discuss the effect of a larger region for the edge, choosing $r>r_-^\prime=1-2/\sqrt{N}$, see Fig. \ref{fig:radial density plot} (right).
The corresponding values for the 0D moments are given in Tab. \ref{tab:scalar moments edge} bottom.
For 2D Poisson we see the smallest effect, where the angular moments $\langle \cos(\phi)\rangle$ and $\langle r\cos(\phi) \rangle$ which are odd powers of cosine are most sensitive (cf. Fig. \ref{fig:radialDensityPoisson}). While also for classes  AI$^\dag$, A, AII$^\dag$ the radial moments are not very sensitive, the odd powers of cosine are very sensitive to the choice of $r_-^\prime$ and almost all change sign, moving towards their bulk value. In conclusion our more conservative choice for the edge region $r_-=1-1/\sqrt{N}$ (motivated from class A, cf. Fig. \ref{fig:radial density plot}), gives a clearer characterisation of the edge universality classes as visible in all complex spacing ratios and their moments.

\subsection{NN and NNN spacing distributions}\label{sec:NN}

In this subsection we turn to numerical results for the NN and NNN spacing distribution at the edge in all 3 ensembles, and compare that to known (analytical and numerical) results  in the bulk of the spectrum. 
To set the stage, we begin with the general definition of the 
NN spacings distribution $p_{\rm NN}^{(N)}(s)$ for all ensembles, at finite $N$. After recalling the known analytical results for NN and NNN for class A and 2D Poisson in the bulk, we present a comparison for all 3 ensembles between bulk and edge. 
A comparison to the conjectured cubic small argument behaviour (times a logarithm for class AII$^\dag$) of $p^{(N)}(s)$ for large-$N$ at bulk and edge in all 3 classes concludes this subsection.\\

The probability that a complex eigenvalue $z_1$ has its NN $z_2$ at radial distance $s$, and all other eigenvalues at further distance, is given by 
\begin{equation}
p_{\rm NN}^{(N)}(s) = \frac{1}{\gamma_{N}} \int_{\mathbb{C}^N} \dif^2z_1 \dots \dif^2 z_N \jpdf^{(N)}(z_1,\dots,z_N) \delta^{(1)}\left(|z_2-z_1|-s\right)  \prod_{j=3}^{N} \heaviside\left(|z_j-z_1|^2-s^2\right). 
 \label{def.gapprob}
\end{equation}
Here, $z_1$ can be chosen in the bulk or at the edge of the spectrum. 
The normalisation constant $\gamma_{N}$ ensures that 
\begin{equation}
\int_0^\infty \dif s\,  p_{\rm NN}^{(N)}(s)=1\ .
\label{def.norm-p}
\end{equation}
In addition, below we normalise the first moment to unity, to fix the scale, 
\begin{equation}
\int_0^\infty \dif s \, s\, p_{\rm NN}^{(N)}(s)=1\ .
\label{def.1stmom}
\end{equation}
In analogy to Section \ref{Sec:AnalyticsA},  for the bulk of the spectrum we may condition the point $z_1=0$ to be at the origin, compare \eqref{def.dist_ofSpacingRatio} vs. \eqref{def.dist_ofCondSpacingRatio}:
\begin{eqnarray}
p_{\rm C, NN}^{(N)}(s) &=& \frac{1}{\gamma_{N,\rm C}} \int_{\mathbb{C}^{N-1}} \dif^2z_2 \dots \dif^2 z_N \jpdf^{(N)}(0,z_2,\dots,z_N) \delta^{(1)}\left(|z_2|-s\right)  \prod_{j=3}^{N} \heaviside\left(|z_j|^2-s^2\right)
\label{eq:NN-A}
\\
 &=& \frac{1}{\gamma_{N,\rm C}} \int_0^{2\pi}\dif \theta_2 \int_{\mathbb{C}^{N-1}} \dif^2 z_3\dots \dif^2 z_N \jpdf^{(N)}(0,s\euler^{\iunit\theta_2},\dots,z_N) \prod_{j=3}^{N} \heaviside\left(|z_j|^2-s^2\right).
 \label{eq:NNN-A}
\end{eqnarray}
Following \cite{GHS}, in the GinUE explicit formulas for NN and NNN spacing distribution were derived at finite-$N$ in class A. The idea was to choose the origin as a generic bulk point, and then use the translational invariance of the bulk in the large-$N$ limit, to obtain a formula valid in the bulk in general. 
In \cite{GHS} the gap probability was derived first, replacing the Dirac-delta in \eqref{eq:NN-A} by a Heaviside function, with one eigenvalue $z_1=0$ at the origin and all others at distance $s$ or larger from the origin, cf. Appendix \ref{App:kernel} for more details. 
Likewise, gap probabilities for higher order gaps can be obtained and 
its derivatives then lead to \eqref{def.gapprob}, and higher NNN spacings, cf. \cite{GHS}. We only quote the results here for the NN and NNN spacing, conditioned to $z_1=0$ in class A at finite $N$:
\begin{align}\label{GinibreSpacing}
	p_{\rm A,C\ \rm NN}^{(N)}(s) \; & = \;
	\left(\prod_{k=1}^{N-1}\frac{\gamma(1 + k, s^2)}{k!}\right)
	\sum_{j=1}^{N-1} \frac{2s^{2j+1}\euler^{-s^2}}{\Gamma(1 + j, s^2)},
	\\
       p_{\rm A,C\ \rm NNN}^{(N)}(s) \; & = \;
	\left(\prod_{k=1}^{N-1}\frac{\gamma(1 + k, s^2)}{k!}\right)
        \sum_{j,k=1;k\neq j}^{N-1}
        \frac{\gamma(1+j, s^2)}{\Gamma(1 + j, s^2)}
	\frac{2s^{2k+1}\euler^{-s^2}}{\Gamma(1 + k, s^2)} .
\label{GinibreSpacingNNN}
\end{align}
Here, $\Gamma(1 + k, s^2)=\int_{s^2}^\infty\dif t\, t^k \euler^{-t}$ and
$\gamma(1 + k, s^2)=\int^{s^2}_0\dif t\, t^k \euler^{-t}$ are the integral representations of the upper and lower incomplete
Gamma-functions, respectively.  These exact relations converge rapidly for large $N$. Below we will plot them using a cutoff at around $N=20$ for a good graphical representation. 
Notice that in \eqref{GinibreSpacing} the first moment still has to be set to unity as in \eqref{def.1stmom}, by rescaling $s\to s\ const.$ with $const.\approx 1.1429$ calculated numerically. This sets the scale and the same rescaling has to be applied to \eqref{GinibreSpacingNNN}. 

Technically speaking, the gap probability leading to \eqref{GinibreSpacing} (and \eqref{GinibreSpacingNNN}) is the product of the exact Fredholm determinant eigenvalues of the kernel of this conditional determinantal point process. In Appendix \ref{App:kernel} we applied the same technique to $|z_1|=1$, conditioned to be at the edge, giving the gap probability through the Fredholm determinant of the conditional kernel. Unfortunately, its eigenvalues are not explicitly known and its expansion does not easily lead to good approximations \cite{BScGA}. However, it can be used to determine the small-argument expansion of $p_{\rm A,C\ NN}^{(N)}(s)$ in class A at the edge which is also cubic, to which we will return below.

For completeness we also give the NN and NNN spacings distribution for 2D Poisson (Poi) in the bulk of the  spectrum,
\begin{align}\label{Eq:Poisson-2dNN}
   p_{\mathrm{Poi,\,NN}}^{\rm bulk}(s) \; & = 
    \; \frac{\pi}{2} s\ \euler^{-\pi s^2/4} ,\\
   p_{\mathrm{Poi,\,NNN}}^{\rm bulk}(s) \; & =
   \; \frac{\pi^2}{8} s^3\ \euler^{-\pi s^2/4},
   \label{Eq:Poisson-2dNNN}
\end{align}
see e.g. \cite{SaRibeiroProsen} for a derivation. Here, the large-$N$ limit and  normalisation \eqref{def.1stmom} have already been applied. Notice that the NN spacing distribution shows a \emph{linear} growth at small $s$ for Poisson, {coming from the area measure}.

\subsubsection{Unfolding at the spectral edge}\label{App:unfold}

Before we present our numerical results for the NN and NNN spacing distributions at the edge, and compare with the known results in the bulk, such as \eqref{GinibreSpacing} -- \eqref{Eq:Poisson-2dNNN} for class A and 2D Poisson, we have to explain how to unfold at the edge of the spectrum. 

In Hermitian RMT this is a standard procedure, cf. \cite{Haake2010}, which is typically applied to the bulk of the spectrum. Also there, for a strongly varying respectively vanishing density at the edge, some care has to be taken, see \cite{ABK2}. In 2D, unfolding is much more sophisticated already in the bulk of the spectrum, see \cite{CP,AKMP,MPW} for different proposals. Let us first describe the problem and then present heuristic arguments that lead us to an unfolding procedure {at the edge}. For that we will use the gap probability $E_k(s)$, that the open disc $D_s(z_0)$ of radius $s$ around an eigenvalue $z_0$ contains $k$ further eigenvalues. Differentiating this for $k=0,1$ then leads to the NN respectively NNN spacing distribution. 

\begin{figure}[h]

    \includegraphics[width=0.42\columnwidth]{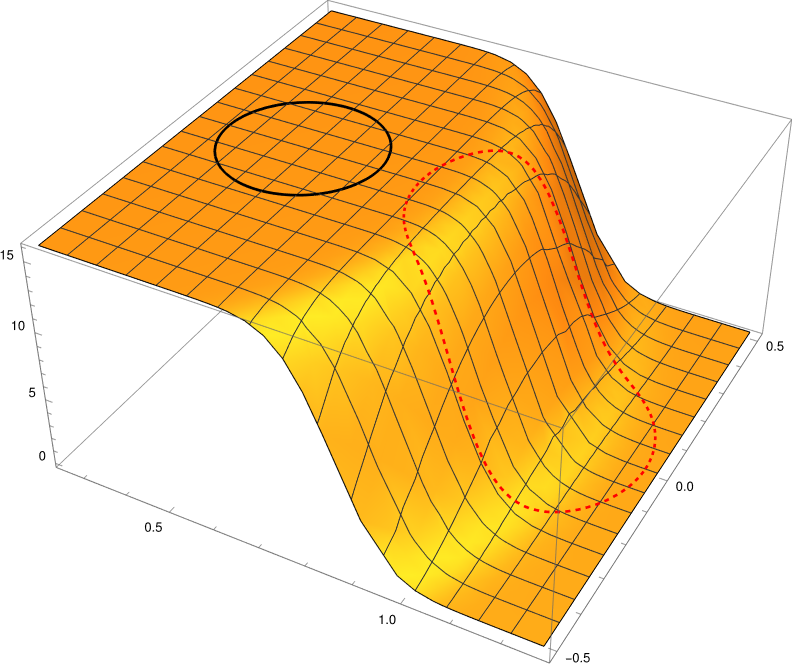}
    \includegraphics[width=0.34\columnwidth]{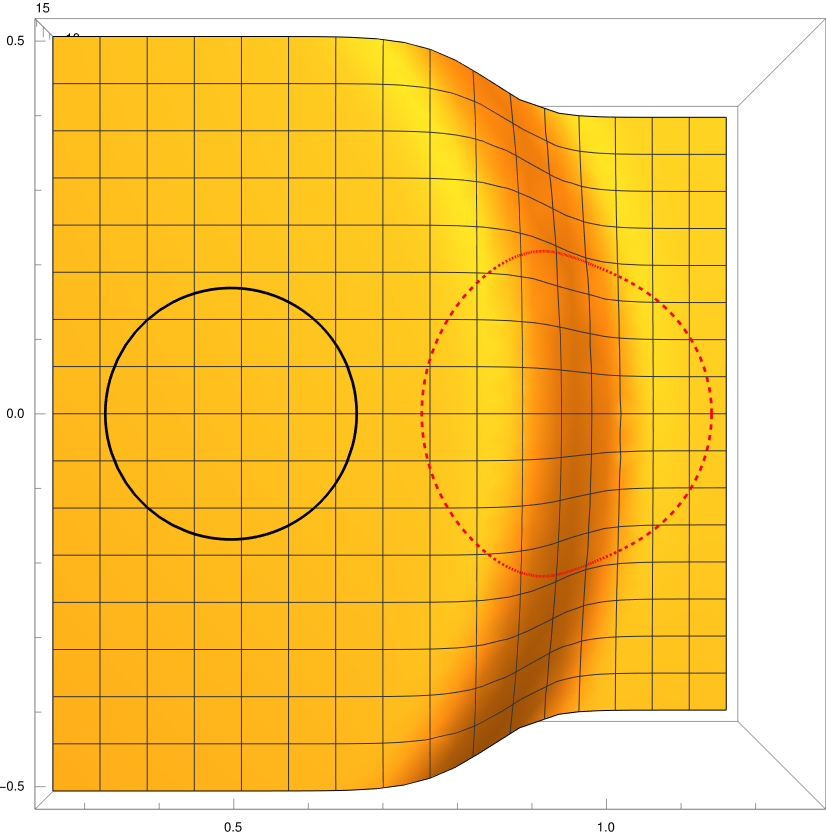}   \caption{
    Magnification  of the edge part (left: side view; right top view) of the spectral density \eqref{R1A} for the GinUE class A at $N=50$, rescaled to have limiting support on the unit disc. The black circle covers the same volume in the bulk in this 3D plot,
     as does the red dashed circle over the edge.}
    \label{plot:DensityVariationsUnderCircles}
\end{figure}

In general, the average spacing between eigenvalues in $D$ dimensions is proportional to the mean density $R(z)^{1/D}$ at that point. Because of that, we cannot compare the NN spacing distribution at points with different densities, in particular between bulk and edge, see   Fig. \ref{plot:DensityVariationsUnderCircles}, and thus we need to unfold. Because at the edge the density is varying very strongly in the scale of the mean level spacing $\sim O(1/\sqrt{N})$, we propose the following procedure which is different from \cite{CP,AKMP,MPW}.

Generally in a point process, given a point at $z_0$ and its $k$-th neighbour $z_k$ in a region $\mathcal{A}$ of constant mean density $R(z_0)$, the expected 
number of points closer than $z_k$ is $R(z_0)$ times the area of the disc of radius $|z_k-z_0|$ around $z_0$. Now imagine a region $\mathcal{B}$ of constant but
different density $R'(z_0')$. A point $z_0'$ and its $k$-th neighbour $z_k'$ shall both be in region $\mathcal{B}$. Then, rather than the distances $|z_k-z_0|$ and $|z_k'-z_0'|$,
the expected numbers of points in the discs should be compared.
In the general case of a point $z_0$ in a region of variable density, this expected number of points within a disc of radius $s$ is the  integral over the density
\begin{equation}
    \label{eq:DensityIntegral}
    m(s) = \int_{\mathbb{D}_s(z_0)} \dif^2 z \ R(z).
\end{equation}
In the  bulk region of constant density $R_{\rm bulk} $, we have $m(s) = R_{\rm bulk} \pi  s^2$. In our case, where the global density is normalised on the unit disc, we have  $R_{\rm bulk}=1/\pi$, and thus $m(s) = s^2$.

For each edge eigenvalue $z_0$ and its NN (NNN etc.),
we take the distance $s$ and calculate $m(s)$ numerically, given the position of the eigenvalue. Then, we calculate an unfolded distance $s'$ from the number
of expected points within the given distance from $z_0$,
\begin{equation}
    s' = \sqrt{m(s)}\ .
\end{equation}
Here, we use the empirical density obtained from the simulations in \eqref{eq:DensityIntegral}.\footnote{Consistently, in the bulk we have $s'=s$ when normalising to $R_{\rm bulk}=1/\pi$.} 
For this purpose, we subdivided the radial range of the sample eigenvalue spectra into
about 1 million equally spaced bins and counted the number of occurring eigenvalues for each bin over all samples. The bins can be thought of as
concentric annuli around the origin in the complex plane. All our ensembles have rotational symmetry for the density, so this is sufficient for integrating each of the discs around edge eigenvalues.
For each annulus, we take the central circle (same distance to inner and outer boundary) and calculate its intersection length with a given disc $D_s(z_0)$ around an edge point $z_0$.
These lengths multiplied by the number of eigenvalues in the bin, and properly normalised, then sum to the integral \eqref{eq:DensityIntegral}, 
as a good approximation for edge eigenvalues.

Note that, compared to \cite{AKMP} where the unfolding is done by rescaling the distance locally, we here perform an integral between the points. This assumes that the shortest distance between the points is a straight line, which is a good approximation. A possible improvement would be to find the NN using geodesic curves, as proposed in \cite{CP}, but that is beyond the scope of this work.

\subsubsection{Comparing NN and NNN spacings at the edge to the bulk
}\label{sec:numNN}

\begin{figure}[h!]
\centering
\hspace{10pt}{Poi} \hspace{210pt}{AI$^\dagger$} \hfill\\
\includegraphics[width=0.49\linewidth]{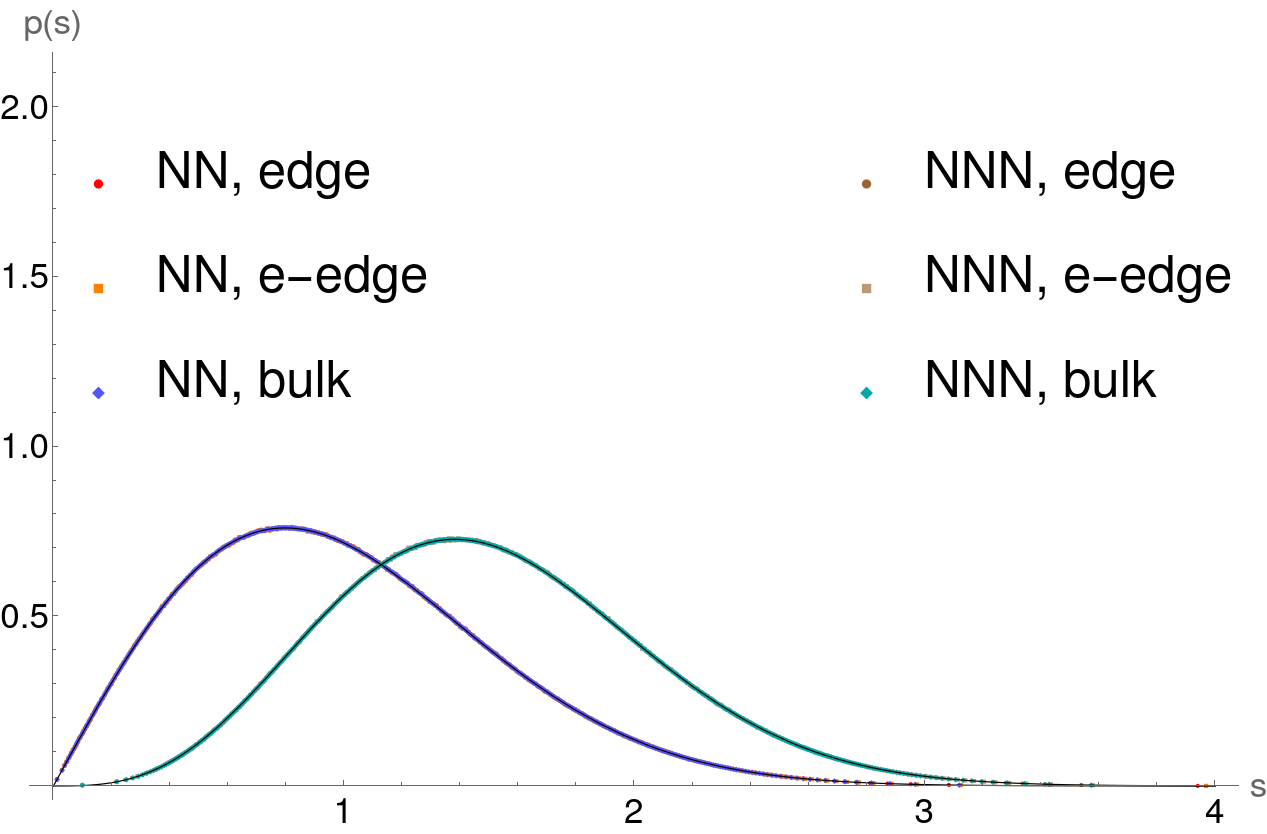} 
\includegraphics[width=0.49\linewidth]{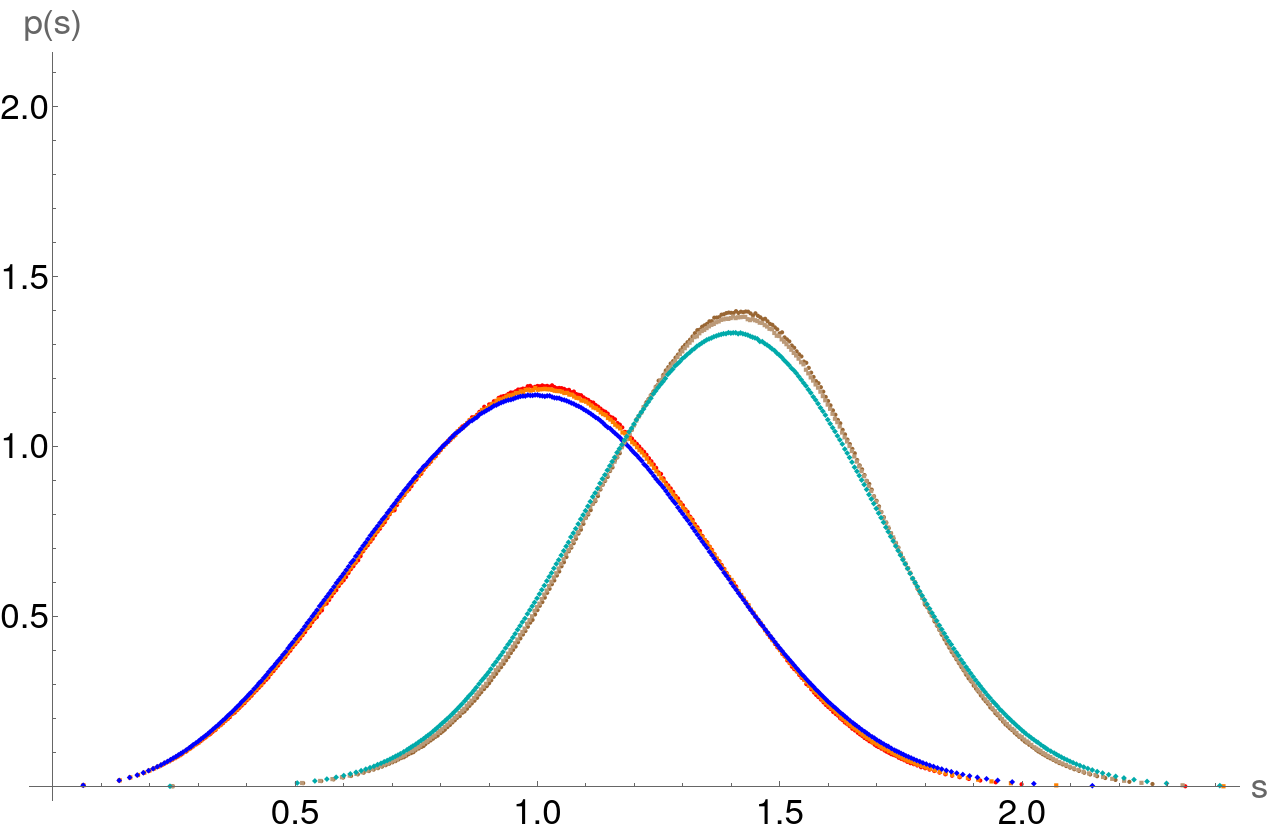} \\
\includegraphics[width=0.49\linewidth]{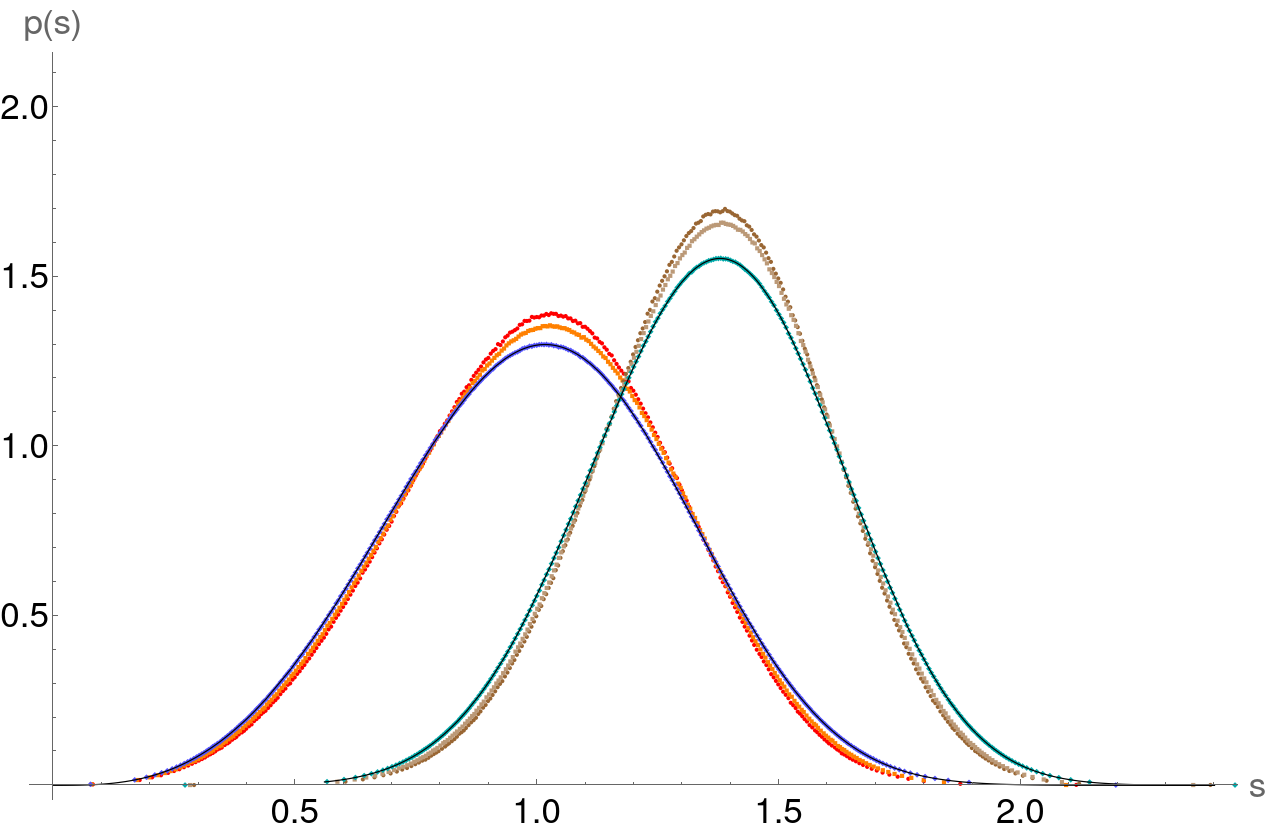} 
\includegraphics[width=0.49\linewidth]{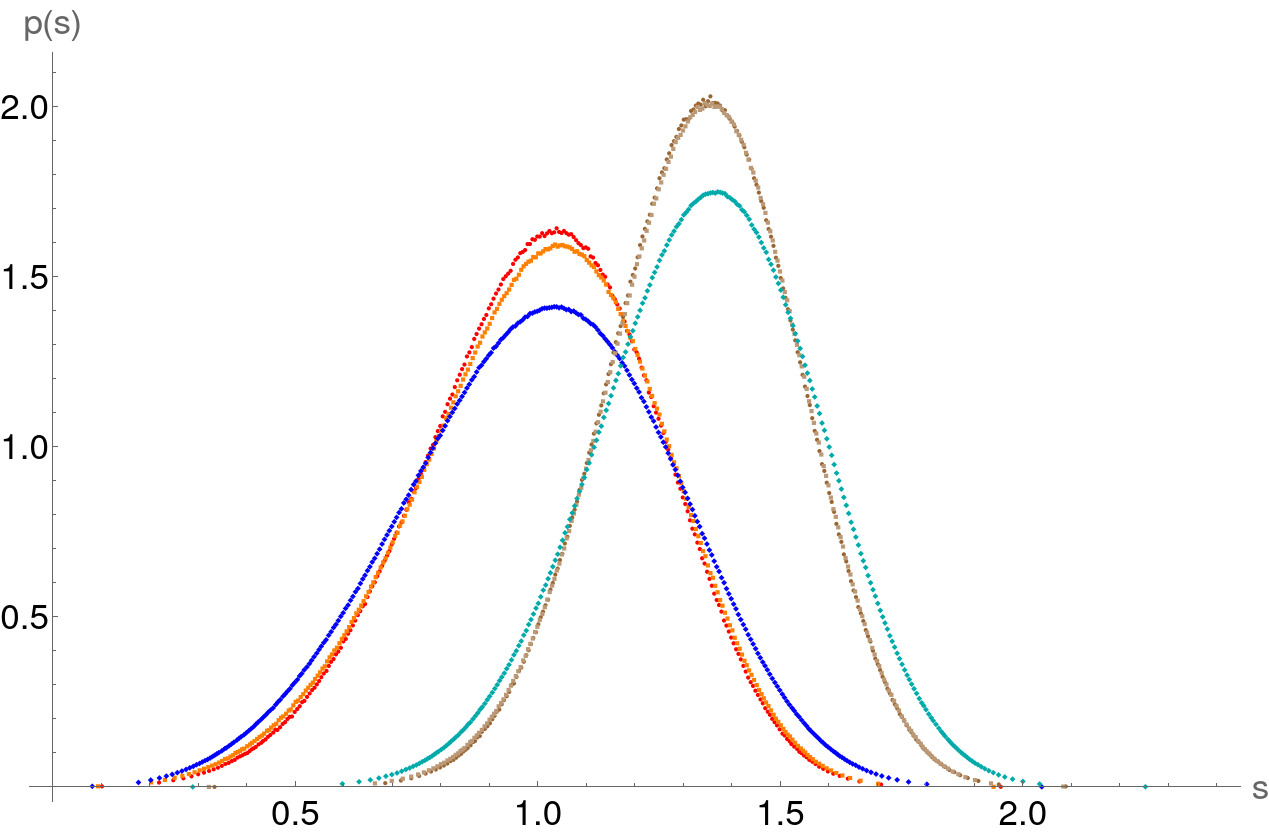}\\
\hspace*{40pt}{A} \hspace*{220pt}{AII$^\dagger$} \hfill
\caption{
Each plot shows the NN (left maxima) and NNN (right maxima) spacing distribution for a given symmetry class from $720\,000$ samples each, with $N=1024$ points (AII$^\dag$ has $2N=1024$):  { Top left 2D Poisson}, with analytic bulk expressions \eqref{Eq:Poisson-2dNN} and
\eqref{Eq:Poisson-2dNNN} (black curves) showing excellent agreement with all bulk {\it and} edge data; {Top right class AI$^\dagger$}; { Bottom left class A}, with analytic bulk expressions
\eqref{GinibreSpacing} and \eqref{GinibreSpacingNNN} (black curves) agreeing excellently with bulk data; { Bottom right class AII$^\dagger$}. We distinguish NN spacing the bulk 
region (blue diamonds, $|z| < r_b=0.8$), edge region (red circles, $|z| > r_-=1-\frac{1}{N} = 0.96875$),
and extended edge region (orange squares, $|z| > r_-^\prime=1-\frac{2}{N} = 0.9375$) from the NNN spacing distributions in the bulk (cyan diamonds), edge (brown circles)
and extended edge (light brown squares). 
Distances are unfolded for all edge regions, and for the bulk region of the 2D Poisson ensemble. 
No unfolding is needed for the bulk region in classes AI$^\dag$, A and AII$^\dag$ due to the flat density.
In each case the distances are scaled such that the first moment of the NN spacing is unity.
}
\label{fig:NearestNeighborSpacings}
\end{figure}

We start with the discussion of the NN and NNN spacing distribution in the 2D Poisson ensemble, 
see Fig. \ref{fig:NearestNeighborSpacings} top left.
All spacing distributions are unfolded and first moment{s} normalised to unity, as described previously. 
It is remarkable to see, that the NN spacing distributions from all three regions, bulk, edge and extended edge agree very well with the analytic prediction for the NN bulk distribution at large $N$. This should perhaps not come as a surprise, as also points close to the edge are independently distributed, and therefore, after proper unfolding and rescaling, should agree with a generic bulk point.

These findings cast some doubts on the claim that in complex spacing ratios unfolding is not necessary, as it drops out in taking the ratio of the difference between NN and NNN. If this was the case, then in Fig. \ref{fig:spacing ratio Gaussian Poisson} (right) at the edge we should also obtain a completely flat density plot, as in the left plot for the bulk. Consequently, it seems that when the density varies on the scale of the mean level spacing $O(1/\sqrt{N})$---which is also the width of the edge region here---the complex spacing ratio apparently fails to properly unfold {completely}.

We should keep that in mind when now comparing the 3 non-Hermitian random matrix symmetry classes. There, in contrast to 2D Poisson, corresponding to $\beta=0$, we also have an additional level repulsion, with increasing {(}effective{)} $\beta$ from classes AI$^\dag$, over class A to class AII$^\dag$.   
As we can see in the corresponding NN and NNN spacing distributions in Fig. \ref{fig:NearestNeighborSpacings} top right to bottom right, here there remains an effect between bulk and edge, increasing in size with effective repulsion $\beta$, after proper unfolding.  
We see a larger maximum at the edge in both the NN and NNN distribution, compared to the bulk. The effect is most prominent in AII$^\dagger$ (Fig. \ref{fig:NearestNeighborSpacings}, lower right),
with the least difference in AI$^\dagger$ (Fig. \ref{fig:NearestNeighborSpacings}, upper right).

We can also see an effect of selecting an extended edge region
$r > r_-^\prime=1 - \frac{2}{\sqrt{N}}$ for AI$^\dagger$ and A, and $r > r_-^\prime =1 - \frac{2}{\sqrt{2N}}$ for AII$^\dagger$. While for class AI$^\dag$ the effect is barely visible, we can see that by extending the edge region, we apparently start to include bulk points, as the NN and NNN distributions then move towards the bulk curve. This is visible for NN in classes A and AII$^\dag$, and for NNN in class A. It is consistent with the observations we made based on moments of the complex spacing ratios in the previous subsection.  

As a conclusion we observe that the difference between bulk and edge universality classes for the NN and NNN spacing distribution is clearly visible in all 3 ensembles, after unfolding. Its effect seems to increases with the repulsion in the 3 ensembles, measured by their respective effective value of $\beta$.

\subsubsection{Small-argument expansion of NN spacing distributions at edge and bulk}\label{sec:NNcubic}

Last but not least we turn to the small argument expansion of all NN spacing distributions. It has been found early on \cite{Grobe Haake}, based on the level crossing of generic $2\times2$ non-Hermitian Hamiltonians without further symmetry, that the repulsion seen at small argument $s\ll1$ is cubic,
\begin{equation}
\label{eq:small-s-A-N=2}
    p_{\rm A\, NN}^{(N=2)}(s)\sim s^3,
\end{equation}
which was claimed to be universal \cite{Grobe Haake}, see also \cite{Haake2010}. In the same paper, is was shown that if in addition the Hamiltonian is {\it complex symmetric}, there is an additional logarithmic repulsion on top of the cubic repulsion, 
\begin{equation}
\label{eq:small-s-AI-N=2}
    p_{\rm AI^\dag\, NN}^{(N=2)}(s)\sim -s^3\log(s).
\end{equation}
This has to be contrasted with 2D Poisson \eqref{Eq:Poisson-2dNN}, where the repulsion in linear
for $s\ll1$:
\begin{equation}
\label{eq:small-s-Poi-bulk}
    p_{\rm Poi\, NN}^{\rm bulk}(s)\sim s,
\end{equation}
which holds in the bulk in the large-$N$ limit.

For the GinUE in class A, the NN spacing distribution was determined in the bulk of the spectrum in \cite{GHS} one year before \cite{Grobe Haake}, confirming that based on \eqref{eq:NN-A}
\begin{equation}
\label{eq:small-s-A-bulk}
    p_{\rm A\, NN}^{\rm bulk}(s)\sim s^3,
\end{equation}
also holds for the limit $N\to\infty$. In Appendix \ref{App:kernel} we will show {analytically} that the same cubic repulsion also holds at the edge of the spectrum in class A, for arbitrary $N$
\begin{equation}
\label{eq:small-s-A-edge}
    p_{\rm A\, NN}^{\rm edge}(s)\sim s^3,
\end{equation}
see \eqref{eq:Esmall-s-A-edge}, after taking the derivative. The constant in front of the cubic term is different though. 

For the symmetry classes AI$^\dag$ and AII$^\dag$ results for the NN spacing distribution are much more recent and valid for a surmise with $N=2$, only. Note, however, that in the non-Hermitian case $N=2$ is not a good approximation \cite{AMP}, which was already stated for class A in \cite{GHS}. In \cite{Hamazaki et al, JPP} \eqref{eq:small-s-AI-N=2} was shown based on an $N=2$ random matrix calculation, whereas for class AII$^\dag$ the cubic repulsion persists at $N=2$ \cite{Hamazaki et al}
\begin{equation}
\label{eq:small-s-AII-N=2}
    p_{\rm AII^\dag\, NN}^{(N=2)}(s)\sim s^3.
\end{equation}

\begin{figure}[h!]
    \centering
    \includegraphics[width = \linewidth]{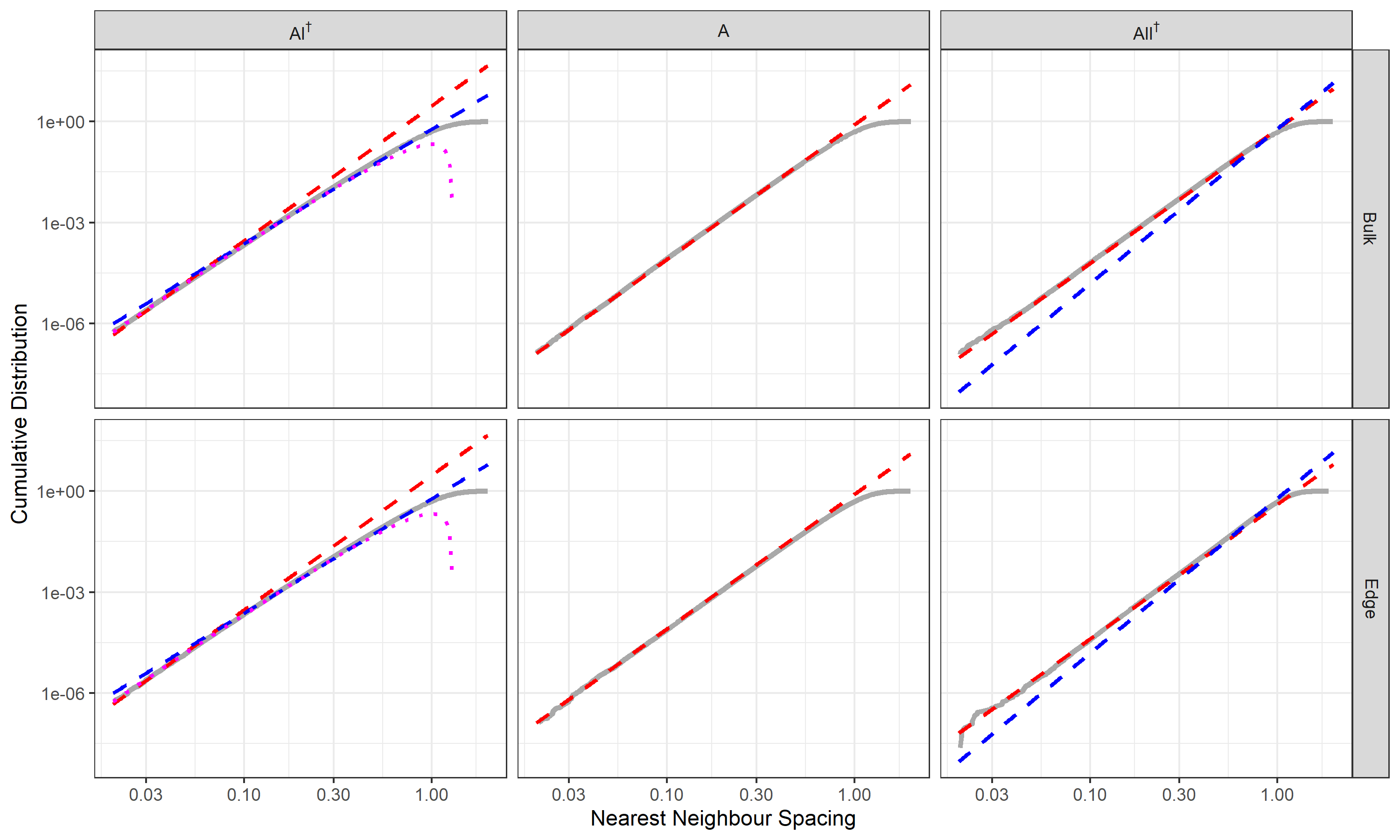}
    \caption{Behaviour of the cumulative distribution of the NN spacing distribution at small $s$, both in the bulk (top curves) and edge (bottom curves). For each ensemble AI$^\dag$, A, and AII$^\dag$ (left to right) we use a log-log scale, for $720\, 000$ ensembles at $N=1024$ ($2N=1024$ for AII$^\dag$), represented by a grey full line. In all plots the cubic repulsion corresponds to the red dashed curve,  where the intercept is fitted in each symmetry class in the bulk and at the edge. For class AI$^\dag$ we also compare with the cubic times logarithmic repulsion \eqref{eq:small-s-AI-N=2} (pink dotted curve). For both classes AI$^\dag$ and AII$^\dag$  we also compare with the small argument behaviour expected based on the effective 2D Coulomb gas description, $\sim s^{\beta+1}$ (blue dashed lines).}
    \label{fig:small}
\end{figure}

In Fig. \ref{fig:small} we compare the small argument asymptotic expansion of the 
cumulative NN spacing distribution,
$E_{\rm NN}(s)=\int_0^s dt\,p_{\rm NN}(t)$, in the bulk (top curves) and at the edge (bottom curves) for all 3 symmetry classes AI$^\dag$, A, and AII$^\dag$ (left to right). This is in order to avoid any bias based on the binning of our data into histograms.

For class A (middle curves) we find the analytically expected agreement with \eqref{eq:small-s-A-bulk} (top) and \eqref{eq:small-s-A-edge} (bottom), up to rather large values of the argument $s<1.0$. 
In class AI$^\dag$ the behaviour is consistent both with a cubic or cubic times logarithmic repulsion, both in the bulk and at the edge. It was already mentioned in \cite{Grobe Haake,Haake2010} that the two cannot be easily distinguished. 
For class AII$^\dag$ we again find a clear agreement with a cubic repulsion up to $s\lessapprox 0.30$, both in the bulk where the agreement extends a bit further and at the edge.

Also shown in these plots is a comparison to a behaviour $\sim s^{\beta+1}$, which would be expected for a true 2D Coulomb gas at inverse temperature $\beta$, {with the additional power +1 coming from the area measure}. While for class A at $\beta=2$ this agrees with the cubic repulsion, for AI$^\dag$ and AII$^\dag$ we only find an agreement with such an assumption for $s\gtrapprox 0.30$ (blue curves) with the corresponding effective values \cite{AMP} at  $\beta=1.4$ in class AI$^\dag$, and  $\beta=2.6$ in class AII$^\dag$, respectively, both in the bulk and at the edge.

In conclusion the universal cubic repulsion seems indeed to be very robust, where our numerical (and analytical) findings extend this universality for all 3 symmetry classes to the edge of the spectrum. It confirms the expected bulk universality for the two symmetry classes AI$^\dag$ and AII$^\dag$ at small $s$.

\section{Conclusions and open problems}\label{Sec:conclusion}

In this article we have studied and compared the complex spacing ratio, nearest-neighbour (NN) and next-to-nearest neighbour (NNN) spacing distribution of three different symmetry classes of non-Hermitian random matrices, that are conjectured to provide the three generic universal scaling limits. In particular, we have shed some light on what characterises these different universality classes at the edge, compared to the bulk of the spectrum. 

In the first part we have analytically studied the complex spacing ratio in symmetry class A of the complex Ginibre ensemble (GinUE), for finite matrix size $N$. Introducing a conditional point process, we have clarified why an uncontrolled approximation used in the literature converges well. For the parameter-dependent elliptic Ginibre ensemble {(eGinUE)} which is Gaussian, we have provided a surmise for the complex spacing ratio for $N=3$ that interpolates towards the Gaussian unitary ensemble (GUE), where it approximates the large-$N$ limit in the bulk very well. {We have also expressed the NN spacing distribution in class A in terms of the Fredholm determinant of a conditional kernel, deriving its cubic vanishing at small argument.}

In the second part, based on extensive numerics we have investigated the edge universality classes of GinUE, complex symmetric (class AI$^\dag$) and complex self-dual (class AII$^\dag$) and compared them to two-dimensional (2D) Poisson statistics of uncorrelated points with a Gaussian distribution. Both for the complex spacing ratios, NN and NNN spacing distributions we have found a clear effect, distinguishing edge from bulk universality classes. This effect increases with the repulsion amongst eigenvalues, measured by an effective $\beta$-value of a 2D Coulomb gas description that was found to hold on local scales {previously} by part of the authors. Here, 2D Poisson corresponds to $\beta=0$ without repulsion, and class AI$^\dag$, A and AII$^\dag$ are ordered according to their increasing values of $\beta=1.4$, 2 and 2.6, respectively.
{For small argument however, the NN spacing distributions of all three classes were shown to be consistent with a universal cubic vanishing, independently of the repulsion, both in the bulk and at the edge.}

For the NN and NNN spacing distribution we provided an unfolding procedure at the edge of the spectrum for radially symmetric densities. Whilst this showed that for 2D Poisson the bulk and edge statistics remain identical, a clear distinction remained between bulk and edge distributions for all 3 symmetry classes. Because for the complex spacing ratio we found a difference also for 2D Poisson, this raises the question if the complex spacing ratio does provide properly unfolded quantities when the density varies on the scale of the mean level spacing, as it happens at the edge. Clearly, further research is needed in this direction. For example at the origin, where products of $m$ non-Hermitian random matrices of class A display a strongly varying density, with distinct universality classes labelled by $m$, it is open if taking a complex spacing ratio will lead to a fully unfolded quantity. 

A further open task is to derive more analytical results for classes AI$^\dag$ and AII$^\dag$, for example for the NN spacing in the bulk, or the distribution of the largest eigenvalue in radius at the edge, given recently developed tools. 
In view of the interpolation between the class A and the GUE that is provided by the eGinUE, and that leads to a transition between the Tracy--Widom and Gumbel distribution for the largest eigenvalue, this will be very interesting to achieve for the other symmetry classes.

\section*{Acknowledgments}
This work was partly funded by the German Research Foundation DFG with grant SFB 1283/2 2021 -- 317210226 (G. Akemann, N. Ayg\"un, P. P\"a{\ss}ler).

\begin{appendix}

\section{Analytical results at the edge for NN spacing in class A}\label{App:kernel}

In this appendix we present the analytical framework for computing the NN spacing distribution in class A at the edge of the spectrum. 
Being a determinantal point process (DPP) we review some of its properties. In particular, we express the gap probability to have an eigenvalue at the edge of the spectrum and its NN at radial distance $s$. It is given by the Fredholm determinant of the conditional kernel,  from which the NN spacing distribution follows by differentiation. 
{This allows us to derive its cubic behaviour at small argument, which is confirmed by the numerics in the main text, see Fig. \ref{fig:small} bottom middle plots.}

The point process in the complex plane of class A, given by the joint density \eqref{eq.jpdf_gen.weight} at $\beta=2$, represents a DPP given as follows:
\begin{equation}
    \jpdf^{(N)}(z_1,\dots,z_N) = \det_{1\leq i,j\leq N}\left[ K_{\rm A}^{(N)}(z_i,\bar{z}_j)\right].
\label{def.dpp}
\end{equation}
The reproducing kernel of planar orthogonal polynomials $P_k(z)$ with respect to the weight function $\omega(z)$, satisfying $\int \dif^2 z\, \omega(z) P_k(z){P_l(\bar{z})}=\delta_{k,l}$, is defined as 
\begin{equation}
K_{\rm A}^{(N)}(z,\bar{v})=\sqrt{\omega(z)\omega(\overline{u})}\sum_{j=0}^{N-1}P_j(z)P_j(\bar{u})\ .
\label{def.kernel}
\end{equation}
The kernel is unique only up to cocycles, $K_{\rm A}^{(N)}(z,\bar{v})\to\frac{f(z)}{f(\bar{u})}K_{\rm A}^{(N)}(z,\bar{v})$, for non-vanishing functions $f(z)$, as they  cancel after taking them out of the determinant in \eqref{def.dpp}. 
Examples include the GinUE and its kernel with respect to $\omega(z)=\exp[-|z|^2]$, where the orthogonal polynomials are monomials:
\begin{equation}
K_{\rm GinUE}^{(N)}(z,\overline{u}) = \frac{1}{\pi} e^{-\frac{1}{2}(|z|^2 + |u|^2)}
\sum_{j=0}^{N-1} \frac{(z \overline{u})^j}{j!} \ .
\label{def.KGin}
\end{equation}
A second example is the corresponding expression for the eGinUE \eqref{eq.weight-eGinUE}, in terms of planar Hermite polynomials, cf. \cite{PdF}. 
In DPP all $k$-point density correlation functions (marginals) are given by the $k\times k$ determinant of the corresponding kernel. In the simplest case $k=1$ we have for the density
\begin{equation}
R_{\rm A}^{(N)}(z) = K_{\rm A}^{(N)}(z,\bar{z}), 
\end{equation}
which is normalised to $N$ here. In the case of the GinUE it reads
\begin{equation}
R_{\rm GinUE}^{(N)}(z) = \frac{1}{\pi} e^{-|z|^2}\esum{N-1}{|z|^2}= \frac{1}{\pi}  Q(N, |z|^2)\ .
\label{eq.R1GinUE}
\end{equation}
In order to converge to the circular law, as shown in Fig. \ref{fig:radial density plot}, we have to rescale the eigenvalues $z\to z\sqrt{N}$. 
For later convenience we defined the truncated exponential $\esum{N}{x}$ inside {\eqref{eq.R1GinUE}} as
\begin{equation}
\esum{N}{x}=\sum_{j=0}^{N} \frac{x^j}{j!} = \euler^x Q(N, x)\ ,
\label{eq.eNQ}
\end{equation}
which can be expressed in {terms} of the normalised upper incomplete Gamma-function $Q(N, x) = \frac{\Gamma(N, x)}{\Gamma(N)}$. This will become convenient when considering asymptotics later.

Let us turn to conditional point processes. As in the derivation of the NN spacing distribution in the bulk \cite{GHS}, or in the complex spacing ratio \eqref{def.dist_ofCondSpacingRatio}, it turns out to be convenient to condition a point to be in the bulk or at the edge. In view of the form of the Vandermonde determinant \eqref{def.vandermonde}, conditioning point $z_1$ to be fixed, while integrating over all remaining points, leads to the following split of variables:
\begin{equation}
|\Delta_{N}(z_1,z_2,\ldots,z_N)|^2 = \left( \prod_{j=2}^N(z_j-z_1)(\overline{z_j}-\overline{z_1})\right)|\Delta_{N-1}(z_2,z_3,\ldots,z_N)|^2.
\end{equation}
This corresponds to the insertion of two complex conjugate characteristic polynomials at argument $z_1$ respectively $\overline{z_1}$, or a modified weight function 
$\omega(z)\to (z-z_1)(\overline{z}-\overline{z_1})\omega(z)$. 
As shown in this example for a more general result, \cite[eq. (3.10)]{AkemannVernizzi2003}, the kernel of a DPP conditioned on one eigenvalue $z_0$ is again a DPP with conditional kernel $K_{\rm A,C}^{(N)}(z,\bar{u})$ given by 
\begin{equation}
    \label{eq:KernelOneInsertion}
    K_{\rm A,C}^{(N)}(z,\bar{u}) =
    \frac{|z-z_0| |u-z_0|}{ (z-z_0)(\overline{u}-\overline{z_0}) }
    \frac{\begin{vmatrix}
        K_{\rm A}^{(N+1)}(z,\bar{u}) & K_{\rm A }^{(N+1)}(z,\overline{z_0}) \\
        K_{\rm A}^{(N+1)}(z_0,\bar{u}) & K_{\rm A}^{(N+1)}(z_0,\overline{z_0})
     \end{vmatrix}}{K_{\rm A}^{(N+1)}(z_0,\overline{z_0})}.
\end{equation}
Compared to \cite{AkemannVernizzi2003} we have multiplied the weight functions into the determinant. 
In particular, the conditional kernel can be expressed in {terms} of the unconditional one, of one degree  higher. For simplicity we drop the dependence on the conditioning 
point $z_0$ of the conditional kernel. This relation is a more general feature, valid when inserting more characteristic polynomials, or equivalently conditioning on more points, see \cite{AkemannVernizzi2003}.

In the case of the GinUE, we obtain from \eqref{eq:KernelOneInsertion} for the conditional kernel
\begin{equation}
    K_{{\rm GinUE, C}}^{(N)}(z,\bar{u}) = \frac{|z-z_0||u-z_0|e^{-\frac{1}{2}(|z|^2+|u|^2)}}{\pi (z_0-z)(\overline{z_0} - \overline{u}) \esum{N+1}{|z_0|^2}} \big(\esum{N+1}{|z_0|^2} \esum{N+1}{z \overline{u}} - \esum{N+1}{z_0 \overline{u}} \esum{N+1}{z \overline{z_0}} \big).
\end{equation}
When investigating the local NN spacing around the conditioning point $z_0$, it will be useful to change variables to $z = z_0 + \xi_1$, $u = z_0 + \xi_2$ with $\xi_1, \xi_2 \in \Cset$. Inserting \eqref{eq.eNQ}, the kernel becomes
\begin{eqnarray}
 &&   K_{{\rm GinUE, C}}^{(N)}(z_0 + \xi_1, \overline{z_0} + \overline{\xi_2}) = \frac{e^{-\frac{1}{2}(|z_0|^2 + z_0 \overline{\xi_1} + \overline{z_0} \xi_1 + |\xi_1|^2 + |z_0|^2 + z_0 \overline{\xi_2} + \overline{z_0} \xi_2 + |\xi_2|^2)}|\xi_1||\xi_2|}{\pi \xi_1 \overline{\xi_2} e^{|z_0|^2} Q(N+1, |z_0|^2)} 
 \nonumber\\
        &&\quad\quad\quad\quad  \times \left( e^{|z_0|^2}Q(N+1,|z_0|^2) e^{|z_0|^2 + z_0 \overline{\xi_2} + \overline{z_0} \xi_1 + \xi_1 \overline{\xi_2}} Q(N+1, |z_0|^2 + z_0 \overline{\xi_2} + \overline{z_0} \xi_1 + \xi_1 \overline{\xi_2}) \right.
        \nonumber\\
        &&\quad\quad\quad\quad  \left. \ \ \ \ - e^{|z_0|^2 + z_0 \overline{\xi_2}} Q(N+1, |z_0|^2 + z_0 \overline{\xi_2}) e^{|z_0|^2 + \overline{z_0} \xi_1} Q(N+1, |z_0|^2 + \overline{z_0} \xi_1) \right)
        \nonumber\\
    &&\quad\quad\quad\quad  \sim \frac{e^{-\frac{1}{2}(|\xi_1|^2+|\xi_2|^2)}}{\pi Q(N+1, |z_0|^2)} 
        \left(e^{\xi_1 \overline{\xi_2}} Q(N+1,|z_0|^2) Q(N+1, |z_0|^2 + z_0 \overline{\xi_2} + \overline{z_0} \xi_1 + \xi_1 \overline{\xi_2})\right. \nonumber\\
        &&\quad\quad\quad\quad  \left. \ \ \ \ - Q(N+1, |z_0|^2 + z_0 \overline{\xi_2}) Q(N+1, |z_0|^2 + \overline{z_0} \xi_1)\right).
 \label{eq:CondGinKernelFinite}
\end{eqnarray}
In the second step we have simplified the expressions and dropped cocycles, indicated by $\sim$.

In the following we are interested in the limiting behaviour at the spectral edge. First, in order to obtain a compactly supported density, we have to rescale $z_0\to z_0\sqrt{N}$. In the limit $N \to \infty$ the edge is then located at $|z_0|=1$. Thanks to the rotational symmetry of the global density, it suffices to 
consider the edge on the real line, $z_0=1$. In \cite{TaoVu} the unconditioned limiting kernel was derived for a general edge point. Here, we will directly derive the conditional kernel, and restrict ourselves to 
\begin{equation}
z_0 = \sqrt{N} + d\ ,
\label{eq.z0d} 
\end{equation}
for an eigenvalue just outside (positive $d$) or inside (negative $d$) the limiting support.
A useful asymptotic for the normalised incomplete Gamma-function follows from \cite[8.11(iv)]{NISTHandbook}
\begin{equation}
    Q(N+1,N+\sqrt{2N}y) \sim \frac{1}{2} \erfc(y) + O\left(\frac{1}{\sqrt{N}}\right).
\end{equation}

Applying this to \eqref{eq:CondGinKernelFinite} we get the conditional kernel at the edge \eqref{eq.z0d} in the large $N$ limit:
\begin{equation}
    \label{eq:CondGinKernelEdge}
    K_{\rm GinUE, C}^{\rm edge}(\xi_1, \overline{\xi_2}) = \frac{e^{-\frac{1}{2}(|\xi_1|^2 + |\xi_2|^2)}}{2 \pi} 
    \left(e^{\xi_1 \overline{\xi_2}} 
    \erfc\left[\frac{\xi_1 + \overline{\xi_2}}{\sqrt{2}} + \sqrt{2} d\right] 
    - \frac{\erfc\left[\frac{\xi_1}{\sqrt{2}} + \sqrt{2} d\right]\erfc\left[\frac{\overline{\xi_2}}{\sqrt{2}} + \sqrt{2} d\right]}{\erfc[{\sqrt{2} d}]}
    \right).
\end{equation}

We turn to the definition of the gap probability in a DPP conditioned to have an eigenvalue at $z_0$, that the disc ${D}_s(z_0)$ of radius $s$ centered at an eigenvalue $z_0$ is empty of eigenvalues.
It is given by the following Fredholm determinant expansion
\begin{equation}
 \label{eq:FredholmExpansion}
    E_{z_0}^{(N)}(s) = 1 + \sum_{l=1}^{N} \frac{(-1)^l}{l!} \int_{\mathbb{D}_s(z_0)^l} \dif^2 z_{1} \dots \dif^2 z_{l} 
 \det_{1\leq i,j\leq l}\left[  K_{\rm A,C}^{(N)}(z_i,\bar{z_j})\right], 
\end{equation}
containing the $l$-point density correlation functions, that we directly give in 
terms of the kernel. 
 The $l=1$ term of this expansion depends 
only on the integral of the density $K_{\rm A,C}^{(N)}(z_1,\overline{z_1})$.
There are two ways to compute the Fredholm determinant \eqref{eq:FredholmExpansion}. When rewriting the determinant of the kernel as an integral operator over $D_s(z_0)$, choosing $z_0=0$ in the bulk, the determinant becomes diagonal, given explicitly by the product of its $N$ eigenvalues. This leads to \eqref{GinibreSpacing}, after differentiation. As was noted already in \cite{GHS} this expansion converges slowly, with $N=20$ giving a good approximation, whilst a surmise $N=2$ does not work. 
A second possibility {is} to use \eqref{eq:FredholmExpansion} as an expansion in $l$. In the bulk of the spectrum this works already quite well for $l=1,2$, however, this is not true at the edge. Because the integrals over higher order $l$-point functions rapidly become intractable, including $l=2$ already, we will be more modest here. It turns out that for the small-argument expansion in $s$, the first term is already sufficient. We will compute it below as a function of the distance to the edge $d$. The resulting quartic behaviour of $E_0(s)$ (cubic for $p_{\rm A,C, NN}(s)$) is compared with numerics for the small-$s$ NN spacing distribution in Subsection \ref{sec:NNcubic}, {see Fig. \ref{fig:small} bottom middle.}

 Let us denote the $l=1$ term in \eqref{eq:FredholmExpansion} by $I_1(s)$, after taking the large-$N$ limit at the edge and using the conditional edge kernel \eqref{eq:CondGinKernelEdge}. The first order Fredholm expansion of $E_0(s)$ is $1 - I_1(s)$, and in polar coordinates we have 
\begin{equation}
\begin{aligned}
    \label{eq:CondGinEdgeFred1}
    I_1(s) & = \int_0^s \dif{r} \, r \int_{-\pi}^{\pi} \dif{\phi} \, K_{\rm GinUE, C}^{\rm edge}(r e^{\iunit \phi}, r e^{-i \phi}) \\
    & = \int_0^s \dif{r} \, r \int_{-\pi}^{\pi} \frac{\dif{\phi}}{2 \pi} 
\left[  1  - 
  \erf\left(\sqrt{2} (r \cos \phi + d)\right)
        - \frac{e^{-r^2}}{\erfc(\sqrt{2}d)} \left( 
1         
        - 
        \erf\left(\frac{r}{\sqrt{2}}e^{\iunit  \phi} + \sqrt{2} d\right)
        \right.\right.\\
    &\hspace{100pt} - 
 \left.\left.   \erf\left(\frac{r}{\sqrt{2}}e^{-\iunit \phi} + \sqrt{2} d\right)
        + 
        \erf\left(\frac{r}{\sqrt{2}}e^{\iunit \phi} + \sqrt{2} d\right) \erf\left(\frac{r}{\sqrt{2}}e^{-\iunit \phi} + \sqrt{2} d\right)
        \right) \right].
\end{aligned}
\end{equation}
Here, we wrote $\erfc(x)=1-\erf(x)$, to use the Taylor series of the entire function $\erf(x)$ \cite[eq. 7.6.1]{NISTHandbook},
\begin{equation}
    \label{eq:ErfTaylor}
    \erf(x) = \frac{2}{\sqrt{\pi}} \sum_{k=0}^{\infty} \frac{(-1)^k x^{2k+1}}{k! (2k+1)},
\end{equation}
to evaluate the terms of \eqref{eq:CondGinEdgeFred1}. While the first term in \eqref{eq:CondGinEdgeFred1} trivially gives $s^2/2$, for the second term, we can do a binomial expansion in the Taylor series and rearrange to get
\begin{equation}
    \erf\left(\sqrt{2} (r \cos \phi + d)\right) = \sqrt{\frac{8}{\pi}} \sum_{m=0}^{\infty} \left(\frac{r \cos \phi}{d}\right)^m \sum_{k=\floor{m/2}}^{\infty} \frac{(-2)^k}{k!(2k+1)} \binom{2k+1}{m} d^{2k+1}.
\end{equation}
The terms with odd $m$ vanish by integration over $\phi$. We rescale the even indices $m \to m/2$, and using $\int_{-\pi}^{\pi} \frac{\dif \phi}{2\pi} (\cos \phi)^{2m} = \frac{1}{4^m} \binom{2m}{m}$,
and some rearrangement of the sum over $k$, to get a confluent Hypergeometric function, we have
\begin{equation}
    \label{eq:PhiIntFrakB}
    \int_{-\pi}^{\pi} \frac{\dif \phi}{2\pi} \erf\left(\sqrt{2} (r \cos \phi + d)\right)
    = 4 d \sqrt{2 \pi} \sum_{m=0}^{\infty} \frac{(-1)^m r^{2m}}{2^m m!} \binom{2m}{m} {}_1F_1\left(\substack{1/2 + m \\ 3/2} ; -2d^2\right).
\end{equation}
For the remaining radial integration we may use
\begin{equation}
    \int_0^s \dif r \, e^{-r^2} r^{2m+1} = \frac{1}{2} \gamma(m+1,s^2),\ m\geq0, 
    \label{eq.radialrem}
\end{equation}
where $\gamma(n,x)$ is the lower incomplete Gamma-function.

The third term in the first line of \eqref{eq:CondGinEdgeFred1} is again trivial, leading to $\euler^{-s^2}/(2 \erfc(\sqrt{2}\,d))$.
The angular integration of the next two terms can be done in parallel. In each term of the Taylor series \eqref{eq:ErfTaylor},
terms containing positive powers of $e^{\pm \iunit\phi}$ drop out,
\begin{equation}
    \label{eq:PhiIntFrakD}
    \int_{-\pi}^{\pi} \frac{\dif \phi}{2\pi}
    \erf\left(\frac{r}{\sqrt{2}}\euler^{\pm\iunit  \phi} + \sqrt{2} d\right)
    = \erf(\sqrt{2}d).
\end{equation}
Because this is radius-independent, the remaining radial integral is then the same as in the previous term.
For the last term in the last line of \eqref{eq:CondGinEdgeFred1}, 
we Taylor expand with \eqref{eq:ErfTaylor} twice, then we expand the binomials and collect terms of equal powers of 
$\euler^{i \phi}$ and $\euler^{-i \phi}$. Using the notation
\begin{equation}
    \delta_m = \begin{cases} 0\text{ for }m\text{ even} \\ 1\text{ for }m\text{ odd}\end{cases},
\end{equation}
after some more calculation we get for the angular integral
\begin{eqnarray}
    &&\int_{-\pi}^{\pi} \frac{\dif \phi}{2\pi}
     \erf\left(\frac{r}{\sqrt{2}}\euler^{\iunit \phi} + \sqrt{2} d\right) \erf\left(\frac{r}{\sqrt{2}}\euler^{-\iunit \phi} + \sqrt{2} d\right)\nonumber\\
    &&= \frac{2}{\pi} \sum_{m=0}^{\infty} r^{2m} \frac{(2d)^{2(1-\delta_m)}}{m!^2} ((m - \delta_m - 1)!!)^2 {}_1F_1\left(\substack{1/2+\floor{m/2} \\ 3/2 - \delta_m}; -2d^2 \right)^2.
        \label{eq:PhiIntFrakF}
\end{eqnarray}
The remaining radial integral follows again from \eqref{eq.radialrem}.
Putting together \eqref{eq:PhiIntFrakB}, \eqref{eq:PhiIntFrakD}, \eqref{eq:PhiIntFrakF} and the other terms of \eqref{eq:CondGinEdgeFred1}, we get from the first 
{order $l=1$} 
of the Fredholm determinant expansion,
\begin{equation}
\begin{aligned}
    \label{eq:GinUEGapProbFred1}
    E_0(s) \approx& \ 1 - \frac{s^2}{2} - (e^{-s^2} - 1) \Big(1 - \frac{1}{2 \erfc(\sqrt{2}d)}\Big) 
    + \frac{2d}{\sqrt{2 \pi}} \sum_{m=0}^{\infty} \frac{s^{2(m+1)}}{(m+1)!} \Big(-\frac{1}{2}\Big)^m \binom{2m}{m} \, {}_1F_1(\substack{1/2+m \\ 3/2}; -2 d^2) \\
    & + \sum_{m=0}^{\infty} \frac{\gamma(m+1,s^2)}{\pi \erfc(\sqrt{2}d)} \frac{(2d)^{2(1-\delta_m)}}{m!^2} ((m-\delta_m - 1)!!)^2 {}_1F_1\left(\substack{1/2 + \floor{m/2} \\ 3/2 - \delta_m}; -2 d^2\right)^2.
\end{aligned}
\end{equation}
If the eigenvalue $z_0$ is exactly at the edge, that is $d=0$, above expression simplifies to
\begin{equation}
    \label{eq:GinUEGapProbFred1Simple}
    E_0(s)|_{d=0} \approx 1 -\frac{s^2+\euler^{-s^2}-1}{2} + \frac{1}{\pi} \sum_{n=0}^{\infty} \frac{\gamma(2n+2,s^2)}{(2^n n! (2n+1))^2}.
\end{equation}
If the full Fredholm expansion \eqref{eq:FredholmExpansion} is differentiated with respect to $s$ and evaluated at $s=0$ for the Taylor series,
notice that for the $k$-th term of the Fredholm expansion, at least the $k$ radial integrals from 0 to $s$ have to be eliminated by differentiation to give a nonzero contribution.
Furthermore, for the $j$-th differentiated integral, the $r_j$ from the two-dimensional area measure becomes $s$, so the resulting factor $s^k$ has to be differentiated as well, raising the requirement to
$2k$ differentiations for a nonzero contribution. Then, for a nonzero determinant, each row in \eqref{eq:FredholmExpansion} has to be differentiated a different number of times in the left kernel argument,
adding $0+1+\dots+(k-1)=\frac{1}{2}k(k-1)$ more differentiations. The differentiation of the $j$-th row results in a phase $\euler^{i \phi_j}$ which makes the angular
integration yield 0; to get a nonzero contribution, these phases must be cancelled by differentiations of columns with respect to the right kernel argument, adding another
$\frac{1}{2}k(k-1)$ differentiations. In total, the $k$-th term of the Fredholm expansion cannot contribute to the coefficient of $s^n$ for $n < 2k +2 \frac{k}{2}(k-1)=k(k+1)$.

For this reason, the $s^4$ coefficient comes only from the first order term $k=1$ of the Fredholm expansion that we computed explicitly.
Thus we can extract the coefficient from \eqref{eq:GinUEGapProbFred1} to get
\begin{equation}
    E_0(s) = 1 - \Big(\frac{\erfc(\sqrt{2}d)}{4} - \frac{\euler^{-4d^2}}{2 \pi \erfc(\sqrt{2}d)} + \frac{d}{\sqrt{2 \pi}} \euler^{-2d^2}\Big) s^4 + O(s^6).
\end{equation}
At $d=0$, this simplifies to
\begin{equation}
\label{eq:Esmall-s-A-edge}
    E_0(s)\Big|_{d=0} = 1 - \frac{\pi-2}{4 \pi} s^4 + O(s^6),
\end{equation}
which has to be compared to the expansion of the gap probability in the bulk region $1 - \frac{1}{2} s^4 + O(s^6)$,
obtained by expanding and integrating \eqref{GinibreSpacing}.

Let us add a comment why we cannot compare the analytically determined $d-$dependent coefficient of the resulting cubic term for the NN spacing distribution $p_{\rm NN}^{\rm edge}(s)$. In \eqref{def.1stmom} we have agreed to rescale its first moment to unity, in order to set the scale in the comparison with the data. This will change the coefficient multiplying the cubic term $\sim s^3$, and since we do not know the full limiting distribution we do not have an analytical prediction for its coefficient.

\end{appendix}


\end{document}